\documentclass{article}
\usepackage{Pegase}
\begin{document}
\begin{center}
  \Huge\bfseries
  The \textsc{Pégase} code
  \\
  of spectrochemical evolution
  \\
  of galaxies
  \\[\baselineskip]
  \LARGE\mdseries
  Version~3{\normalsize.0.0}
  \\
  \vfill
  \centerline{\includegraphics{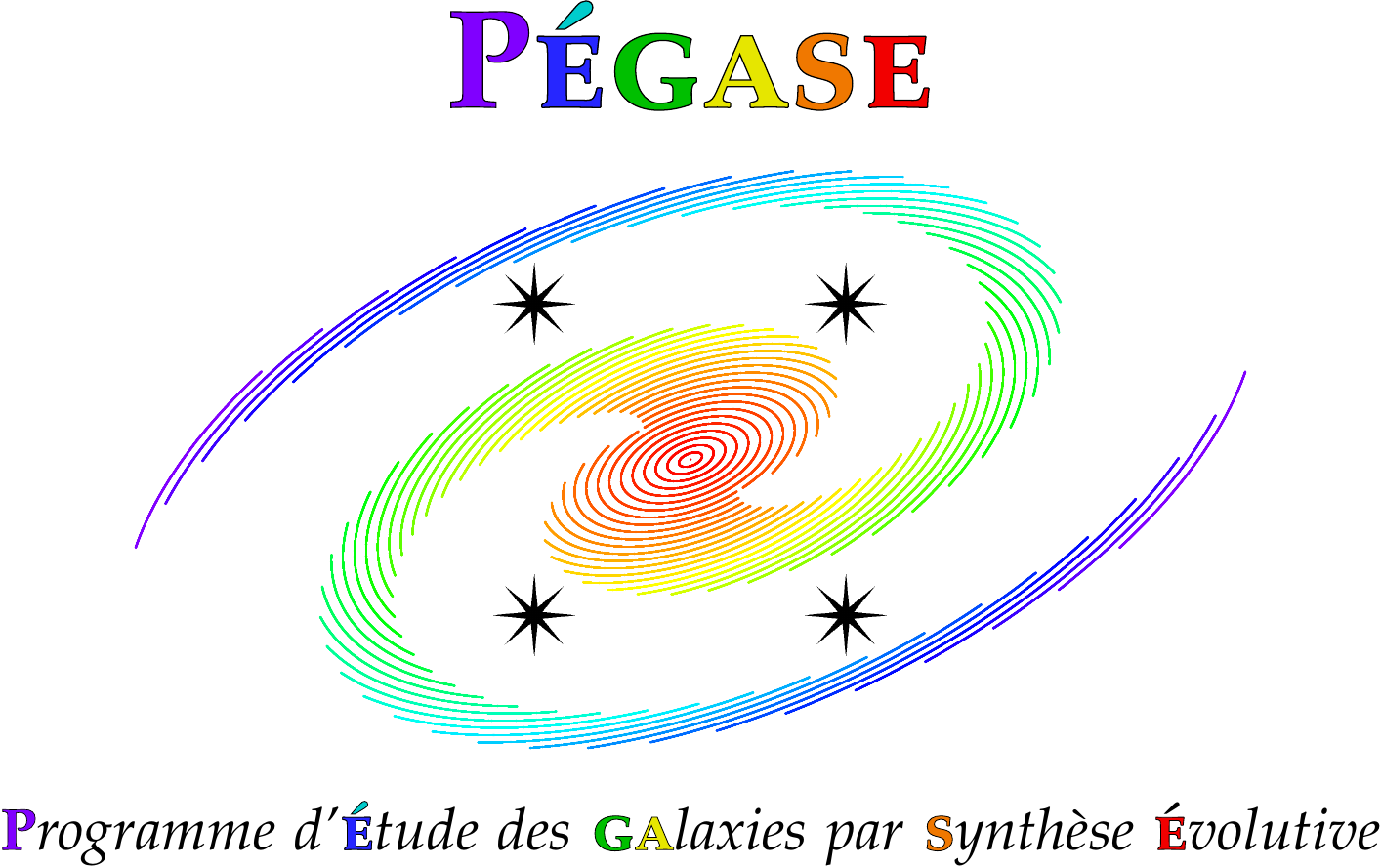}}
  \vfill
  \LARGE{\bfseries
  Documentation and complements}
  \\[\baselineskip]
  \Large
  Michel \textsc{Fioc}%
  \address{\textreferencemark}{\email{Michel.Fioc@iap.fr}.
    Institut d'astrophysique de Paris \&
    Sorbonne université, France.
  }
  \&
  Brigitte \textsc{Rocca}-\textsc{Volmerange}%
  \address{\textreferencemark\kern1pt\textreferencemark}{\email{Brigitte.Rocca@iap.fr}.
    Institut d'astrophysique de Paris \& université Paris-sud, France.}
  \\[\baselineskip]
  2019-02-06
\end{center}
%
\startcontents[sommaire]
\stopcontents[sommaire]
\thispagestyle{empty}
\clearpage
\addcontentsline{toc}{section}{Short table of contents}
\printcontents[sommaire]{sommaire}{0}
{%
  \setcounter{tocdepth}{1}%
}
\setcounter{tocdepth}{5}
\vfill
\resumecontents[sommaire]
\section*{Purpose of the code,
  reference, license and contact}
\addcontentsline{toc}{section}{Purpose of the code,
  reference, license and contact}
\Pegase.3%
\footnote{%
  Short form for \mention{version~3 of \Pegase}.
  \mention{\Pegase\unskip} is a French acronym for
  \mention{\textbf{P}\emph{rogramme d'\textbf{\textsc{é}}tude
      des \textbf{\textsc{ga}}laxies par \textbf{\textsc{s}}ynthèse
      $\mathrm{\acute{\textbf{\textsc{e}}}}$volutive}}
  (\mention{Program for the study of galaxies by evolutionary synthesis} in English;
  formerly, the \mention{P} was for \mention{\textbf{P}\emph{rojet}}\slash
  \mention{Project}).
  The acronym is pronounced [pegaz] in International Phonetic Alphabet
  transcription.

  The recommended spelling is \curlyDOQ\verb+\textsc{P\'egase}+\curlyDCQ\ (in \LaTeXe\ form)
  or, if small capitals are not available, \curlyDOQ\verb+P\'egase+\curlyDCQ.
  In contexts where diacritics are problematic (\eg in filenames),
  the code's name may be written
  \curlyDOQ\verb+Pegase+\curlyDCQ, \curlyDOQ\verb+PEGASE+\curlyDCQ 
  or \curlyDOQ\verb+pegase+\curlyDCQ, in order of decreasing preference.

  \mention{Pégase} is also the French name of \emph{Pegasus},
  a constellation dominated by an easily recognizable asterism,
  the Great Square (hence the four stars in the logo).
}
is a Fortran~95 code modeling the spectral evolution of galaxies
from the far-ultraviolet to submillimeter wavelengths.
It also follows the chemical evolution of their stars, gas and dust.
\medskip

The present document is the user's guide of the code.
It also provides complements to the article
\begin{center}
  {\bfseries\itshape
    \citet{article_Pegase.3}}, in press in \emph{Astronomy and Astrophysics}
\end{center}
(denoted by \mention{\AApaper}
hereafter), which should be cited when referring to \Pegase.3.
\medskip

This code is distributed under version~2.1 of the CeCILL license.
English and French versions of the latter are provided with this documentation in directory \codefile{doc_dir/} (see \fullref{sec:instal}) and are also available at
\webaddress{https://cecill.info}.
In practice, this license is equivalent to the \textsc{Gnu} General Public License and is compatible with it.
\medskip

To be informed of future developments of the code, ask questions
or make comments or suggestions, contact Michel Fioc at
\[
  \email{pegase@iap.fr}.
\]
\section*{Overview}
\addcontentsline{toc}{section}{Overview}
For a given \emph{scenario}, \ie a set of parameters defining the
history of mass assembly, the star formation law, the initial mass
function\textellipsis, \Pegase.3
consistently computes the following:
\begin{itemize}
\item
  \noUCase the star formation, infall, outflow and supernova rates from $0$ to $20$~Gyr;
\item
  \noUCase the stellar metallicity, the abundances of main elements in the gas and the composition of dust;
\item
  \noUCase   the unattenuated stellar spectral energy distribution (SED);
\item
  \noUCase the nebular SED, using nebular continua and emission lines precomputed
  with code \textsc{Cloudy} \citep{Cloudy};
\item
  \noUCase the attenuation in star-forming clouds and the diffuse interstellar medium, 
  by absorption and scattering on dust grains, of the stellar and nebular SEDs. 
  For this, the code uses grids of the transmittance for spiral and spheroidal galaxies.
  We precomputed these grids through Monte Carlo simulations of radiative transfer 
  based on the method of virtual interactions;
\item
  \noUCase   the re-emission by grains of the light they absorbed, taking into account
  stochastic heating.
\end{itemize}

The main innovation compared to \Pegase.2 \citep{Pegase.2, Pegase.1}
is the modeling of dust emission and its evolution.
While version~2 of the code computed spectra from the far-ultraviolet to the
near-infrared%
\footnote{In the presence of dust, \Pegase.2 energy distributions were not significant at longer wavelengths.},
\Pegase.3 extends the wavelength range through the mid-~and far-infrared
up to the submillimetric domain.
The computation of nebular emission has also been entirely upgraded to take
into account metallicity effects and infrared lines.

Other major differences are that complex scenarios of evolution (derived
for instance from cosmological simulations), with several
episodes of star formation, infall or outflow, may now be implemented,
and that the detailed evolution of the most important elements \dashOpen not only the
overall metallicity\dashClos\ is followed.
More details are provided in \AApaper and \fullref*{sec:modeling} (p.~\pageref{sec:modeling} \emph{et seq.}).
%
\section{Getting started}
\subsection{Installation%
  \label{sec:instal}}
The code is freely available at
\begin{center}
  \leavevmode
  \llap\guillemotleft
  \href{http://www2.iap.fr/users/fioc/Pegase/Pegase.3/}{{\ttfamily http://www2.iap.fr/users/fioc/Pegase/Pegase.3/}}%
  \guillemotright\footnote{See \fullref{app:typo}, for a description of the typographical
    conventions used in this documentation.}%
\end{center}
and
\webaddress[http://www2.iap.fr/pegase/]{www2.iap.fr/pegase/}.
Download file \mention*{\codefile{Pegase.3.0.0\-.tar.gz}} and type in a terminal%
\footnote{We assume hereafter that the user works with
  a command-line interface on a \textsc{Linux}/\textsc{Unix} operating system.}%
\begin{center}
  \mention{\shell{tar xvf Pegase.3.0.0.tar.gz}}%
\end{center}
in the directory where the archive has been saved.
This extracts all the files in directory \mention*{\codefile{Pegase.3.0.0/}}.

This directory contains the following subdirectories:
\begin{description}
\item[\codefile{ages_dir/}]
  files determining the ages at which executables print their outputs;
\item[\codefile{bin_dir/}]
  \codefile{Makefile}, executable files, log files and all the files created
  by the compilation;
\item[\codefile{calib_dir/}]
  filter passbands, spectra of reference stars
  and output file\index{calib.txt@\codefile{calib.txt}} of code \codefile{calib}\index{calib@\codefile{calib}};
\item[\codefile{Cloudy_dir/}]
  nebular continua and a selection of emission lines computed by \textsc{Cloudy} \citep{Cloudy}. 
  Emission lines are listed in file \codefile{list_neb_lines.txt};
\item[\codefile{colors_dir/}]
  files\index{file of colors} produced by code \codefile{colors}\index{colors@\codefile{colors}};
\item[\codefile{doc_dir/}]
  this documentation and French and English versions of the CeCILL license (in HTML and text format);
\item[\codefile{dust_dir/}]
  input files related to dust properties;
\item[\codefile{grain_SED_dir/}]
  files of grain SEDs%
  \index{file of grain SEDs}
  optionally produced by code \codefile{spectra}\index{spectra@\codefile{spectra}};
\item[\codefile{grain_temp_dir/}]
  files of grain temperatures%
  \index{file of grain temperatures}
  optionally produced by code \codefile{spectra}\index{spectra@\codefile{spectra}};
\item[\codefile{IMFs_dir/}]
  initial mass functions;
\item[\codefile{RT_dir/}]
  input files related to radiative transfer;
\item[\codefile{scenarios_dir/}]
  files%
  \index{file of scenarios}
  created by the user and containing the scenarios used as input by
  code \codefile{spectra}\index{spectra@\codefile{spectra}};
\item[\codefile{source_dir/}]
  Fortran source files specifically related to \Pegase;
\item[\codefile{spectra_dir/}]
  main output files%
  \index{file of spectra}
  produced by code \codefile{spectra}\index{spectra@\codefile{spectra}};
\item[\codefile{SSPs_dir/}]
  files%
  \index{files of SSPs}
  containing properties of single stellar populations
  and produced by code \codefile{SSPs}\index{SSPs@\codefile{SSPs}};
\item[\codefile{stel_lib_dir/}]
  libraries of stellar spectra;
\item[\codefile{tracks_dir/}]
  stellar evolutionary tracks;
\item[\codefile{util_dir/}]
  Fortran source files of general interest;
\item[\codefile{yields_dir/}]
  yields of elements produced by stars.
\end{description}

All the text and Fortran files are in UTF-8 format.
%
\subsection{Compilation}
\label{sec:compil}
\subsubsection{Executables and their input and output data files}
Go in subdirectory \mention*{\codefile{bin_dir/}} of \codefile{Pegase.3.0.0/}
and type \mention{\shell{make}} in a terminal to
compile the code according to the instructions given in file
\mention*{\codefile{Makefile}}%
\footnote{%
  This file is written for the \textsc{Gnu}~\shell{make} command.%
}.
This creates the following executable files in \mention*{\codefile{bin_dir/}}:
\begin{description}
\item[\codefile{SSPs}]\index{SSPs@\codefile{SSPs}}
  code computing several evolving properties of single stellar populations (\mention{SSPs})
  \dashOpen but not their spectral energy distribution.
  \codefile{SSPs}\index{SSPs@\codefile{SSPs}} creates a file, called hereafter a
  \mention*{\Emph{set of SSPs}}\index{set of SSPs}, listing the names of the other output files
  produced during the run;
\item[\codefile{spectra}]\index{spectra@\codefile{spectra}}
  code processing one or more scenarios of galaxy evolution and computing,
  for each of them,
  the SED and various other properties of the modeled galaxy as a function of time.

  For each scenario, the main output of \codefile{spectra}\index{spectra@\codefile{spectra}} is a \mention*{\Emph{file of spectra}}\index{file of spectra}
  containing aforementioned data.
  Optionally, one or two auxiliary files are also produced for stochastically heated dust grains:
  \begin{itemize}
  \item
    \noUCase a \mention*{\Emph{file of grain temperatures}}\index{file of grain temperatures}, which provides the temperature probability distribution of individual grains of various species and sizes;
  \item
    \noUCase a \mention*{\Emph{file of grain SEDs}}\index{file of grain SEDs}, which contains their SEDs.
  \end{itemize}

  As input, \codefile{spectra}\index{spectra@\codefile{spectra}} needs a \mention*{\Emph{file of scenarios}}\index{file of scenarios}
  written by the user and defining, for each of the scenarios it contains,
  the parameters of the modeled galaxy%
  \footnote{This file points in particular to a set of SSPs\index{set of SSPs}.};
\item[\codefile{calib}]\index{calib@\codefile{calib}}
  code computing several properties for a series of filters, in particular
  the in-band fluxes of reference stars used to calibrate magnitudes;
\item[\codefile{colors}]\index{colors@\codefile{colors}}
  code computing the colors and other properties of a galaxy for a given scenario of evolution.
  The output is written in a \mention*{\Emph{file of colors}}\index{file of colors}.

  Input: file of spectra\index{file of spectra} produced by \codefile{spectra}
  for this scenario.
  Code \codefile{colors}\index{colors@\codefile{colors}} also uses the output file of \codefile{calib}\index{calib@\codefile{calib}};
\item[\codefile{plot_spectra}]\index{plot_spectra@\codefile{plot_spectra}}
  code plotting the evolving SED of a galaxy for one or more scenarios.

  Inputs: files of spectra\index{file of spectra} produced by \codefile{spectra} for these scenarios;
\item[\codefile{plot_grain_temp}]\index{plot_grain_temp@\codefile{plot_grain_temp}}
  code plotting the temperature probability distribution of stochastically heated dust grains for a
  given scenario.

  Input: file of grain temperatures\index{file of grain temperatures} produced by \codefile{spectra} for this scenario;
\item[\codefile{plot_grain_SED}]\index{plot_grain_SED@\codefile{plot_grain_SED}}
  code plotting the SED of individual grains for a given scenario.

  Input: file of grain SEDs\index{file of grain SEDs} produced by \codefile{spectra} for this scenario.
\end{description}
\subsubsection{Compiler}
The default compiler is \textsc{GFortran}.
To use \textsc{IFort} instead, comment the line
\mention{\codetext{COMPILER = gfortran}} in \codefile{Makefile}
(\ie, prefix this line with a \mention{\codetext{$\diese$}})
and uncomment the line \mention{\codetext{$\diese$ COMPILER = ifort}}.

To use another Fortran~95 compiler, comment the lines 
\mention{\codetext{COMPILER = gfortran}} and \mention{\codetext{COMPILER = ifort}},
add a line such as \mention{\codetext{COMPILER = \usertext{compil_command}}} and modify the
options of compilation (variable \mention{\codetext{OPTIONS}}) in accordance.
(Actually, the options defined for \textsc{GFortran} and \textsc{IFort}
are for debugging, and removing them might speed up the execution.)

If you have run \shell{make} for some Fortran compiler already and you want to
switch to another one, once you have changed \codetext{COMPILER} in \codefile{Makefile},
type first \mention{\shell{make clean}} in a terminal, then
\mention{\shell{make}}.
(The command \shell{make clean} deletes all the object and module files, 
as well as the executables; otherwise, \shell{make} will not recompile the code.)
\subsubsection{Graphical library for \codefile{plot_spectra}, \codefile{plot_grain_temp} and \codefile{plot_grain_SED}}
\index{plot_spectra@\codefile{plot_spectra}}%
\index{plot_grain_temp@\codefile{plot_grain_temp}}%
\index{plot_grain_SED@\codefile{plot_grain_SED}}%
The codes \mention*{\codefile{plot_spectra}}\index{plot_spectra@\codefile{plot_spectra}}, \mention*{\codefile{plot_grain_temp}}\index{plot_grain_temp@\codefile{plot_grain_temp}} and
\mention*{\codefile{plot_grain_SED}}\index{plot_grain_SED@\codefile{plot_grain_SED}} require the \mention*{\textsc{PGPlot}} graphical library
\citep{PGPLOT}.
The location of this library on the author's system is set by the variable
\mention*{\codetext{PGPLOT_LIB}} of \codefile{Makefile},
both for \textsc{IFort} and for \textsc{GFortran}.
To use \codefile{plot_spectra}\index{plot_spectra@\codefile{plot_spectra}}, \codefile{plot_grain_temp}\index{plot_grain_temp@\codefile{plot_grain_temp}} or \codefile{plot_grain_SED}\index{plot_grain_SED@\codefile{plot_grain_SED}} on your system,
you will most certainly have to redefine \codetext{PGPLOT_LIB}.
You may also replace \textsc{PGPlot} instructions (lines beginning with
\mention{\codetext{call pg}}) in \codefile{plot_spectra.f90}, \codefile{plot_grain_temp.f90}
or \codefile{plot_grain_SED.f90}
(all in directory \mention*{\codefile{source_dir/}})
by equivalents in your favorite graphical language.
\subsubsection{Home directory and directory separator}
To be able to refer to your home directory through the 
symbol \mention{\codetext{\homedir}}, you need to provide the path
from the root directory down to the home directory%
\footnote{Some compilers, \eg \textsc{GFortran}, do not associate
  \mention{\codetext{\homedir}} to the home directory.}.
For this, edit file \codefile{util_dir/mod_dir_access.f90}
and follow the instructions in this file to 
assign parameter \codetext{home_dir} to the correct value for this path.

On non-\textsc{Linux}/\textsc{Unix} systems, you may also need to change 
the value of the directory separator (parameter \codetext{dir_sep} in 
the same file.)

If you change any of these parameters, 
recompile the code with \shell{make}.
\subsubsection{Real numbers}
Except for plots with \textsc{PGPlot}, the range and precision used for real numbers are set by parameter \codetext{CDR}.
By default, \codetext{CDR} gives double precision reals.
If memory is missing, you may select single precision reals by replacing \mention{\codetext{DPR}} with \mention{\codetext{SPR}} 
in the statement \mention{\codetext{integer, parameter :: CDR = DPR}} of file \codefile{util_dir/mod_types.f90}.
Recompile the code with \shell{make} afterwards.
%
\subsection{Execution}
We recommend that, to become acquainted with the code, the user first run once
\codefile{SSPs}\index{SSPs@\codefile{SSPs}}, 
\codefile{spectra}\index{spectra@\codefile{spectra}} and 
\codefile{colors}\index{colors@\codefile{colors}} as in the examples below.
\subsubsection{Running \codefile{SSPs}}
\label{sec:run_SSPs}%
\index{SSPs@\codefile{SSPs}}%
\codefile{SSPs}\index{SSPs@\codefile{SSPs}} computes from stellar evolutionary tracks of various
metallicities the evolution of an SSP for a given initial mass function.
It calculates isochrones (\ie the locus of stars having the same age but different masses)
in the Hertzsprung \& Russell diagram and apportions them
among the elements of a library of stellar spectra.
\codefile{SSPs}\index{SSPs@\codefile{SSPs}} also computes from a set of stellar yields 
the ejection rate in the interstellar medium, by an SSP, of gas enriched in metals
(globally and for each main element) and of circumstellar dust, as well as
the number of supernovae and the mass locked in compact stellar remnants.
\begin{RENVOI}
  \renvoi\fullref{sec:SSPs_inputs}, for details on the inputs to \codefile{SSPs}\index{SSPs@\codefile{SSPs}}.
\end{RENVOI}

To run \codefile{SSPs}\index{SSPs@\codefile{SSPs}}, just type \mention{\shell{./SSPs}} 
in \codefile{bin_dir/} and answer the questions.

Here is a minimal example%
\footnote{The symbol \mention*{\character{\skipped}} denotes lines written on the terminal
  but skipped here;
  \mention*{\character{\return}} means that the \mention{\keystroke{Return}}\,/\,\mention{\keystroke{Enter}}
  key has been pressed, which selects the default answer;
  \mention*{\character{$\blacktriangleright$}} and \mention*{\character{$\blacktriangleleft$}} distinguish outputs to the terminal
  and inputs from the keyboard.
}:
\ScreenOutputBegin[;]{%
  \skipped\\Initial mass function? \skipped}
\KeyboardInput{\ignorespaces}
\ScreenOutput[;]{Lower mass of the IMF (in solar masses)? \skipped}
\KeyboardInput{\ignorespaces}
\ScreenOutput[;]{Upper mass of the IMF (in solar masses)? \skipped}
\KeyboardInput[;]{\ignorespaces}
\ScreenOutput[;]{\skipped\\Set of evolutionary tracks? \skipped}
\KeyboardInput[;]{\ignorespaces}
\ScreenOutput[;]{\skipped\\Set of yields for high-mass stars? \skipped}
\KeyboardInput[;]{\ignorespaces}
\ScreenOutput[;]{\skipped\\Set of stellar libraries? \skipped}
\KeyboardInput[;]{\ignorespaces}
\ScreenOutput[;]{\skipped\\File defining the output ages of code "SSPs"? \skipped}
\KeyboardInput[;]{\ignorespaces}
\ScreenOutput{%
  \skipped\\Identifier prefixed to the output files of code "SSPs"?}
(provide a string of characters. For the first run, we recommend to type
\mention{\shell{example}} because this is the identifier used in the file input
to \codefile{spectra} in Sec.~\ref{sec:run_spectra});
\KeyboardInputEnd[.]{example}

Once \codefile{SSPs}\index{SSPs@\codefile{SSPs}} has finished running, several output files\index{files of SSPs} have been created in \codefile{SSPs_dir/}.
The master file \dashOpen the \mention{set of SSPs}\index{set of SSPs}\dashClos is named \mention{\codefile{\usertext{prefix}_SSPs.txt}}, where \mention*{\usertext{prefix}}%
\footnote{We use an italic font to distinguish placeholders.}
is the value of the identifier requested in the last question above%
\footnote{Creating a specific directory would clutter less \codefile{SSPs_dir/},
  but it is impossible in standard Fortran~95 to do this from within the code.
  The Fortran~2008 \codetext{execute_command_line} subroutine 
  or system-specific procedures are needed for this; 
  we plan to implement them in future versions of \Pegase.
  Nonetheless, if the prefix ends with a directory separator 
  (symbol \mention{\codetext{/}} on \textsc{Linux}/\textsc{Unix} 
  or \mention{\codetext{\textbackslash}} on \textsc{Windows}), the code
  will ask you to create the directory by yourself;
  once done, il will write the files of SSPs (without the leading 
  \mention{\usertext{prefix_}} in their names) in this directory.}.
The set of SSPs\index{set of SSPs} (\mention*{\codefile{example_SSPs.txt}} in the example) lists the names of the associated output files\index{files of SSPs}
(one per file of stellar evolutionary tracks, each one corresponding to an initial metallicity;
their name begins also with \mention{\codetext{\usertext{prefix}_}}).
These files contain all the information needed by \codefile{spectra}\index{spectra@\codefile{spectra}}
to compute the evolving SED of a single stellar population at
any reasonable metallicity;
they are not meant to be read by the user, but are a prerequisite for \codefile{spectra}\index{spectra@\codefile{spectra}}.

All the sets of SSPs produced by \codefile{SSPs}\index{SSPs@\codefile{SSPs}} are listed in \codefile{SSPs_dir/list_SSPs_sets.txt}.
\subsubsection{Running \codefile{spectra}}
\label{sec:run_spectra}%
\index{spectra@\codefile{spectra}}%
For each galaxy, a scenario of evolution must be given.
Scenarios are read from a file containing the parameters of
one or more scenarios%
\footnote{%
  In \Pegase.2, this file was created by code \mention*{\codefile{scenarios}}
  through an interactive dialog.
  Because of the large number of parameters, this would have been very
  inconvenient in \Pegase.3, and the file of scenarios\index{file of scenarios} must now be written
  as a series of statements with a text editor.
  Although the user is less guided, it makes it much easier to correct mistakes
  and to update the file.
}.
These parameters are organized in trees; many have a default value.
\begin{RENVOI}
  \renvoi\fullref{sec:scenarios}, for a complete description of the contents of a file
  of scenarios\index{file of scenarios};
  \renvoi\fullref{app:trees_tables}, to consult the trees of parameters
  and the tables providing their type, possible values and, if any, their default value.
\end{RENVOI}

As an example, let us use the file \mention*{\codefile{example_scenarios.txt}}%
\footnote{%
  This file is available in \codefile{scenarios_dir/} and is described in
  \fullref{sec:example}.%
}
(this requires that \codefile{example_SSPs.txt} has first been
created by \codefile{SSPs}\index{SSPs@\codefile{SSPs}}; see \fullref*{sec:run_SSPs}).
Type \mention{\shell{./spectra}} in \codefile{bin_dir/}, and when asked
\ScreenOutputBegin[,]{Name of the file of scenarios? \skipped}
provide this name%
\footnote{%
  \label{fn:path}%
  If the file of scenarios\index{file of scenarios} is not in \codefile{scenarios_dir/}, either
  its path relative to \codefile{scenarios_dir/}
  or its absolute path must be given.

  The same holds for the files of spectra%
  \index{file of spectra} input to \codefile{colors}\index{colors@\codefile{colors}} and
  \codefile{plot_spectra}\index{plot_spectra@\codefile{plot_spectra}} with respect to
  directory \codefile{spectra_dir/},
  the files of grain temperatures and SEDs input to 
  \codefile{plot_grain_temp}\index{plot_grain_temp@\codefile{plot_grain_temp}}%
  \index{file of grain temperatures}
  and \codefile{plot_grain_SED}\index{plot_grain_SED@\codefile{plot_grain_SED}}%
  \index{file of grain SEDs}
  with respect to
  \codefile{grain_temp_dir/} and \codefile{grain_SED_dir/}, and for the output file%
  \index{file of colors}
  of \codefile{colors}\index{colors@\codefile{colors}} with respect
  to \codefile{colors_dir/}.
}:
\KeyboardInputEnd[.]{example_scenarios.txt}

For each scenario, \codefile{spectra}\index{spectra@\codefile{spectra}} creates a \mention{file of spectra}\index{file of spectra} 
containing, among other data, the SED of the galaxy as a function of time
(file \mention{\codefile{example_spectra1.txt}} for the first scenario in the example considered here).

If requested (see \fullref{sec:output_grain}), \codefile{spectra}\index{spectra@\codefile{spectra}} also outputs for each scenario the temperature probability distribution
and the spectral energy distribution of individual dust grains
in a \mention{file of grain temperatures}\index{file of grain temperatures}
and a \mention{file of grain SEDs}\index{file of grain SEDs},
respectively.

Unless another directory is specified, all the files of spectra (\resp grain temperatures, grain SEDs) 
are written in \codefile{spectra_dir/}
(\resp \codefile{grain_temp_dir/}, \codefile{grain_SED_dir/}).
If their names are not given in the file of scenarios\index{file of scenarios}, or
if they are already attributed, \codefile{spectra}\index{spectra@\codefile{spectra}} automatically assigns names 
to them.
The files of spectra\index{file of spectra} (\resp grain temperatures%
\index{file of grain temperatures},
grain SEDs%
\index{file of grain SEDs})
are listed in the file \codefile{spectra_dir/list_spectra.txt}%
\index{list_spectra.txt@\codefile{list_spectra.txt}} (\resp \codefile{grain_temp_dir/list_grain_temp.txt}%
\index{list_grain_temp.txt@\codefile{list_grain_temp.txt}},
\codefile{grain_SED_dir/list_grain_SED.txt}%
\index{list_grain_SED.txt@\codefile{list_grain_SED.txt}}).

A \Emph{log file}\index{log file} is also created in subdirectory
\codefile{log_spectra_dir/} of \codefile{bin_dir/}.
Its name is built by prefixing the time-stamp mentioned in \fullref{sec:spectra_file} and the character \mention{\codetext{_}} to the name of the file of scenarios.
\subsubsection{Running \codefile{calib}}
Code \codefile{calib}\index{calib@\codefile{calib}} produces 
the file \mention{\codefile{calib.txt}} in \codefile{calib_dir/}, 
which \codefile{colors}\index{colors@\codefile{colors}} needs 
to compute in-band fluxes and magnitudes.

\codefile{calib}\index{calib@\codefile{calib}} is automatically run whenever you compile all the codes with a simple \mention{\shell{make}} command.
You will also need to run it again every time you add filters or change them (all filters are listed in
\mention*{\codefile{calib_dir/list_filters.txt}});
to do this, type \mention{\shell{./calib}} in \codefile{bin_dir/}.
\subsubsection{Running \codefile{colors}}
\label{sec:run_colors}%
\index{colors@\codefile{colors}}%
Code \codefile{colors}\index{colors@\codefile{colors}} computes colors, mass-to-light ratios, equivalent widths
of emission lines, \etc, for a given scenario.
Type \mention{\shell{./colors}} in \codefile{bin_dir/} to run it, 
and when asked
\ScreenOutputBegin[,]{Name of the input file (file of spectra)? \skipped}%
follow the instructions to select the name of the file of spectra\index{file of spectra}
corresponding to the scenario.
For instance, type
\KeyboardInput[.]{example_spectra1.txt}%

Code \codefile{colors}\index{colors@\codefile{colors}} then asks
\ScreenOutput[.]{Name of the output file? \skipped}
Answer
\KeyboardInputEnd{\ignorespaces}
to select the default name.
This creates the file \mention{\codefile{colors_\usertext{name}}}
(\mention{\codefile{colors_example_spectra1.txt}} in the example) in \codefile{colors_dir/},
where \mention{\usertext{name}} is the name of the file of spectra\index{file of spectra} (stripped from its path, if present).

You may choose another name than the default one for the output file of \codefile{colors}\index{colors@\codefile{colors}}.
Unless a path is specified, it will be written in \codefile{colors_dir/}.
\subsubsection{Running \codefile{plot_spectra}}
\index{plot_spectra@\codefile{plot_spectra}}%
Code \codefile{plot_spectra}\index{plot_spectra@\codefile{plot_spectra}} reads one or more files of spectra%
\index{file of spectra}. 
For each age, it plots or overplots the SEDs on a $\log$-$\log$ plot,
with the wavelength $\lambda$ (in $\AA$) on the $x$-axis and $\lambda*\norm L_\lambda$
on the $y$-axis, where $\norm L_\lambda$ (in $\mathrm{erg*s^{-1}*\AA^{-1}}/\Mref$)
is the continuum monochromatic luminosity (spectral power), per unit wavelength,
normalized to the mass $\Mref$ of the system (see \fullref{sec:norm}).

\Pegase\ does not compute the profile of emission lines, only their integrated luminosity.
For the plots it creates, \codefile{plot_spectra}%
\index{plot_spectra@\codefile{plot_spectra}} considers that all lines have a Gaussian profile.
The standard deviation $\sigma_\iline$, in wavelength, of the $\iline$-th line
is related 
to the full width at half-maximum%
\footnote{The default value of $W$ is the parameter \codetext{FWHM_v_def} 
  in \codefile{plot_spectra.f90};
  if you change it, recompile with \shell{make}.
  You may also enter another value of $W$ while running \codefile{plot_spectra}.}
of the velocity distribution, $W$, by 
\begin{equation}
  \sigma_\iline = \frac{\lambda_\iline*W}{2*\sqrt{2*\ln 2}*c},
\end{equation}
where $\lambda_\iline$ is the wavelength of the line and $c$ is the speed of light.
Emission lines are represented as spikes of height 
\begin{equation}
  \frac{\lambda_\iline*\norm\mathcal{L}_\iline}{\sqrt{2*\pi}*\sigma_\iline},
\end{equation}
where $\norm\mathcal{L}_\iline$ is the normalized integrated luminosity of the line (in $\mathrm{erg*s^{-1}}/\Mref$).
Note that the line profiles themselves are not shown.

To run \codefile{plot_spectra}\index{plot_spectra@\codefile{plot_spectra}}, type \mention{\shell{./plot_spectra}} in
\codefile{bin_dir/} and follow the instructions to select the files of spectra%
\index{file of spectra}, change the value of $W$
and define the bounds of the plot.
\subsubsection{Running \codefile{plot_grain_temp} and \codefile{plot_grain_SED}}
\index{plot_grain_temp@\codefile{plot_grain_temp}}%
\index{plot_grain_SED@\codefile{plot_grain_SED}}%
Code \codefile{plot_grain_temp}\index{plot_grain_temp@\codefile{plot_grain_temp}} plots the temperature probability distribution of individual dust grains for a given scenario;
\codefile{plot_grain_SED}\index{plot_grain_SED@\codefile{plot_grain_SED}} plots their spectral energy distribution.

To run \codefile{plot_grain_temp}\index{plot_grain_temp@\codefile{plot_grain_temp}} (\resp \codefile{plot_grain_SED}\index{plot_grain_SED@\codefile{plot_grain_SED}}), type \mention{\shell{./plot_grain_temp}}
(\resp \mention{\shell{./plot_grain_SED}}) in
\codefile{bin_dir/}, provide the name of a file of grain temperatures%
\index{file of grain temperatures} (\resp of grain SEDs%
\index{file of grain SEDs}) when requested
and follow the instructions.
%
\section{Modeling}
\label{sec:modeling}
The modeling of galaxies by \Pegase.3 is described in \AApaper.
We repeat here only what is required for the consistency of this documentation
and what the user needs to know to run the code.
We also add a few technical details, especially on the meaning of some
parameters.
\subsection{The system: zones}
\subsubsection{Zones}
\label{sec:zones}
The \Emph{system} considered in code \codefile{spectra}\index{spectra@\codefile{spectra}} 
is made of several distinct zones:
\begin{itemize}
\item
  \noUCase   the \Emph{galaxy} proper, modeled as one zone,
  where all the processing occurs, and from where all
  the light comes.
  Initially, the galaxy contains only interstellar gas (or nothing if entirely formed by subsequent infall),
  but no stars or compact stellar remnants%
  \footnote{See \fullref*{sec:gal_comp} for definitions of the various components.};
  this gas is primordial by default.
  The galaxy contains initially no dust either, unless the gas metallicity has been set to
  some non-null value and the basic model of dust evolution is adopted (\fullref{sec:dust_evol_basic});
\item
  \noUCase zero, one or more \Emph{reservoirs}.
  These contain only gas (primordial by default),
  which may fall onto the galaxy via \Emph{infall}.
  As modeled, nothing may enter them;
\item
  \noUCase   the part of the \Emph{intergalactic medium} (IGM) produced by \Emph{outflows}
  ($={}$galactic winds) of interstellar matter (gas and dust) from the galaxy only.
  Nothing may exit from the IGM.
\end{itemize}

The mass assembly history of the galaxy is determined by the relative initial
masses of the galaxy and of the reservoirs, and by the infall and outflow rates.
One has
\begin{equation}
  \Mgal(t) = \Mgal(0) + \int_{t'=0}^t \bigl(\IfR[t'] - \OfR[t']\bigr)*\df t',
\end{equation}
where $\Mgal(t)$ is the mass of the galaxy at age $t$, and
$\IfR$ and $\OfR$ are the total infall and outflow rates.
One or more instantaneous or extended, overlapping or consecutive
\Emph{episodes of infall} or \Emph{outflow} may occur.
\begin{RENVOI}
  \renvoi\fullref{sec:reserv_infall_param}, and \fullref{app:reserv_infall_list}, to set the parameters for reservoirs and
  infall episodes;
  \renvoi\fullref{sec:outflow_param}, and \fullref{app:outflow_list}, to set the parameters for outflow episodes.
\end{RENVOI}

\subsubsection{System's mass and normalized quantities}
\label{sec:norm}
Many quantities are \Emph{normalized} to the constant \Emph{mass of the system},
\begin{equation}
  \Mref \egdef \Mgal(0) + \sum_{j=1}^{\nres} \Mres_j(0)
  = \Mgal(t) + \sum_{j=1}^{\nres} \Mres_j(t) + \int_{t'=0}^t \OfR(t')*\df t',
\end{equation}
where
$\nres$ is the number of reservoirs,
$\Mres_j$ is the mass of the $j$-th reservoir
and $\int_{t'=0}^t \OfR(t')*\df t'$
is the mass increment of the IGM due to outflows from
the galaxy.
\emph{Normalized quantities are denoted by a hat accent hereafter.}
For instance, $\smash{\norm\Mgal}(t) = \Mgal(t)/\Mref$.
See \fullref{app:normalization},
for explanations on how to relate normalized quantities computed by the code to observed ones.
\subsubsection{Components}
\label{sec:gal_comp}
The galaxy contains \Emph{stars}, an \Emph{interstellar medium} (ISM) and
\Emph{compact stellar remnants} (\ie\ white dwarfs, neutron stars and black
holes, but excluding stellar ejecta).
By \mention{stars}, we usually mean both stars proper
(\ie\ regulated by thermonuclear reactions), called \Emph{live stars}
hereafter, and \Emph{inert} (\eg\ substellar) \Emph{objects}.
Note that, although the latter are assumed to be dark and unevolving,
they lock some matter during star formation, which affects the history
of the galaxy and thus, indirectly, its spectral energy distribution.
We also neglect the emission of light by late white dwarfs, neutron stars
and the surroundings of black holes.

By \Emph{gas} and \Emph{dust}, unless otherwise specified,
we mean the gas and the dust in the ISM only:
not that in stars,
their circumstellar envelopes, the reservoirs or the IGM.
The stars and the interstellar gas are mainly made of hydrogen and helium.
As is traditional in astrophysics, all other elements are called \Emph{metals}.
Except for a negligible
fraction of hydrogen atoms, dust consists only in metals.
We give the abundances of elements in terms of mass fraction, not
of relative numbers of atoms;
in particular, what we call \Emph{metallicity} of a component is the overall mass fraction of metals
in that component,
\begin{equation}
  Z  =\frac{M(\txt{metals})}{M(\txt{H}) + M(\txt{He}) + M(\txt{metals})}.
\end{equation}
\subsubsection{Regions}
To compute the attenuation of the light by dust grains and the re-emission of the
latter, we distinguish two \Emph{regions} in the interstellar medium:
\begin{itemize}
\item
  \noUCase \Emph{the diffuse ISM} (DISM);
\item
  \noUCase   \Emph{star-forming clouds} (SFC).
\end{itemize}
\subsection{Single stellar populations}
\label{sec:SSPs}
\begin{RENVOI}
  \renvoi\fullref{sec:SSPs_inputs}.
\end{RENVOI}

The stellar content of a galaxy is a mixture of
\Emph{single stellar populations} (SSP), \ie ensembles of stars formed
instantaneously, simultaneously and with the same initial chemical composition.
The monochromatic luminosity of an SSP (per unit of initial mass of the SSP)
at age~$t'$ and with an initial chemical composition
$\{\chi^{}_0\} \egdef \{\chi^{}_0(\txt{H}), \chi^{}_0(\txt{He}), \ldots\}$ is
\begin{equation}
  \label{eq:integr_isoch}
  \Lmon^\SSP(t', \{\chi^{}_0\}) =
  \int \Lmon^\STAR(m, t', \{\chi^{}_0\})*\df n,
\end{equation}
where
$\Lmon^\STAR(m, t', \{\chi^{}_0\})$ is the monochromatic luminosity at age~$t'$
of a star of initial mass~$m$ and initial composition~$\{\chi^{}_0\}$,
and $\df n$ is the number of stars, per unit of initial mass of the SSP,
born with a mass in $\interv[m, m+\df m[$%
\footnote{%
  Binary stars are not taken into account in the spectral evolution
  (but they have an impact on the chemical evolution because of the model
  assumed for type~Ia supernovae; see \fullref{sec:chem_evol}).}.
This number is computed from the \Emph{initial mass function} (IMF) $\phi$ as
\begin{equation}
  \label{eq:def_IMF}
  \df n = \phi(m)*\df(\ln m),
\end{equation}
where $\phi$ is normalized, \ie\
\begin{equation}
  \label{eq:norm_IMF}
  \int m*\phi(m)*\df(\ln m) = 1.
\end{equation}
\begin{RENVOI}
  \renvoi\fullref{sec:IMF}, to choose the IMF used in \codefile{SSPs}\index{SSPs@\codefile{SSPs}};
  \renvoi\fullref{sec:SSPs_ages}, to choose the output ages of \codefile{SSPs}\index{SSPs@\codefile{SSPs}}.
\end{RENVOI}

In the code,
$\Lmon^\STAR$
is computed from
\Emph{stellar evolutionary tracks} and a \Emph{library of stellar spectra} as
\begin{equation}
  \Lmon^\STAR(m, t', \{\chi^{}_0\}) = \Lbolstar(m, t', \{\chi^{}_0\})*
  \lmon^\STAR(\{\chisurf\}, \Teff, g),
\end{equation}
where $\Lbolstar$
is the bolometric luminosity of the star,
\ie\ the \emph{amplitude} of the stellar spectrum, and
the function $\lambda \mapsto \lmon^\STAR$ is the \emph{shape} of this spectrum.
This shape depends only on the surface composition $\{\chisurf\}$
of the star, its effective temperature $\Teff$
and its surface gravity $g$%
\footnote{We neglect the effect of rotation and stellar winds on the spectrum.}.

$\Lbolstar$, $\Teff$, $g$ and $\{\chisurf\}$ are given by
stellar evolutionary tracks as a function of $m$ and $t'$.
They also depend on the initial composition $\{\chi^{}_0\}$, but because tracks are
computed only as a function of the initial metallicity~$Z$,
with hydrogen and helium abundances determined by $Z$ and
fixed relative ratios of abundances for metals%
\footnote{We consider only tracks at $[\alpha/\txt{Fe}] = 0$.},
we use $Z$ as a substitute for $\{\chi^{}_0\}$.
\begin{RENVOI}
  \renvoi\fullref{sec:tracks}, to choose the stellar evolutionary tracks
  used in \codefile{SSPs}\index{SSPs@\codefile{SSPs}}.
\end{RENVOI}

The relative monochromatic luminosities $\lmon^\STAR(\{\chisurf\}, \Teff, g)$
are obtained by interpolating the shapes
$\lambda\mapsto\lmon_i^\LSS$ of the elements of a library of stellar spectra
as a function of the chemical composition%
\footnote{%
  Such libraries usually provide spectra only as a function of $Z$
  (or, equivalently, $[\txt{Fe}/\txt{H}]$).
  Because the evolution of the surface composition is negligible, except maybe
  during the latest phases, we take the initial value of $Z$ instead of
  $\{\chisurf\}$.},
of $\Teff$ and $g$:
$\lmon^\STAR = \sum_i \alpha_i*\lmon_i^\LSS$\,, where the weights~$\alpha_i$ are independent
of $\lambda$, non negative and $\sum_i \alpha_i = 1$%
\footnote{We never extrapolate the spectra because it might lead to negative
  fluxes:
  when the metallicity of the star is outside the range of the library's
  spectra, we take the closest one;
  we do the same for the gravity and the temperature, but give a larger weight
  to the latter.
  In the best case, we keep the eight nearest points bracketing
  the point $(Z, \Teff, g)$ in the $(Z^\LSS, \Teff^\LSS, g^\LSS)$
  space, so at most eight of the $\alpha_i$ are non null.}.

The relative monochromatic luminosity $\lmon_i^\LSS$ of an element of the library is calculated as
\begin{equation}
  \label{eq:lmon}
  \lmon_i^\LSS = \Lmon_i^\LSS\mkern3mu\left/\int \Lmon_i^\LSS*\df\lambda\right.,
\end{equation}
where $\Lmon_i^\LSS$ is the absolute monochromatic luminosity of the element,
and $\int \Lmon_i^\LSS*\df\lambda$ is its bolometric luminosity
\unskip\footnote{%
  For spectra available on too narrow a wavelength range, in particular
  observed ones, the integral in \fullref*{eq:lmon} underestimates the bolometric luminosity.
  One has then to rely on bolometric corrections
  (see~\fullref{app:bolom_corr}), but these may be
  inconsistent with the spectra.}.
\begin{RENVOI}
  \renvoi\fullref{sec:stel_lib}, to choose the libraries of stellar spectra
  used in \codefile{SSPs}\index{SSPs@\codefile{SSPs}}.
\end{RENVOI}

$\Lmon^\SSP$ is computed by code \codefile{SSPs}\index{SSPs@\codefile{SSPs}} from \fullref*{eq:integr_isoch}.
To do this, \codefile{SSPs} derives from
the evolutionary tracks the \Emph{isochrone} of an SSP at age $t$,
\ie the locus of all the stars in the
$(\Lbolstar, \Teff, g)$ space.
\subsection{Star formation and chemical evolution}
\label{sec:chem_evol}
The unattenuated stellar monochromatic luminosity of a galaxy at age~$t$
and wavelength~$\lambda$ is
\begin{equation}
  \label{eq:CSP}
  L_{\lambda}^{\stel, \unatt}(t) =
  \int_{t'=0}^t \SFR(t-t')*\Lmon^\SSP\bigl(t', Z[t-t']\bigr)*\df t',
\end{equation}
where $\SFR(t-t')$ is the \Emph{star formation rate} (including inert
objects) at age $t-t'$ and
$Z(t-t')$ is the metallicity of the interstellar medium at that age.
One or more instantaneous or extended, overlapping or consecutive
\Emph{episodes of star formation} may occur.
\begin{RENVOI}
  \renvoi\fullref{sec:SF_param}, and \fullref{app:SF_list}, to set the parameters of
  star formation episodes.
\end{RENVOI}

Stars eject matter in the ISM through stellar winds and when they explode
as supernovae.
In low-mass stars, stellar winds occur mainly during the Red (or \mention{First})
Giant Branch (RGB) and the Asymptotic Giant Branch (AGB) phases; the final
remnant is a white dwarf%
\footnote{%
  In the evolutionary tracks based on \citet{Groenewegen+deJong} that we use
  for the thermally pulsing AGB (TPAGB) phase,
  a core-collapse may however happen if the core mass exceeds the Chandrasekhar
  mass.
}.
High-mass stars undergo strong stellar winds during their whole life and
usually end as core-collapse supernovae (type~II, Ib or Ic);
the final remnant is a neutron star or a black hole.
In all cases, when computing the chemical evolution of a galaxy,
we assume that the ejection happens only at the end of the life of stars:
this is justified by the short life of high-mass stars compared to the age of a galaxy
and by the late onset of intense winds in low-mass stars.

According to the favorite model for type~Ia supernovae, these occur
in close binaries where the primary star becomes a CO white dwarf:
because the mass ejected by the secondary falls on the primary, the latter
may reach the Chandrasekhar mass and become a type~Ia supernova;
no compact remnant is then left.
We use the prescriptions of \citet{MG86} to model the rate of type~Ia
supernovae and the ejecta of close binaries%
\footnote{%
  The constants \codetext{min_mass_SNIa} and \codetext{close_bin_gamma} correspond,
  respectively, to the parameters $M_{\mathrm{Bm}}$ and $\gamma$ defined in sec.~2.1
  of \citet*{MG86}.
  They are set in \codefile{source_dir/mod_SSPs_constants.f90} and may be modified.
  To take a modification into account, recompile the code (with \shell{make};
  see \fullref{sec:compil}) and run \codefile{SSPs}\index{SSPs@\codefile{SSPs}} again (cf~\fullref{sec:run_SSPs}).%
}.
Note that, beside this, all stars are considered as isolated.
The fraction of close binaries is set by parameter \codetext{close_bin_frac}
(see \fullref{sec:close_bin_frac}).

The chemical evolution of a galaxy is determined by the history of the
rates $\SFR$, $\IfR$ and $\OfR$ of star formation, infall and outflow.
In particular, the evolution of the mass $\MISM$ of the interstellar medium is
given by
\begin{equation}
  \label{eq:MISM}
  \Deriv{\MISM}{t} = -\SFR(t) + \IfR(t) -\OfR(t)
  + \int_{t'=0}^t \SFR(t-t')*\EjR^\SSP\bigl(t', \{\chi^{}_0\}[t-t']\bigr)*\df t',
\end{equation}
where
$\EjR^\SSP(t', \{\chi^{}_0\}[t-t'])$ is the mass ejection rate of matter into the
ISM at age $t'$
by the stars of an SSP formed with the initial composition $\{\chi^{}_0\}(t-t')$,
per unit of initial mass of the SSP.

For the evolution of the mass of metals in the ISM, we have
\begin{equation}
  \Deriv{(\MISM*\ZISM)}{t} = -(\SFR*\ZISM)(t)
  + \sum_{j=1}^{\nres} \IfR_j(t)*\Zin_j -(\OfR*\ZISM)(t)
  + \int_{t'=0}^t \SFR(t-t')*\EjRZ^\SSP\bigl(t', \{\chi^{}_0\}[t-t']\bigr)*\df t',
\end{equation}
where $\ZISM$ is the mass fraction of metals in the ISM,
$\Zin_j$ is the same quantity in reservoir~$j$, $\IfR_j$ is the infall rate from
reservoir~$j$, and $\EjRZ^\SSP$ is the mass ejection
rate of metals by an SSP.
The code computes the evolution of the abundances of He, C, N, O, Ne, Mg,
Si, S, Ca and Fe in the ISM with similar equations.
In these, we assume that the composition of galactic outflows is the same
as that of the ISM and that stellar ejecta are instantaneously and
homogeneously mixed with the ISM.
\subsection{Stellar yields}
\label{sec:yields}
\begin{RENVOI}
  \renvoi\fullref{sec:SSPs_yields}, to choose the yields
  used in \codefile{SSPs}\index{SSPs@\codefile{SSPs}}.
\end{RENVOI}

The ejection rate in \fullref*{eq:MISM} is given by
\begin{equation}
  \label{eq:ej_SSP}
  \EjR^\SSP(t', \{\chi^{}_0\}) = \int \ejR(m, t', \{\chi^{}_0\})*\phi(m)*\df(\ln m),
\end{equation}
where $\ejR(m, t', \{\chi^{}_0\})$ is the mass ejection rate at age $t'$ of a star
of initial mass $m$ and initial composition $\{\chi^{}_0\}$.
As mentioned in previous section, we assume that the ejection occurs at once, at the end of the life of the star,
\ie\ at the end of the TPAGB phase for low-mass stars
or when they explode for those becoming supernovae:
\begin{equation}
  \label{eq:ej_m}
  \ejR(m, t', \{\chi^{}_0\}) = \mej(m, \{\chi^{}_0\})*\dirac(t'-t_{\txt{end}}[m, \{\chi^{}_0\}]),
\end{equation}
where $t_{\txt{end}}(m, \{\chi^{}_0\})$ is the life duration of the star
and $\dirac$ is the Dirac delta \mention{function}.
Note that \emph{recycling is not instantaneous}.
As usual and for want of better, we use $Z$ as a substitute for $\{\chi^{}_0\}$
and interpolate the total ejected mass $\mej$ as a function of $m$ and $Z$
from a \Emph{set of yields}.

Sets of yields also provide either the \Emph{gross yield} or the
\Emph{net yield} for various elements:
the gross yield of any element~$i$ is the total mass $\meji$ of~$i$ ejected by a star in the ISM;
the net yield $\meji^{\txt{net}}$ is related to the gross yield by
\begin{equation}
  \meji = \meji^{\txt{net}} + \chi_{0, i}^{}*\mej\,,
\end{equation}
where $\chi_{0, i}^{}$ is the initial abundance of~$i$ in the star,
$\chi_{0, i}^{}*\mej$ is the portion of the initial mass released in the ISM and
originally in the form of~$i$,
and $\meji^{\txt{net}}$ is the additional ejected mass of~$i$.
(The same holds for all the metals together: $\mej_Z = \mej_Z^{\txt{net}} + Z*\mej$.)

The ejection rate of element~$i$ in the ISM by an SSP is
\begin{equation}
  \EjR_i^\SSP(t', \{\chi^{}_0\}) = \int \ejR_i(m, t', \{\chi^{}_0\})*\phi(m)*\df(\ln m).
\end{equation}
Because sets of yields are only available for fixed initial compositions,
which cannot be interpolated (for all elements simultaneously) to the
composition~$\{\chi^{}_0\}$ of the ISM when stars form, it is more accurate,
in order to compute $\EjR_i^\SSP$\kern2pt,
to interpolate the net yields rather than the gross yields as a function
of the initial stellar metallicity~$Z$, and to use $\{\chi^{}_0\}$ for the
original contribution.
Using the assumption expressed by \fullref*{eq:ej_m},
\ie\ $\ejR_i(m, t', \{\chi^{}_0\}) = \meji(m, \{\chi^{}_0\})*\dirac(t'-t_{\txt{end}}[m, \{\chi^{}_0\}])$,
one has then
\begin{equation}
  \EjR_i^\SSP(t', \{\chi^{}_0\})
  \approx \int \meji^{\txt{net}}(m, Z)*\dirac(t'-t_{\txt{end}}[m, Z])
  *\phi(m)*\df(\ln m) + \chi_{0, i}^{}*\EjR^\SSP(t', Z).
  \label{eq:ej_i_SSP}
\end{equation}
\subsection{Dust grains}
\subsubsection{Grain size distribution}
\label{sec:GSD}
Two \Emph{families} of dust grains are considered:
silicates and carbonaceous grains.
The latter family is itself subdivided in three \Emph{grain species}:
graphites, neutral PAHs (Polycyclic Aromatic Hydrocarbons) and ionized PAHs.
The silicate family is a grain species by itself.

The size distributions of grains and the relative masses of the various species
within a given family are constant in the code.
Three models are provided currently for these quantities:
\begin{itemize}
\item
  \noUCase the \mention{\textsc{bare\_gr\_s}} model of \citet{ZDA} (file \codefile{ZDA.txt} in
  \codefile{dust_dir/});
\item
  \noUCase   the model number~$7$ of table~1 of \citet{Weingartner+Draine} used in
  \citet{Li+Draine} (file \codefile{LWD.txt});
\item
  \noUCase   the outdated model, without PAHs, of \citet{MRN} (file \codefile{MRN.txt}).
\end{itemize}
\begin{RENVOI}
  \renvoi\fullref{sec:GS_param}, and \fullref{app:dust_transfer_list}.
\end{RENVOI}

The files containing the size distributions also point to files
giving the optical properties (absorption and scattering opacities, asymmetry
parameter) of individual grains as a function of wavelength
(see \fullref{sec:opt_prop}).
The wavelenth range is $\interv[10*\AA, 1*\textup{mm}]$, with
resolutions $\lambda/\Delta\lambda \approx 17$ for graphites and silicates
and $\lambda/\Delta\lambda \approx 90$ for PAHs.
\vfill\break
\subsubsection{Dust evolution}
\label{sec:dust_evol}
\begin{RENVOI}
  \renvoi\fullref{sec:dust_evol_param}, and \fullref{app:dust_evol_list}.
\end{RENVOI}

While size distributions of grains and relative masses of dust species
within a given family are constant, the overall masses in the ISM
of the silicate and carbonaceous dust families evolve with time.
Two mutually exclusive models are implemented in \codefile{spectra}\index{spectra@\codefile{spectra}}:
\leaderouv
\par\nobreak\makeatletter\@afterheading\makeatother
\paragraph{Basic model}
\label{sec:dust_evol_basic}
In this model, the mass of dust is simply proportional to the mass of its
constituents in the ISM:
one does not consider the processes by which dust is produced and destroyed.
The mass of carbonaceous grains in the ISM is given by
\begin{equation}
  \label{eq:M_carb_formed_anew}
  M_{\txt{carb}}^{\ISM}(t) = \delta_{\txt{carb}}^{\ISM} * M_{\txt{C}}^{\ISM}(t),
\end{equation}
where $M_{\txt{C}}^{\ISM}$ is the mass of carbon in the ISM (including dust grains)
and $\delta_{\txt{carb}}^{\ISM}$\,, the depletion factor for carbon atoms, is constant.
(The mass of hydrogen atoms in carbonaceous grains is neglected.)

The mass of silicate grains in the ISM is given by
\begin{equation}
  \label{eq:M_sil_formed_anew}
  M_{\txt{sil}}^{\ISM}(t) = \delta_{\txt{sil}}^{\ISM} *
  \sum_{i=1}^{n_\txt{sil}} {M_i^{\ISM}(t) * (1 + \OToSil*A_{\txt{O}}/A_i)},
\end{equation}
where the $n_{\txt{sil}}$ elements referred to by index~$i$
are Mg, Si, S, Ca and Fe;
$M_i^{\ISM}$ is the mass of element~$i$ in the ISM (including dust grains);
$A_i$, its atomic mass;
$A_{\txt{O}}$, that of oxygen;
and $\OToSil$ is the number of atoms
of oxygen in silicate dust per atom of Mg, \etc.
Two assumptions are made in this equation:
\begin{itemize}
\item
  \noUCase   the same constant depletion factor $\delta_{\txt{sil}}^{\ISM}$ applies to each of the $n_{\txt{sil}}$ elements%
  \footnote{The same assumption is made for the various $\delta_\txt{sil}$ used
    in Dwek's model.};
\item
  \noUCase   the number of oxygen atoms in the ISM is always sufficient to keep
  $\OToSil$ constant%
  \footnote{The same assumption is made regarding dust accretion
    in Dwek's model.
    For circumstellar dust production, we assume that the number of
    free (not locked in CO) oxygen atoms in the environment
    is sufficient to keep $\OToSil$ constant.}.
\end{itemize}
\paragraph{Dwek's model}
\label{sec:dust_evol_Dwek}
This model is based on \citet{Dwek98} and tries to describe the formation of dust in the
late phases of stellar evolution, its destruction in the ISM by the blast waves of
supernovae and the accretion of dust constituents on grains already present in the ISM.
\subparagraph{Circumstellar dust production}
Here, dust is produced in the following environments:
\begin{description}[.\space]
\item[\UCase in the winds of mass-losing stars]
  Carbon and oxygen atoms are assumed to be locked preferentially
  in CO molecules.
  If the number of carbon atoms in the overall%
  \footnote{%
    It would be more physical to compare the ejection \emph{rates} $\ejR_{\txt{C}}$ and $\ejR_{\txt{C}}$ of carbon
    and oxygen rather than the integrated quantities $\mej_{\txt{C}}$ and $\mej_{\txt{C}}$\textellipsis{}

    \nopagebreak
    Note also that, to compute the amount of dust produced in circumstellar
    environments, we need the gross yields for each stellar mass,
    not the net yields weighted by the number of dying stars;
    \fullref*{eq:ej_i_SSP} is therefore not appropriate here.
  } ejecta of a star is
  larger than the number of oxygen atoms, only carbonaceous dust is produced.
  The mass of carbonaceous grains formed by a star is then
  \begin{equation}
    \label{eq:LMW_carb}
    m_{\txt{ej}, \txt{carb}} = \delta_{\txt{carb}}^{\textsc{hmw}/\textsc{lmw}} * (\mej_{\txt{C}}
    - \mej_{\txt{O}} * A_{\txt{C}}/A_{\txt{O}}),
  \end{equation}
  where $\mej_{\txt{O}} * A_{\txt{C}}/A_{\txt{O}}$ is the mass locked in CO,
  $\delta_{\txt{carb}}^{\textsc{hmw}}$ is the depletion factor in the winds of
  high-mass stars and $\delta_{\txt{carb}}^{\textsc{lmw}}$ the one in low-mass stars.

  If the number of carbon atoms is less than that of oxygen ones,
  only silicate dust is produced.
  The mass of silicate dust produced by a star is then
  \begin{equation}
    \label{eq:LMW_sil}
    m_{\txt{ej}, \txt{sil}} = \delta_{\txt{sil}}^{\textsc{hmw}/\textsc{lmw}} *
    \sum_{i=1}^{n_\txt{sil}} {\mej_i * (1 + \OToSil*A_{\txt{O}}/A_i)},
  \end{equation}
  where $i$ designates the same elements as in \fullref*{eq:M_sil_formed_anew};
\item[\UCase in the ejecta of supernovae]
  No CO is formed (among other reasons, because C and O are in separate shells), so
  \begin{equation}
    \label{eq:SN_carb}
    m_{\txt{ej}, \txt{carb}} = \delta_{\txt{carb}}^{\CCSN/\SNIa} * \mej_{\txt{C}}\,,
  \end{equation}
  where $\delta_{\txt{carb}}^{\CCSN}$ is the depletion factor in the ejecta of
  core-collapse supernovae and $\delta_{\txt{carb}}^{\SNIa}$ the one in type~Ia supernovae.
  Similarly,
  \begin{equation}
    \label{eq:SN_sil}
    m_{\txt{ej}, \txt{sil}} = \delta_{\txt{sil}}^{\CCSN/\SNIa} *
    \sum_{i=1}^{n_\txt{sil}} {\mej_i * (1 + \OToSil*A_{\txt{O}}/A_i)}.
  \end{equation}
\end{description}
As for all other stellar ejecta, we assume that dust grains produced
in circumstellar environments are formed instantaneously when the star dies.
They are then immediately scattered through the whole ISM%
\footnote{%
  We therefore neglect the effect of circumstellar dust grains on the light
  of the stars they surround.}.
\subparagraph{Dust destruction in the ISM}
Dust is also destroyed by supernovae.
The destruction rate of the mass $M_{\txt{carb}}^{\ISM}(t)$ of carbonaceous dust
in the ISM is computed as
\begin{equation}
  \label{eq:dust_destr}
  \left(\Deriv{M_{\txt{carb}}^{\ISM}}{t}\right)_{\txt{destr}} =
  - \dot n_{\SN}(t) * m_{\txt{swept}} * \frac{M_{\txt{carb}}^{\ISM}(t)}{M_\ISM(t)},
\end{equation}
where
$\dot n_{\SN}(t)$ is the number rate of supernovae, and the constant $m_{\txt{swept}}$ is the mass
of ISM swept by the blast of a single supernova.
Idem for silicate dust.
\subparagraph{Dust accretion in the ISM}
The accretion rate of carbon on carbonaceous grains already in the ISM
is modeled as
\begin{equation}
  \label{eq:carb_accr}
  \left(\Deriv{M_{\txt{carb}}^{\ISM}}{t}\right)_{\txt{accr}} =
  \left(1-\frac{M_{\txt{carb}}^{\ISM}(t)}{M_{\txt{C}}^{\ISM}(t)}\right)
  *\frac{M_{\txt{carb}}^{\ISM}(t)}{\tau_{\txt{carb}}^{\txt{accr}}},
\end{equation}
where $\tau_{\txt{carb}}^{\txt{accr}}$ is a constant timescale and
$1-M_{\txt{carb}}^{\ISM}(t)/M_{\txt{C}}^{\ISM}(t)$ is the fraction of ISM carbon atoms
in the gaseous phase, thus able to accrete on dust grains.
Similarly, for silicates,
\begin{equation}
  \label{eq:sil_accr}
  \left(\Deriv{M_{\txt{sil}}^{\ISM}}{t}\right)_{\txt{accr}} =
  \left(1-\frac{M_{\txt{sil}}^{\ISM}(t)}{\sum_{i=1}^{n_\txt{sil}} {M_i^{\ISM}(t) * (1 + \OToSil*A_{\txt{O}}/A_i)}}\right)
  *\frac{M_{\txt{sil}}^{\ISM}(t)}{\tau_{\txt{sil}}^{\txt{accr}}}.
  \postdisplaypenalty=10000
\end{equation}
\par\leaderfer

While the model where dust is produced in circumstellar environments, destroyed and
accreted is more physical, it involves a lot of parameters, the value of which is rather uncertain.
The basic model is therefore the default one.
The parameters for these two models are listed in \fullref{sec:dust_evol_param}.
\subsubsection{Optical properties}
\label{sec:opt_prop}
The optical properties of dust species are computed as a function
of wavelength from their size distributions (see \fullref{sec:GSD})
and from the optical properties of individual grains
as follows.
The absorption opacity in surface per unit mass (\ie~the mass absorption coefficient)
of dust species~$i$ at wavelength~$\lambda$ is
\begin{equation}
  \kappa_{i, \lambda}^\txt{abs} =
  \int_a \pi* a^2* Q_{i, \lambda}^\txt{abs}(a)*\frac{\df n_i}{\df a}*\df a,
\end{equation}
where $\pi* a^2* Q_{i, \lambda}^\txt{abs}(a)$ is the absorption cross-section of a grain
with radius~$a$, and $\df n_i/\df a$ is the number of grains per unit size
per unit mass of dust
(i.e.,
\begin{equation}
  \int_a \frac{4}{3}*\pi* a^3*\varrho_i*\frac{\df n_i}{\df a}*\df a = 1,
\end{equation}
where $\varrho_i$ is the inner mass density of a grain).
Similarly for the scattering opacity $\kappa_{i, \lambda}^\txt{sca}$ (with
$Q_{i, \lambda}^\txt{abs}(a)$ replaced by $Q_{i, \lambda}^\txt{sca}(a)$).
The extinction opacity is given by
\begin{equation}
  \kappa_{i, \lambda}^\txt{ext} = \kappa_{i, \lambda}^\txt{abs} + \kappa_{i, \lambda}^\txt{sca}\,,
\end{equation}
the albedo by
\begin{equation}
  \omega_{i, \lambda} = \kappa_{i, \lambda}^\txt{sca}\left/\kappa_{i, \lambda}^\txt{ext}\right.
\end{equation}
and the asymmetry parameter by
\begin{equation}
  g_{i, \lambda} = \frac{\int_a \pi* a^2* Q_{i, \lambda}^\txt{sca}(a)*c_{i, \lambda}(a)*
    (\df n_i/\df a)*\df a}{\kappa_{i, \lambda}^\txt{sca}},
\end{equation}
where $c_{i, \lambda}(a)$ is the average cosinus of the scattering angle.
The quantities $Q_{i, \lambda}^\txt{abs}(a)$, $Q_{i, \lambda}^\txt{sca}(a)$ and $c_{i, \lambda}(a)$ are
taken from \citet{Laor+Draine}, \citet{Draine+Lee} and \citet{Li+Draine}
(files \codefile{opt_prop_graphites.txt} for graphites, \codefile{opt_prop_silicates.txt} for silicates,
\codefile{opt_prop_neutral_PAHs.txt} for neutral PAHs
and \codefile{opt_prop_ionized_PAHs.txt} for ionized ones, all in \codefile{dust_dir/}).
\subsection{Attenuation by grains in the diffuse ISM}
\label{sec:diffuse_attenuation}
\begin{RENVOI}
  \renvoi\fullref{sec:dust_transfer_param}, and \fullref{app:dust_transfer_list}.
\end{RENVOI}

The attenuation of the galaxy SED by grains in the diffuse medium is
extensively described for \Emph{geometries} appropriate to spiral and spheroidal
galaxies in sec.~4.1 and 4.2, respectively, of \AApaper;
all the notations used in Sec.~\ref{sec:spiral} and \ref{sec:spher} below
are defined therein.

In addition to these two realistic geometries,
a third one is still available for reasons
of compatibility with previous versions of the code:
the \Emph{infinite slab} in which stars and dust are well mixed.

See \fullref{sec:geometry_param} to select the geometry.
\subsubsection{Spiral galaxies}
\label{sec:spiral}
Grids of the transmittance for the spiral disk and bulge are provided
in the files \codefile{disk_transmit.txt} and
\codefile{bulge_transmit.txt} of directory \codefile{RT_dir/}, both as
a function of the viewing angle (inclination) toward the galaxy and averaged
over all inclinations.

The default values of the bulge-to-total mass ratio $\bulgeToTot$ and
of $\Mref^\spir/R_\DD^2$%
\footnote{\label{fn:unnormalized}%
  $\Mref^\spir/R_\DD^2$\kern1pt, $\Mref^\sph/\Rcore^2$ and $\alpha_\txt{slab}$
  are the only unnormalized quantities in the code.
},
where $\Mref^\spir$ is the mass of the system
and $R_\DD$ is the characteristic radius of the dust disk, are given in
\fullref*{tab:default_spiral_param}.
They are derived from the values
provided in table~I-2 of \citeauthor{B+T} (\citeyear{B+T}; BT)
for the Milky Way and
may be changed (see parameters \codetext{bulge_tot_ratio},
\codetext{M_sys_spiral} and \codetext{expo_radius}, \fullref{sec:dust_transfer_param}).
\begin{table}[H]
  \caption{\label{tab:default_spiral_param}%
    Default structural parameters for spirals\captionPoint}
  \begin{center}
    \begin{tabular}{ll}
      \hline
      $\bulgeToTot = 1/7$
      &
      \begin{tabular}[t]{@{}l@{}}
        \Tstrut
        From the values for
        $L_V(\txt{disk})$ ($1.2\times10^{10}*L_V(\odot)$) and
        \\
        $L_V(\txt{bulge})$ ($2\times10^9*L_V(\odot)$) in BT's model of the Milky Way.
        \Bstrut
      \end{tabular}
      \tabularnewline
      $\Mref^\spir/R_\DD^2 
      = 2915*M_\odot/\txt{pc}^2$
      &
      \begin{tabular}[t]{@{}l@{}}
        \Tstrut
        From $\bulgeToTot$,
        the value of $R_\DD/R^\SD_\stel$ in table~1 of \AApaper\ and
        \\
        the values for
        $M_\txt{tot}(\txt{disk})$ ($6\times10^{10}*M_\odot$) and
        $R^\SD_\stel$ ($3.5*\txt{kpc}$)
        \\
        in BT's model of the Milky Way.
        \Bstrut
      \end{tabular}
      \tabularnewline
      \hline
    \end{tabular}
  \end{center}
\end{table}
\subsubsection{Spheroidal galaxies}
\label{sec:spher}
Grids of the transmittance for spheroidal galaxies
are provided in file \codefile{RT_dir/King_transmit.txt}.

For the mass of the system, $\Mref^\sph$, we use the current mass of stars in
model~\emph{b} of \citet{TM95},
neglecting thus the matter expelled by the galaxy in the intergalactic medium
since it formed.
The mass of the system is then
\begin{equation}
  \label{eq:M_sph_stell}
  \Mref^\sph = 4*\pi*\Rcore^3*\mu_{\stel, 0}^\sph*
  \int_{s=0}^{\mkern2mu s_\txt{t}} {s^2*(1+s^2)^{-3/2}*\df s}
  = 4*\pi*\Rcore^3*\mu_{\stel, 0}^\sph*\left(\operatorname{arsinh}s_\txt{t}
    - s_\txt{t}/\mkern-3mu\sqrt{\strut\smash{1+s_\txt{t}^2}}\right),
\end{equation}
where $s_\txt{t} \egdef \Rtrunc/\Rcore$, $\Rcore$ and $\Rtrunc$ are the core
and truncation radii, and $\mu_{\stel, 0}^\sph$ is the central stellar mass density.
The default value of $\Mref^\sph/\Rcore^2$%
\footnoteref{fn:unnormalized}
is $7.73\times10^6*\Msol*\txt{pc}^{-2}$;
it may be changed (see parameters \codetext{M_sys_spher} and
\codetext{core_radius}, \fullref{sec:dust_transfer_param}).
\subsubsection{Infinite slab}
\label{sec:slab}
The column mass density of dust through the slab and perpendicularly to it
is derived from the column number density of hydrogen, computed as
\begin{equation}
  \label{eq:slab}
  \sigma_\txt{H}^\txt{slab} = \alpha_\txt{slab}*\norm M_\ISM.
  \postdisplaypenalty=10000
\end{equation}
Grids of the transmittance for this geometry are provided as a function
of inclination in file \codefile{RT_dir/slab_transmit.txt}.
The default value of $\alpha_\txt{slab}$ may be changed
(see parameter \codetext{slab_factor}, \fullref{sec:dust_transfer_param}).
\subsection{Dust emission by grains in the diffuse medium}
\label{sec:diffuse_emission}
\begin{RENVOI}
  \renvoi\fullref{sec:dust_transfer_param}, and \fullref{app:dust_transfer_list}.
\end{RENVOI}
\subsubsection{Interstellar radiation field}
The computation of the mean interstellar radiation field in the diffuse
ISM, $\langle u_\lambda\rangle$, is described in sec.~5.1 of \AApaper.
Note that, dividing the numerator and denominator of eq.~\AApaper(25)
by $\Mref$, we may express
$\langle u_\lambda\rangle$ exclusively in terms of the normalized
unattenuated luminosity $\norm L^0_{\lambda}$ and of the normalized dust mass $\norm M_\dust$ provided by the code:
\begin{equation}
  \label{eq:mean_ISRF}
  \langle u_\lambda\rangle = \frac{\norm L^0_{\lambda}*(1-\overline\Theta_\lambda)}{
    c*\kappa^{\txt{abs}}_{\lambda}*\norm M_\dust},
\end{equation}
where $\overline\Theta_\lambda$ is the inclination-averaged transmittance,
$c$ is the speed of light and $\kappa^{\txt{abs}}_{\lambda}$
is the absorption opacity per unit mass of dust.
(Note, however, that $\overline\Theta_\lambda$ depends on the unnormalized optical depth
$\tau_\lambda^\txt{ext}$ through the ratio $\Mref^\spir/R_\DD^2$ or $\Mref^\sph/\Rcore^2$ (\fullref{fn:unnormalized}), so $\langle u_\lambda\rangle$ and the dust emission spectrum too.)

\subsubsection{Stochastic heating}
\label{sec:stoch_heat}
Because of their large cross section, big grains are permanently hit by a huge number of photons
and always have temperatures close to their equilibrium value.
On the other hand, small grains are seldom hit and undergo strong temperature fluctuations:
they spend most of the time at low temperatures but sometimes
reach very high temperatures for short durations.
The computation of the temperature probability distribution of grains due
to stochastic heating and of the resulting emission is explained in sec.~5.2 of
\AApaper.
To switch off stochastic heating, see \fullref{sec:stoch_heat_param},
of this documentation.

As we do not evolve the size distribution of grains, we do not take into account the destruction
of grains which may occur during temperature spikes.
For information only, we however count separately the energy emitted by grains at temperatures
larger than the \Emph{sublimation temperature} (one for all carbonaceous grains, one for silicates,
whatever their size and species; see \fullref{sec:sublim_param} to change
the sublimation temperatures).
\subsubsection{Self-absorption}
\label{sec:self_abs}
By default, we neglect \Emph{self-absorption}, \ie\ we assume that the ISM,
whether the diffuse one or the star\babelhyphen{nobreak}forming
clouds, is transparent to the emission of dust grains in the same medium%
\footnote{To be clearer, grains in the diffuse ISM (DISM)
  absorb and scatter the
  light from stars in the DISM, but also the light emitted and processed
  in star-forming clouds (SFCs), since these are embedded in the DISM.
  On the other hand, because SFCs are small compared to the
  DISM, grains in clouds process only the stellar and nebular photons emitted
  within SFCs, but not those coming from the DISM.}.
The monochromatic luminosity leaving the region (the galaxy, if the region is
the diffuse medium) is then
\begin{equation}
  \label{eq:noSA}
  L_\lambda^\noSA = \overline\Theta_\lambda*L_\lambda^\unatt + L_\lambda^{\dust, \noSA},
\end{equation}
where $\overline\Theta_\lambda$ is the inclination-averaged transmittance, through
dust in the medium, of the unattenuated luminosity $L_\lambda^\unatt$ (stellar, mainly),
and $L_\lambda^{\dust, \noSA}$ is the emission by dust if self-absorption is neglected.

To assess the effects of self-absorption, we propose to model it in the
following way:
\begin{itemize}
\item
  \UCase the transmittance for dust-emitted photons is parametrized
  as
  \begin{equation}
    \Theta_\lambda^\dust = \overline\Theta_\lambda^{\mkern9mu\gamma},
  \end{equation}
  where the power
  $\gamma$ (parameter \codetext{self_abs_power}; see \fullref{sec:dust_transfer_param}) is a constant in $\interv[0, 1]$.
  The default case, $\gamma = 0$, corresponds to no self-absorption
  ($\Theta_\lambda^\dust = 1$).
  On the other hand, for very dusty environments,
  the mean free path of the original photons absorbed by dust
  is so small that one may consider that they are absorbed \mention{on the spot};
  in this approximation, dust-emitted photons are thus transmitted with the
  same factor $\overline\Theta_\lambda$ than original photons at the same wavelength,
  so taking $\gamma = 1$ should be appropriate%
  \footnote{Note that using $\gamma = 1$ for environments with little dust
    does not change the results by much, compared to the no self-absorption
    case: since $\overline\Theta_\lambda \approx 1$, one has then
    $\Theta_\lambda^\dust \approx 1$ too, especially at the wavelengths where
    dust grains emit.};
\item
  \UCase moreover, since most of the photons re-emitted by grains are in the infrared,
  the fraction of this dust-emitted energy which is absorbed again by grains is
  much smaller than for the original photons.
  We therefore assume that the temperature distribution of dust grains,
  and thus the \emph{shape} of the dust-emitted SED, is unchanged by the
  subsequent processing.
\end{itemize}
With these assumptions, \fullref*{eq:noSA} becomes
\begin{equation}
  \label{eq:SA}
  L_\lambda^\SA =
  \overline\Theta_\lambda*L_\lambda^\unatt + \alpha*\overline\Theta_\lambda^{\mkern9mu\gamma}*L_\lambda^{\dust, \noSA},
\end{equation}
where $L_\lambda^\SA$ is the monochromatic luminosity, with \mention{self-absorption},
emerging from the region,
and $\alpha$ is a constant (\ie, independent of wavelength) introduced to ensure the conservation of
the energy: indeed, self-absorption redistributes
the SED of dust grains at other wavelengths but conserves their overall
bolometric luminosity.
The factor $\alpha$ is thus fixed by the equation
\begin{equation}
  \int \alpha*\overline\Theta_\lambda^{\mkern9mu\gamma}*L_\lambda^{\dust, \noSA}*\df\lambda = \int L_\lambda^{\dust, \noSA}*\df\lambda.
\end{equation}

See also~\fullref{app:self_abs}.
\subsection{Star-forming regions and nebular emission}
\label{sec:cloud_neb}
\begin{RENVOI}
  \renvoi\fullref{sec:cloud_neb_param}, and \fullref{app:cloud_list}.
\end{RENVOI}

The modeling of star-forming regions and of the effects of dust on the stellar
and nebular light they emit, in particular in the Lyman continuum,
as well as the computation with version~c17.01 of \textsc{Cloudy}
\citep{Cloudy} of the nebular emission in star-forming clouds
and in the diffuse ISM,
are extensively described in sec.~6 and app.~1 of \AApaper.
Let us just remind that the fraction $\varphi(t')$ of stars aged~$t'$ still in their birth cloud
is modeled as
\begin{equation}
  \label{eq:cloud_frac}
  \varphi(t') = \varphi_0*(1-t'/\theta)^\beta,
\end{equation}
that all stars form in clusters and that all these clusters have the same initial stellar mass.
%
\section{Computing the properties of single stellar
  populations with \codefile{SSPs}}
\label{sec:SSPs_inputs}%
\index{SSPs@\codefile{SSPs}}%
Code \codefile{SSPs}\index{SSPs@\codefile{SSPs}} computes as a function of age the isochrones
of single stellar populations for several initial metallicities;
it then apportions their light among the elements of a library of stellar
spectra.
The code also outputs the mass of compact stellar remnants produced
by an SSP, the amount of matter ejected by stars into the ISM, the number
of supernovae, and the radiation rate of Lyman continuum photons%
\footnote{%
  \label{fn:SSPs_outputs}%
  Although they are in text format,
  the files produced by \codefile{SSPs}\index{SSPs@\codefile{SSPs}} are not aimed to be read by
  the user and do not contain spectral energy distributions.
  To compute these quantities for an SSP, first run \codefile{SSPs}\index{SSPs@\codefile{SSPs}}, then
  \codefile{spectra}\index{spectra@\codefile{spectra}} for an episode of instantaneous star formation.
}.

To produce these outputs, \codefile{SSPs}\index{SSPs@\codefile{SSPs}} needs the inputs described below.
\subsection{Initial mass function}
\label{sec:IMF}
\begin{RENVOI}
  \renvoi\fullref*{eq:def_IMF} and \fullref{eq:norm_IMF}, for the definition and normalization of the initial mass function~$\phi$.
\end{RENVOI}

The initial mass functions (IMF) mentioned in \fullref{tab:IMFs}, are provided with the code.
The files associated to these IMFs are gathered in \codefile{IMFs_dir/} and are listed in
\codefile{list_IMFs.txt};
the first IMF in the latter file is the default one.
References are given in the headers of the IMF files.

The form of IMFs marked as \mention{analytical} is actually not provided in the corresponding file,
but is hard\--coded in \codefile{source_dir/mod_IMF.f90}.
In particular, the log-normal IMF is given (see eq.~(30) in \citet{MS79}) by
\begin{equation}
  \phi(m) = C_0*\exp\bigl(-C_1*[\log_{10}(m/\Msol) - C_2]^2\Bigr).
\end{equation}
If you select this IMF when running \codefile{SSPs}\index{SSPs@\codefile{SSPs}}, you will be requested to enter values for $C_1$ and $C_2$
($C_0$ is fixed by the normalization).
You may also use the default values, $C_1 = 1.09$ and $C_2 = -1.02$,
corresponding to \citet{MS79}'s fit of the solar neighborhood at $12*$Gyr with a constant star formation rate.

All other IMFs are modeled as \emph{continuous} piecewise power-law functions
of $m$
(\ie, for $m$ in the bin $\interv[m_i, m_{i+1}[$, $\phi(m) \propto m^{s_i}$,
where the slope $s_i$ is constant in the bin).
In all cases, $\phi$ is automatically normalized by code \codefile{SSPs}\index{SSPs@\codefile{SSPs}}.

To add other IMFs, see \fullref{app:add_IMFs}.

When running \codefile{SSPs}\index{SSPs@\codefile{SSPs}}, it is possible to supersede the default
values for the lower and upper masses of the IMF defined in
\codefile{source_dir/mod_SSPs_constants.f90} or the \codefile{IMF_\userfile{*}.txt} files.
Note also that the IMF used in \codefile{SSPs}\index{SSPs@\codefile{SSPs}} relates only to \emph{live stars}:
it is possible to add a population of \emph{inert} (\ie, dark and unevolving)
\emph{objects} to the
evolutionary scenario of \codefile{spectra}\index{spectra@\codefile{spectra}}
via the parameter \codetext{SF_inert_frac} (see \fullref{sec:SF_param}).

\begin{table}[H]
  \caption{\label{tab:IMFs}Initial mass functions provided with the code\captionPoint}
  \begin{center}
    \begin{tabular}{ll>{\Tstrut}l}
      \hline
      \multicolumn{1}{@{}>{\Tstrut}c<{\Bstrut}@{}}{File}
      & \multicolumn{1}{@{}c@{}}{Form of the IMF}
      & \multicolumn{1}{@{}c@{}}{Reference} \\
      \hline\hline
      \codefile{IMF_Rana_Basu.txt} & \UCase analytical & \citet{RB}. \\
      \codefile{IMF_Chabrier_2003.txt} & \UCase analytical & \citet{Chabrier2003}. \\
      \codefile{IMF_Chabrier_2005.txt} & \UCase analytical &  \citet{Chabrier}. \\
      \codefile{IMF_Ferrini.txt} & \UCase analytical &  \citet[p.~520]{Ferrini+90}.\\
      \codefile{IMF_Kennicutt.txt} & \UCase piecewise power-law &  \citet{Kennicutt83}. \\
      \codefile{IMF_Kroupa.txt} & \UCase piecewise power-law &  \citet{Kroupa}. \\
      \codefile{IMF_log_normal.txt} & \UCase analytical &  Log-normal IMF (see below). \\
      \codefile{IMF_Miller_Scalo.txt} & \UCase piecewise power-law &  \citet{MS79}. \\
      \codefile{IMF_Salpeter.txt} & \UCase piecewise power-law &  \citet{Salpeter}. \\
      \codefile{IMF_Scalo86.txt} & \UCase piecewise power-law &  \citet{Scalo86}. \\
      \codefile{IMF_Scalo98.txt} & \UCase piecewise power-law &  \citet{Scalo98}. \\
      \codefile{IMF_Kroupa1.5.txt} & \UCase piecewise power-law &
      \begin{tabular}[t]{@{}l@{}}
        \citet{Kroupa}, but with\\%
        a steeper slope at high mass.\Bstrut%
      \end{tabular}%
      \Bstrut
      \\%
      \hline
    \end{tabular}
  \end{center}
\end{table}

As implemented, the IMF is independent of time. It is however possible
to have an IMF evolving with the initial metallicity of SSPs, and thus
indirectly with time. See \fullref{app:evolve_IMF}, for this.
\subsection{Stellar evolutionary tracks}
\label{sec:tracks}
In the code, \Emph{files of stellar evolutionary tracks}
provide the evolution of the bolometric
luminosity, effective temperature and surface gravity of a star as a function
of its age for a range of initial masses and a single initial metallicity.
\Emph{Sets of stellar evolutionary tracks} gather files with various
metallicities.
Sets of tracks are in \codefile{tracks_dir/} and are listed in
\codefile{list_tracks_sets.txt};
the first set in this file is the default one.

The default set with current settings, \codefile{tracks_set+.txt},
is based on the tracks computed by the Padova group in the 1990's
and is described in sec.~2.2.2 of \AApaper.
At $Z=0.1$, pseudo-tracks for masses larger than $9*M_{\odot}$ have been
computed from the corresponding masses in the $Z=0.02$ and $Z=0.05$ sets.
For stars undergoing the helium flash, the zero-age main sequence (ZAMS) tracks
are connected to the zero-age horizontal branch tracks with the same core mass,
assuming a Reimers law \citep{Reimers1975} for the mass loss along the first giant branch with $\eta=0.4$
\citep{Renzini1981}.

Hydrogen burning post-AGB and CO white dwarf tracks from \citet{Bloecker}
($m/\Msol \in \{0.605,\ 0.625,\ 0.696,\ 0.836,\ 0.940\}$),
\citet{Schonberner83} ($m/\Msol \in \{0.546,\ 0.565\}$;
extrapolated with the $0.456~\Msol$-track from
\citet{Koester+Schoenberner}) and \citet{Paczynski} ($m = 1.2~\Msol$)
are then connected.
\subsection{Stellar yields}
\label{sec:SSPs_yields}
\begin{RENVOI}
  \renvoi\fullref{sec:yields}.
\end{RENVOI}

The sets of yields we use are described in sec.~2.2.3 of \AApaper.

The ejecta of low-mass stars are from \citet{Marigo2001}.
The gross and net yields are given, respectively, in files
\codefile{gross_ejecta_LMW.txt} and \codefile{net_ejecta_LMW.txt} of
directory \codefile{yields_dir/}, as are all the files mentioned in this section.

For high-mass stars, the default yields, from \citet{Portinari}, are
in the files \codefile{\userfile{*}_ejecta_HMW_P.txt} for the wind phase and
\codefile{\userfile{*}_ejecta_CCSN_P.txt}
for the core-collapse phase.
The explosive yields of model~B of \citet{Woosley+Weaver}
may also be selected;
they are in the files \codefile{\userfile{*}_ejecta_CCSN_WW.txt}
(the mass loss before the supernova is not taken into account in these).

To compute the net yields (files \codefile{net_\userfile{*}})
from the gross ones (files \codefile{gross_\userfile{*}})
or the reverse, we need the
initial abundances in the star.
These are provided for hydrogen, helium and all the
metals together by \citet{Marigo2001} and \citet{Portinari}, but not for
individual metals.
For these, we scale the solar abundances of \citet{Anders+Grevesse}
(file \codefile{Anders_Grevesse_1989.txt}) to $Z$.

We do the same for the metal yields of \citet{Woosley+Weaver}.
For $Z < Z_\odot$, however, these authors do not provide the initial
abundances of hydrogen and helium either.
We thus compute the initial mass fraction of helium, $Y$,
assuming a constant relative enrichment of helium with respect to metals:
\begin{equation}
  \frac{Y-Y_\txt{prim}}{Z-Z_\txt{prim}} = \frac{Y_\odot-Y_\txt{prim}}{Z_\odot-Z_\txt{prim}},
\end{equation}
where $Y_\txt{prim} = 0.23$ and
$Z_\txt{prim} \approx 0$ are the primordial abundances of helium and metals
(just after the Big Bang), and the solar abundances $Y_\odot$ and $Z_\odot$ are taken from
\citet{Anders+Grevesse}.
The initial mass fraction $X$ of hydrogen is then given by $X = 1 - Y - Z$.

The gross yields of model~W7 of \citet{Thielemann+}, used to compute
the ejecta of type~Ia supernovae,
are provided in file \codefile{W7.txt}.
The net yields are computed by \codefile{SSPs}\index{SSPs@\codefile{SSPs}}
with the same procedure as for \citet{Woosley+Weaver}.
\subsection{Stellar spectra}
\label{sec:stel_lib}
In the code, \Emph{files of stellar spectra}
provide the SEDs of stars
as a function of wavelength
for a range of effective temperatures and surface gravities
and for a single initial metallicity.
The emission rates of Lyman continuum photons, derived from these SEDs and used to model nebular emission, are also supplied.
The SEDs and rates are normalized to the bolometric luminosity of the star.

Files of spectra are gathered in \Emph{sets of stellar spectra}
to cover the Hertzsprung \& Russell diagram at the
metallicities possibly occurring during the evolution of a galaxy.
All the sets are in \codefile{stel_lib_dir/}
and are listed in \codefile{list_stel_lib_sets.txt};
the first set in the latter file is the default one (\codefile{BaSeL2.2_Rauch_set1.txt} currently).

Each set of stellar spectra is made of two libraries:
\begin{itemize}
\item
  \noUCase  BaSeL's spectra for stars with an effective temperature
  $\Teff < 50 \mkern2mu 000*\txt{K}$.
  The wavelength range is $\interv[91*\AA,\allowbreak 160*\micron]$,
  with a variable resolution decreasing from a few $\AA$ in the far-UV to
  $20*\AA$ in the visible and to $20*\micron$ in the far-IR/submm;
\item
  \noUCase the spectra from \citet{Rauch}, rebinned to the wavelengths of BaSeL,
  for hotter stars.
  These spectra are only available at $[\txt{Fe}/\txt{H}] \in \{-1, 0\}$.
\end{itemize}

The BaSeL library is based on theoretical spectra (\citet{Kurucz}, mostly).
These are corrected with wave\-length\--dependent factors to fit observed colors.
The library comes in two versions:
\begin{itemize}
\item
  \noUCase   v2.2%
  \footnote{Several variants and combinations of the BaSeL-2.2 library are
    actually provided, for reasons explained in the headers of the
    \codefile{BaSeL2.2_Rauch_set\userfile{*}.txt} files.}
  \citep{Lejeune+98}, where the same correction factors,
  obtained from solar-metallicity calibrations of observed colors
  as a function of effective temperature and surface gravity,
  are applied to all metallicities;
\item
  \noUCase v3.1 WLBC99 \citep{Westera}, where correction factors are,
  somewhat indirectly, derived from fits
  to observed color\,-\,mag\-nitude diagrams of star clusters
  with subsolar metallicities.
\end{itemize}
As recognized by \citet{Westera}, when combined with theoretical isochrones,
the WLBC99 library does not improve the fits to observed color\,-\,magnitude
diagrams, compared to BaSeL-2.2.
Because the latter covers a larger range of metallicities, we use it as our default.

\subsection{Output ages of \codefile{SSPs}}
\label{sec:SSPs_ages}%
\index{SSPs@\codefile{SSPs}}%
The span in initial masses, phases and metallicities
of the default set of stellar
evolutionary tracks allows to follow the evolution of a galaxy from
$0$ to $20*\txt{Gyr}$.

The output ages of \codefile{SSPs}\index{SSPs@\codefile{SSPs}} are determined from a file.
The default file, \codefile{SSPs_ages.txt}, has a time resolution
degrading from $1*\txt{Myr}$ for ages less than $30*\txt{Myr}$ to
$100*\txt{Myr}$ above $10*\txt{Gyr}$.
Other files may be used instead, to enhance the time resolution for instance,
but all of them must be listed in
\codefile{ages_dir/list_SSP_ages.txt} and be located in the same directory.
Look at the comments in \codefile{SSP_ages.txt} and
\codefile{list_SSP_ages.txt} to define another file of output ages for
\codefile{SSPs}\index{SSPs@\codefile{SSPs}} and add it to the list.
Note that the first number in the file of output ages
is the \Emph{convolution time-step},
\ie~the time-step used in \codefile{spectra}\index{spectra@\codefile{spectra}} to evolve the system
(in particular, to convolve the properties of SSPs with the star formation
history).
%
\section{Defining scenarios of evolution for \codefile{spectra}}
\index{spectra@\codefile{spectra}}%
%
\label{sec:scenarios}%
\subsection{Preliminaries}
\label{sec:prelim_scenarios}%
\subsubsection{Types of statements}
A file of scenarios\index{file of scenarios} describes one or more scenarios and consists in a list
of statements written following the syntax detailed in \fullref{app:syntax}.
There are two kinds of statements:
\begin{description}[,\space]
\item[assignments]%
  \ie all the statements 
  which explicitely set (with \mention{\codetext{=}}) the values of the \Emph{parameters} defining a scenario.

  The basic form of an assignment is 
  \mention{\codetext{\usertext{key}~= \usertext{val}}},
  where \usertext{key} is the name of a single scalar parameter and
  \usertext{val} its value.
  To assign the $k$-th element
  of an array of parameters \usertext{key}, use the syntax
  \mention{\codetext{\usertext{key}($k$)~= \usertext{val}}}.
  For the first element,
  you may more conveniently write
  \mention{\codetext{\usertext{key}~= \usertext{val}}}
  instead of 
  \mention{\codetext{\usertext{key}(1)~= \usertext{val}}}.

  {\emergencystretch=1em For more details, in particular about complex assignments involving arrays
  of parameters, see \fullref{app:key_val};\par}
\item[commands]
  \ie\ all other statements.

  Some commands implicitely assign (without \mention{\codetext{=}}) parameters to their default values
  (the \codetext{reset_\usertext{*}} statements, for instance);
  other ones control the back and forth between the reading of scenarios
  and the computation of the evolution for the last read scenario.

  In particular, any scenario must be separated from the previous one in the
  file of scenarios\index{file of scenarios} by the command \codetext{return}: this returns control
  to the main procedure in \codefile{spectra}\index{spectra@\codefile{spectra}}, which then processes the
  current scenario and computes the corresponding evolution and spectra.
\end{description}

All the parameters and commands are described in Sec.~\ref{sec:SSPs_param} to~\ref{sec:other_statements}, p.~\pageref{sec:SSPs_param} to \pageref{sec:other_statements}.
\subsubsection{General rules applying to parameters}
Complete lists of the parameters, with their default value if any, are given
in Tables~\ref{tab:SSPs_chemical} to~\ref{tab:other_param}, p.~\pageref{tab:SSPs_chemical} to~\pageref{tab:other_param}.

\begin{figure}[H]
  \centerline{\includegraphics{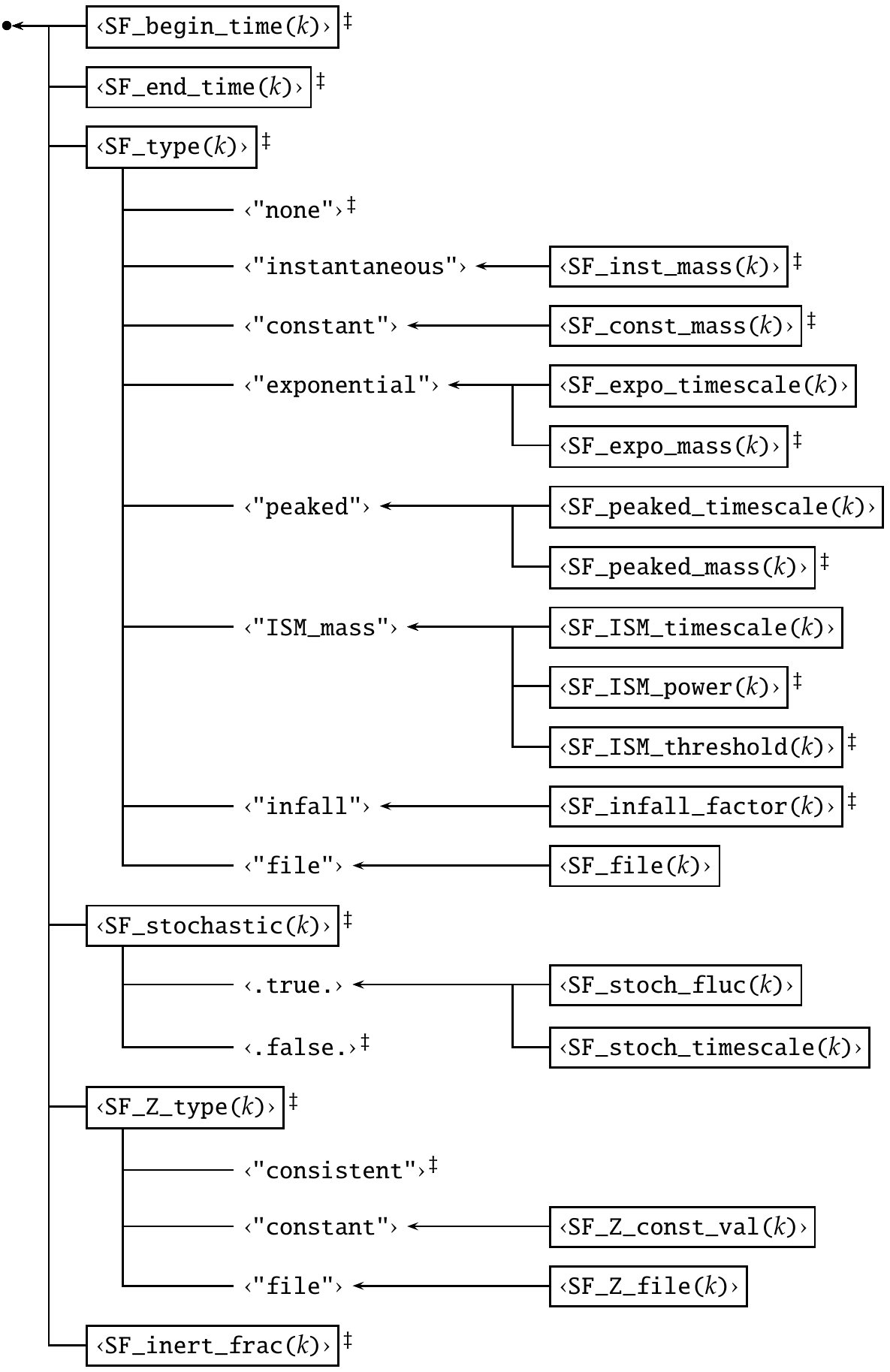}}
  \caption{\label{fig:SF}%
    Tree of parameters related to the $k$-th episode of star formation\captionPoint}
\end{figure}

\begin{figure}[H]
  \centerline{\includegraphics{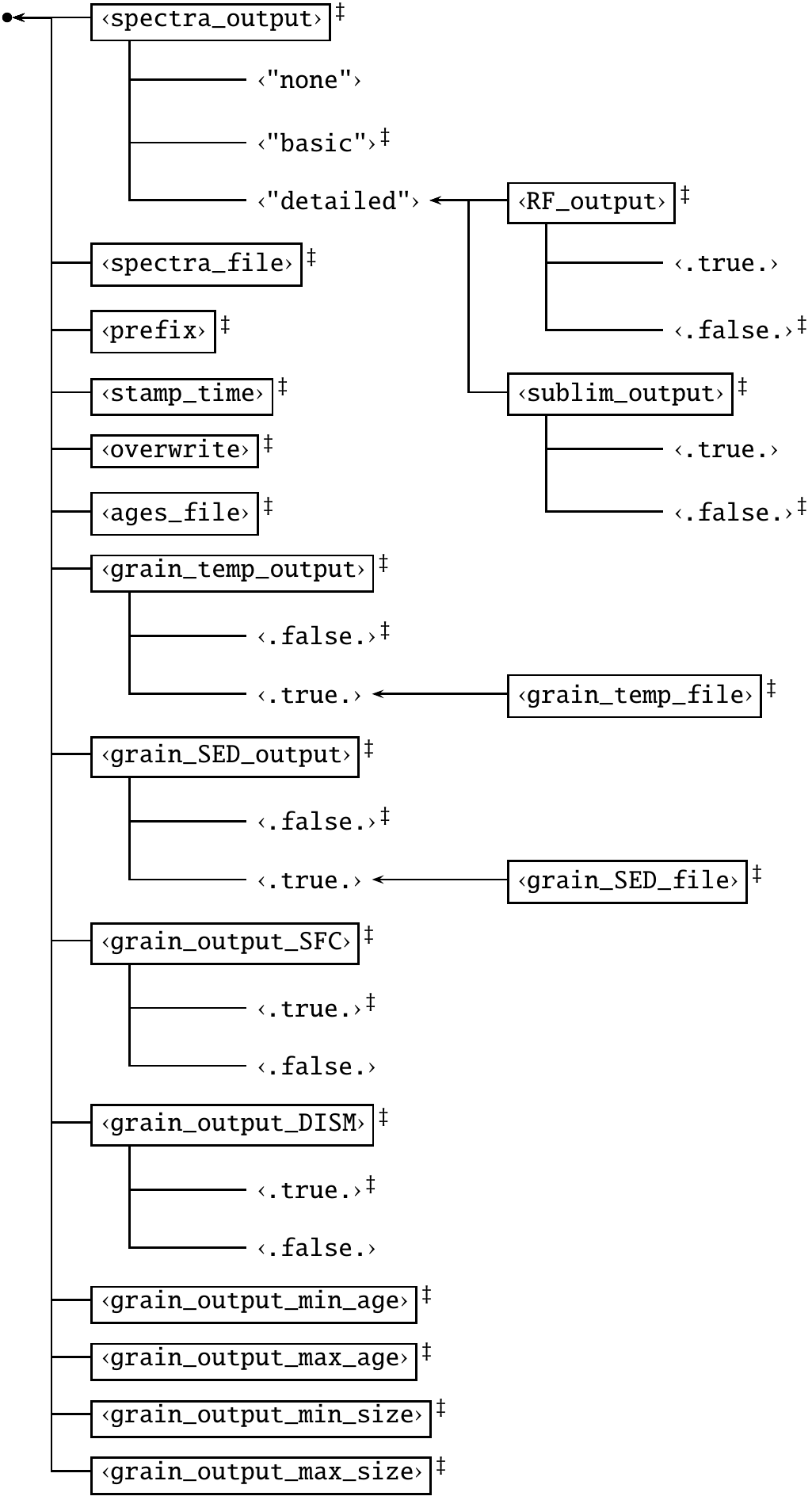}}
  \caption{\label{fig:output_tree}%
    Tree of parameters related to output files\captionPoint}
\end{figure}

Related parameters are organized in trees
(see Fig.~\ref{fig:reserv_infall_tree} to~\ref{fig:output_tree},
p.~\pageref{fig:reserv_infall_tree} to~\pageref{fig:output_tree}).
Detailed explanations on how to read these trees are given at the beginning of \fullref{app:trees_tables}.

Parameters obey two main rules:
\begin{enumerate}
\item
  \textbf{Explicitely setting the value of a parameter implicitely assigns all
    its \mention{ancestors} to consistent values.}

  There is therefore no need to provide all the parameters in a tree.

  As an example, let us consider the tree related to star-forming clouds
  and nebular emission
  (see \fullref{fig:cloud_neb_tree}). 
  If you write \mention{\codetext{neb_emis_const_frac~= 0.5}} in the file
  of scenarios, then the parent of \codetext{neb_emis_const_frac},
  \codetext{neb_emis_type}, is automatically set to
  \codetext{"constant"},
  which in turn
  sets its grandparent, \codetext{nebular_emission}, to \codetext{.true.}.
  You still have to provide the values of other parameters
  (\codetext{cloud_init_frac}, \etc), unless
  you are satisfied with their default values;
\item 
  \textbf{Parameters which are not assigned in a scenario, whether explicitely
    or implicitely, retain the values they had in the previous scenario
    or remain undefined if they have no default value and have never been
    assigned.}

  This rule holds for all parameters%
  \footnote{%
    \label{fn:rule2_exceptions}%
    The only exceptions are the names of the output files 
    of \codefile{spectra}\index{spectra@\codefile{spectra}} for the current 
    scenario (\codetext{spectra_file}%
    \index{file of spectra}
    and, if required, \codetext{grain_temp_file}%
    \index{file of grain temperatures}
    and \codetext{grain_SED_file}%
    \index{file of grain temperatures}).},
  including those not relevant for the current scenario.

  For example (see~\fullref{fig:reserv_infall_tree}), 
  if \codetext{infall_expo_\-time\-scale} was defined 
  in previous scenario (the initial value of this parameter is undefined), 
  its value is irrelevant in the
  current scenario if \codetext{infall_type} is now set to \codetext{"constant"}, but will be used
  again in the next scenario if \codetext{infall_type}
  is then set to \codetext{"exponential"}. 
\end{enumerate}

If you are lost at some point, you can use one of the
\codetext{reset_\usertext{*}} commands (see \fullref{sec:reset}),
where \usertext{*} stands for \codetext{reserv_infall},
\codetext{SF}, \etc,
to reset all the parameters related to \usertext{*} to their
default values, if they exist, or to undefined otherwise.

You can also write \codetext{echo}\index{echo@\codetext{echo}} at any point
in the file of scenarios\index{file of scenarios} to write the values of all the
parameters at that point, both on the screen and in the log file\index{log file}
(see \fullref{sec:run_spectra}).
%
\subsection{An example of file of scenarios}
\label{sec:example}
Before we describe in detail how to write a file of scenarios\index{file of scenarios} and
the meaning of all parameters, let us look at an example, file
\codefile{example_scenarios.txt} (available in \codefile{scenarios_dir/}).
This file contains three scenarios.
\subsubsection{First scenario (instantaneous burst of star formation)}
\begin{lstlisting}[name=example, frame=trl]
SSPs_set = "example_SSPs.txt"/*\label{line:SSPs_set}*/
/*\label{line:blank}*/
! Instantaneous burst of star formation:/*\label{line:comment}*/
SF_type = "instantaneous"/*\label{line:instantaneous}*/
spectra_file = "example_spectra1.txt"/*\label{line:spectra1}*/
return/*\label{line:return}*/

\end{lstlisting}
Explanations:
\begin{description}
  \item[Line~\ref{line:SSPs_set}]
    the IMF, yields, evolutionary 
    tracks and stellar spectra defined in 
    \codefile{\usertext{example}_SSPs.txt} will be used;
  \item[Lines~\ref{line:blank} and~\ref{line:comment}]
    skipped (blank lines and comments);
  \item[Line~\ref{line:instantaneous}]
    the scenario consists in a single instantaneous burst of star formation;
  \item[Line~\ref{line:spectra1}]
    the name of the file of spectra produced for this scenario is \userfile{example_spectra1.txt};
  \item[Line~\ref{line:return}]
    reading of the scenario is done. 
    Return to \codefile{spectra}\index{spectra@\codefile{spectra}} and compute the
    evolution of the modeled galaxy for the first scenario.
\end{description}
All other parameters have their default values:
the whole mass is  initially in the galaxy and in the form of zero-metallicity gas;
the burst occurs at age~$0$ and consumes all this gas to form zero-metallicity stars;
there is neither infall nor outflow;
nebular emission and dust effects are not considered.
\subsubsection{Second scenario ($\approx$~spiral galaxy)}
The parameters of the second scenario are defined by the following lines:
\begin{lstlisting}[name=example, frame=rl]
! Spiral galaxy:
reserv_init_mass = 1./*\label{line:reserv_init_mass}*/
infall_expo_timescale = 1000/*\label{line:infall_expo_timescale}*/
SF_ISM_timescale = 3000/*\label{line:SF_ISM_timescale}*/
nebular_emission = .true./*\label{line:nebular_emission}*/
extinction = .true./*\label{line:extinction}*/
inclin_averaged = .true./*\label{line:inclin_averaged}*/
dust_emission = .true./*\label{line:dust_emission}*/
spectra_file = "example_spectra2.txt"/*\label{line:spectra2}*/
return

\end{lstlisting}
Explanations:
\begin{description}
  \item[Line~\ref{line:reserv_init_mass}]
    the galaxy accretes gas from a reservoir (the first one by default)
    with a mass \codetext{reserv_init_\-mass}${}\times\Mref$.
    As $\codetext{reserv_Z} = 0$ by default and
    $\codetext{reserv_init_\-mass} = 1$, 
    the galaxy forms entirely from the zero-metallicity gas 
    in the reservoir;
  \item[Line~\ref{line:infall_expo_timescale}]
    this explicitely sets 
    \codetext{infall_expo_timescale} to $1000$~Myr and, implicitely, all the 
    parameters above it, along the path to the top of the tree, 
    to consistent values:
    \codetext{infall_type} is therefore set \codetext{"exponential"};
    parameters in other branches, 
    \eg \codetext{infall_begin_time},
    are untouched and keep their value if it has been
    defined previously, explicitely or implicitely, or if there
    is a default one. 
    Consequently, the infall rate is exponentially decreasing with a timescale
    of $1$~Gyr from $0$ to $20$~Gyr;
  \item[Line~\ref{line:SF_ISM_timescale}]
    this sets the star formation rate to proportional to the 
    mass of ISM with a timescale of $3$~Gyr. 
    The implicit $\codetext{SF_type} = \codetext{"ISM_mass"}$ activated by 
    \mention{\codetext{SF_ISM_timescale = 3000}} supersedes the 
    \mention{\codetext{SF_type = "instantaneous"}} of the previous scenario;
  \item[Line~\ref{line:nebular_emission}]
    nebular emission will be computed;

  \item[Line~\ref{line:extinction}]
    extinction by dust is taken into account;

  \item[Line~\ref{line:inclin_averaged}]
    dust attenuation for a disk galaxy,  averaged over all inclinations, is applied to the SEDs;
  \item[Line~\ref{line:dust_emission}]
    dust emission spectrum is added.
\end{description}
Except \codetext{spectra_file}, all the parameters defined previously in the 
file, \eg \codetext{SSPs_set},
retain their value, unless modified explicitely or implicitely.

\subsubsection{Third scenario (spiral galaxy + late burst)}
\begin{lstlisting}[name=example, frame=rlb]
! Spiral galaxy with a late burst:
SF_type(2) = "constant"/*\label{line:constant}*/
SF_const_mass(2) = 1.e-1/*\label{line:intensity}*/
SF_begin_time(2) = 10000/*\label{line:begin}*/
SF_end_time(2) = 11000/*\label{line:end}*/
spectra_file = "example_spectra3.txt"
\end{lstlisting}
These lines define a second star formation episode occuring
from $10$~Gyr (line~\ref{line:begin}) to $11$~Gyr (line~\ref{line:end}), 
with a constant rate (line~\ref{line:constant}) 
and involving a mass of $0.1\,\Mref$ (line~\ref{line:intensity}). 
This episode is \emph{added} to the one defined by scenario number~$2$.
Note that the first episode of star formation could have been defined
with the statement \mention{\codetext{SF_ISM_timescale(1) = 3000}} instead of 
\mention{\codetext{SF_ISM_timescale = 3000}} on line~\ref{line:SF_ISM_timescale}.\relax
%
\subsection{Description of main parameters}
\label{sec:descr_param}%
Default values are indicated for some parameters only.
For others, see~\fullref{app:trees_tables}.
%

\subsubsection{Parameters related to cosmology}
\label{sec:cosmo_param}
\begin{RENVOI}
\renvoi\fullref{app:cosmo_list}.
\end{RENVOI}

\paragraph{\codetext{Omega_m},
  \codetext{H_0},
  \codetext{form_redshift}}
\index{Omega_m@\codetext{Omega_m}}%
\index{H_0@\codetext{H_0}}%
\index{form_redshift@\codetext{form_redshift}}%
\index{redshift}%

\codetext{Omega_m} is the current value
of the ratio $\Omega_\txt{m}$ of the mean density of matter in the Universe to 
the critical density of the latter; 
\codetext{H_0} is the current value of the Hubble constant in $\mathrm{km*s^{-1}*Mpc^{-1}}$.

These cosmological parameters are used to compute the redshift of the galaxy as a function of its age
and of \codetext{form_redshift}, its formation redshift \citep[p.~317]{Peebles}.
A flat universe with a cosmological constant $\Omega_\Lambda$ is assumed in this calculation
($\Omega_\txt{m} + \Omega_\Lambda = 1$);
the density of radiation is neglected with regard to that of matter.
Redshifts are printed in the main output file of \codefile{spectra}\index{spectra@\codefile{spectra}}.


\paragraph{\codetext{CBR} \notInToc{(default: \codetext{.false.})}}
\index{CBR@\codetext{CBR}}%

The redshift is also used to compute the temperature of the cosmic blackbody.
The heating of dust by the cosmic background radiation field is considered, in addition 
to the radiation field emitted by the galaxy, if and only if \codetext{CBR} is \codetext{.true.}.
%
\subsubsection{Single stellar populations: \codetext{SSPs_set} \notInToc{(no default value)}}
\label{sec:SSPs_param}
\index{SSPs_set@\codetext{SSPs_set}}%
\begin{RENVOI}
  \renvoi\fullref{sec:run_SSPs}, and \fullref{app:SSPs_chemical}.
\end{RENVOI}

\codetext{SSPs_set} holds the name of the set of SSPs%
\footnote{\label{fn:once}%
  For computational reasons, it is recommended to set the parameters
  \codetext{SSPs_set}, \codetext{close_bin_frac} and
  \codetext{grains_file} (or \codetext{grains_file_SFC} and 
  \codetext{grains_sizes_DISM}; see~\fullref{sec:GS_param}) at most once, 
  in the first scenario. 
  Note that \codetext{SSPs_set} must be provided at least one since there is 
  no default value for this parameter.},
the master file
produced by \codefile{SSPs}\index{SSPs@\codefile{SSPs}}.
(This file points to auxiliary files containing
the properties of SSPs with a single initial metallicity \dashOpen one per file of evolutionary tracks.)
%

\subsubsection{Chemical evolution parameters: \codetext{ISM_init_Z}, \codetext{close_bin_frac}}
\label{sec:chemical_param}%
\index{ISM_init_Z@\codetext{ISM_init_Z}}%
\label{sec:close_bin_frac}%
\index{close_bin_frac@\codetext{close_bin_frac}}%
\begin{RENVOI}
  \renvoi\fullref{sec:chem_evol}, and \fullref{app:SSPs_chemical}.
\end{RENVOI}

\codetext{ISM_init_Z} is the initial value of the metallicity of the interstellar medium in the galaxy.

\codetext{close_bin_frac} is the fraction of close binaries%
\footnoteref{fn:once}.
This parameter is used to compute the rate
and ejecta of type~Ia supernovae.
%
\vfill\break
\subsubsection{Reservoir and infall parameters}
\label{sec:reserv_infall_param}
\begin{RENVOI}
  \renvoi\fullref{sec:zones}, and \fullref{app:reserv_infall_list}.
\end{RENVOI}


\paragraph{%
  \codetext{reserv_init_mass($j$)},
  \codetext{reserv_Z($j$)}  \notInToc{(defaults: $0$, $0$)}%
}
\index{reserv_init_mass@\codetext{reserv_init_mass}}%
\index{reserv_Z@\codetext{reserv_Z}}%
\codetext{reserv_init_mass($j$)} is the initial normalized mass $\norm\Mres_j$ of reservoir~$j$%
\footnote{\label{fn:max_epis}%
  The maximal numbers of reservoirs and of infall, star formation and outflow episodes
  are set (to $10$ by default) in \codefile{source_dir/mod_spectra_constants.f90} by parameters \codetext{max_dim_reserv},
  \codetext{max_dim_infall_epis}, \codetext{max_dim_SF_epis}
  and \codetext{max_dim_outflow_epis}, respectively.
  If needed, change these values and recompile the code with \shell{make}
  (see~\fullref{sec:compil}).};
\codetext{reserv_Z($j$)} is its constant metallicity.

The normalized initial mass of the galaxy is computed as
\begin{equation}
  \norm\Mgal(t = 0) = 1 - \sum_j \norm\Mres_j(t = 0).
\end{equation}
By default, $\norm\Mres_j(t = 0) = 0$ for all reservoirs,
so $\norm\Mgal(t = 0) = 1$.

If the values of the parameters \codetext{reserv_init_mass($j$)} are such that
$\sum_j \norm\Mres_j(t = 0) > 1$, then $\norm\Mgal(t = 0)$ is
set to $0$, and the values used (\emph{in this scenario only})
for the $\norm\Mres_j(t = 0)$
are scaled by a common factor to ensure this%
\footnote{\label{fn:warn}%
  A warning will be printed at the end of the file of spectra%
  \index{file of spectra},
  in the log file%
  \index{log file}
  and, if $\codetext{verbosity} \ge 0$%
  \index{verbosity@\codetext{verbosity}} (see~\fullref{sec:verbosity}),
  on the screen.
  See also \fullref{tab:output_warn}.
}.


\paragraph{Infall episodes, \codetext{infall_source($k$)}}
\index{infall_source@\codetext{infall_source}}%
More than one episode of infall%
\footnoteref{fn:max_epis}
on the galaxy can occur, and this from any reservoir.
The parameter \codetext{infall_source($k$)} is the
index of the reservoir from which the infalling gas in episode~$k$ comes.
By default, all infall episodes remove gas from the first reservoir.

The mass removal rate from reservoir~$j$ at age~$t$ is
\begin{equation}
  \dot\Mres_j(t) = -\sum_{\hbox to0pt{\hss$\scriptstyle\substack{\text{all }k\text{ such}\\\text{that }r_k=j}$\hss}} \IfR_k(t),
\end{equation}
where $\IfR_k(t)$ is the infall rate of episode~$k$, and $r_k$ is the
value of \codetext{infall_source($k$)}.

If the mass removed from reservoir~$j$ in one convolution time-step,
$-\dot\Mres_j(t)*\Delta t$,
is larger than $\Mres_j(t)$, the infall rates of all the episodes drawing
from reservoir~$j$ are scaled down at $t$ by the same factor so that
$-\dot\Mres_j(t)*\Delta t = \Mres_j(t)$\footnoteref{fn:warn}.

For each $k$, the value of $\IfR_k(t)$ is computed using the values
of the parameters defined in the following paragraphs.


\paragraph{\codetext{infall_begin_time($k$)},
  \codetext{infall_end_time($k$)}}
\index{infall_begin_time@\codetext{infall_begin_time}}%
\index{infall_end_time@\codetext{infall_end_time}}%
Whatever the value of $\IfR_k(t)$ computed using other parameters,
\begin{equation}
  \forall~t \not\in \interv[\ti_k, \tf_k[,\quad \IfR_k(t) = 0,
\end{equation}
where $\ti_k$ and $\tf_k$ are the values of \codetext{infall_begin_time($k$)}
and \codetext{infall_end_time($k$)}.


\paragraph{%
  $\codetext{infall_type($k$)} = \codetext{"none"}$ \notInToc{(default)}%
}
\index{infall_type@\codetext{infall_type}|(}%
\index{infall_type@\codetext{infall_type}!none@\codetext{\dq none\dq}}%
\kern-\baselineskip
\begin{equation}
  \forall~t,\quad \nIfR_k(t) = 0.
\end{equation}


\paragraph{$\codetext{infall_type($k$)} = \codetext{"instantaneous"}$:
  \codetext{infall_inst_mass($k$)}}
\index{infall_type@\codetext{infall_type}!instantaneous@\codetext{\dq instantaneous\dq}}%
\index{infall_inst_mass@\codetext{infall_inst_mass}}%
\kern-\baselineskip
\begin{equation}
  \forall~t\in\interv[\ti_k, \tf_k[,\quad \nIfR_k(t) = f_k*\dirac(t-\ti_k),
\end{equation}
where $f_k$ is the value of \codetext{infall_inst_mass($k$)},
$\ti_k$ is that of \codetext{infall_begin_time($k$)},
and $\dirac$ is the Dirac distribution.
(In practice, the infall event occurs at the nearest convolution time.)


\paragraph{$\codetext{infall_type($k$)} = \codetext{"constant"}$:
  \codetext{infall_const_mass($k$)}}
\index{infall_type@\codetext{infall_type}!constant@\codetext{\dq constant\dq}}%
\index{infall_const_mass@\codetext{infall_const_mass}}%
\kern-\baselineskip
\begin{equation}
  \forall~t \in \interv[\ti_k, \tf_k[,\quad \nIfR_k(t) = \frac{f_k}{\tf_k-\ti_k},
\end{equation}
where $f_k$ is the value of \codetext{infall_const_mass($k$)}.


\paragraph{%
  $\codetext{infall_type($k$)} = \codetext{"exponential"}$:
  \codetext{infall_expo_timescale($k$)}, \raggedallowbreak
  \codetext{infall_expo_mass($k$)}%
}
\index{infall_type@\codetext{infall_type}!exponential@\codetext{\dq exponential}}%
\index{infall_expo_timescale@\codetext{infall_expo_timescale}}%
\index{infall_expo_mass@\codetext{infall_expo_mass}}%
\kern-\baselineskip
\begin{equation}
  \forall~t \in \interv[\ti_k, \tf_k[,\quad
  \nIfR_k(t) = \frac{f_k}{\valabs{\tau_k}}*\neper^{-(t-\ti_k)/\tau_k},
\end{equation}
where $f_k$ is the value of \codetext{infall_expo_mass($k$)}
and $\tau_k$ that of \codetext{infall_expo_timescale($k$)}.
A positive value of $\tau_k$ corresponds to a decreasing infall rate;
a negative value, to an increasing one.%


\paragraph{%
  $\codetext{infall_type($k$)} = \codetext{"reserv_mass"}$:
  \codetext{infall_reserv_timescale($k$)}, \raggedallowbreak
  \codetext{infall_reserv_power($k$)}%
}
\index{infall_type@\codetext{infall_type}!reserv_mass@\codetext{\dq reserv_mass\dq}}%
\index{infall_reserv_timescale@\codetext{infall_reserv_timescale}}%
\index{infall_reserv_power@\codetext{infall_reserv_power}}%
\kern-\baselineskip
\begin{equation}
  \forall~t \in \interv[\ti_k, \tf_k[,\quad
  \nIfR_k(t) = \frac{\left(\norm\Mres_j[t]\right)^{\alpha_k}}{\tau_k},
\end{equation}
where $j = \codetext{infall_source($k$)}$,
$\norm\Mres_j(t)$ is the normalized mass of reservoir~$j$ at age $t$,
$\tau_k$ is the value of \codetext{infall_reserv_timescale($k$)},
and $\alpha_k$ is that of \codetext{infall_reserv_power($k$)}.


\paragraph{%
  $\codetext{infall_type($k$)} = \codetext{"file"}$:
  \codetext{infall_file($k$)}%
}
\index{infall_type@\codetext{infall_type}|)}%
\index{infall_type@\codetext{infall_type}!file@\codetext{\dq file\dq}}%
\index{infall_file@\codetext{infall_file}}%
For all $t \in \interv[\ti_k, \tf_k[$, $\nIfR_k(t)$ is interpolated
from the values read in a file.
The name of the latter is stored in parameter \codetext{infall_file($k$)}%
\footnote{\label{fn:path2}%
  If this file is not in \codefile{scenarios_dir/}, its absolute path
  or its path relative to \codefile{scenarios_dir/} must be given.}.

The first lines of the file may be blank lines or comments beginning
with a \mention{\codetext{!}}.
On the following lines, the first column will be the age in $\unit{Myr}$,
and the second one, the normalized infall rate at that time.
%

\subsubsection{Star formation parameters}
\label{sec:SF_param}
\begin{RENVOI}
  \renvoi\fullref{sec:chem_evol}, and \fullref{app:SF_list}.
\end{RENVOI}

More than one episode of star formation%
\footnoteref{fn:max_epis}
can occur. 
Each episode is characterized by its rate at age $t$, $\SFR_k(t)$.
The total star formation rate is
\begin{equation}
\SFR(t) = \sum_k \SFR_k(t).
\end{equation}
For each $k$, the value of $\SFR_k(t)$ is computed using the values of the 
parameters defined in this section.

If the mass of stars formed in one convolution time-step, $\SFR(t)*\Delta t$, 
is larger than the mass of the ISM
at that age, $\MISM(t)$, all the $\SFR_k(t)$ are scaled down by the same factor
so that $\SFR(t)*\Delta t = \MISM(t)$\footnoteref{fn:warn}. 

 
\paragraph{%
   \codetext{SF_begin_time($k$)},
   \codetext{SF_end_time($k$)}%
}
\index{SF_begin_time@\codetext{SF_begin_time}}%
\index{SF_end_time@\codetext{SF_end_time}}%
Whatever the value of $\SFR_k(t)$ computed using other parameters,
\begin{equation}
\forall~t \not\in \interv[\ti_k, \tf_k[,\quad \SFR_k(t) = 0,
\end{equation}
where $\ti_k$ and $\tf_k$ are the values of \codetext{SF_begin_time($k$)}
and \codetext{SF_end_time($k$)}.


\paragraph{%
  $\codetext{SF_type($k$)} = \codetext{"none"}$  \notInToc{(default)}%
}
\index{SF_type@\codetext{SF_type}|(}%
\index{SF_type@\codetext{SF_type}!none@\codetext{\dq none\dq}}%
\kern-\baselineskip
\begin{equation}
\forall~t,\quad \nSFR_k(t) = 0.
\end{equation}

 
\paragraph{%
  $\codetext{SF_type($k$)} = \codetext{"instantaneous"}$:
  \codetext{SF_inst_mass($k$)}%
}
\index{SF_type@\codetext{SF_type}!instantaneous@\codetext{\dq instantaneous\dq}}%
\index{SF_inst_mass@\codetext{SF_inst_mass}}%
\kern-\baselineskip
\begin{equation}
\nSFR_k(t) = f_k*\dirac(t-\ti_k),
\end{equation}
where $f_k$ is the value of \codetext{SF_inst_mass($k$)}, 
$\ti_k$ is that of \codetext{SF_begin_time($k$)},
and $\dirac$ is the Dirac distribution.
(In practice, the star formation event occurs at the nearest convolution time.)

 
\paragraph{%
  $\codetext{SF_type($k$)} = \codetext{"constant"}$:
  \codetext{SF_const_mass($k$)}%
}
\index{SF_type@\codetext{SF_type}!constant@\codetext{\dq constant\dq}}%
\index{SF_const_mass@\codetext{SF_const_mass}}%
\kern-\baselineskip
\begin{equation}
\forall~t\in\interv[\ti_k, \tf_k[,\quad \nSFR_k(t) = \frac{f_k}{\tf_k-\ti_k},
\end{equation}
where $f_k$ is the value of \codetext{SF_const_mass($k$)}.


\paragraph{%
  $\codetext{SF_type($k$)} = \codetext{"exponential"}$:
  \codetext{SF_expo_timescale($k$)}, 
  \codetext{SF_expo_mass($k$)}%
}
\index{SF_type@\codetext{SF_type}!exponential@\codetext{\dq exponential\dq}}%
\index{SF_expo_timescale@\codetext{SF_expo_timescale}}%
\index{SF_expo_mass@\codetext{SF_expo_mass}}%
\kern-\baselineskip
\begin{equation}
\forall~t\in\interv[\ti_k, \tf_k[,\quad 
\nSFR_k(t) = \frac{f_k}{\valabs{\tau_k}}*
\exp\left(-\frac{t-\ti_k}{\tau_k}\right),
\end{equation}
where $f_k$ is the value of \codetext{SF_expo_mass($k$)} 
and $\tau_k$ that of \codetext{SF_expo_timescale($k$)}.
A positive value of $\tau_k$ corresponds to a decreasing star formation rate; 
a negative value, to an increasing one.


\paragraph{%
  $\codetext{SF_type($k$)} = \codetext{"peaked"}$:
  \codetext{SF_peaked_timescale($k$)}, 
  \codetext{SF_peaked_mass($k$)}%
}
\index{SF_type@\codetext{SF_type}!peaked@\codetext{\dq peaked\dq}}%
\index{SF_peaked_timescale@\codetext{SF_peaked_timescale}}%
\index{SF_peaked_mass@\codetext{SF_peaked_mass}}%
\kern-\baselineskip
\begin{equation}
\forall~t\in\interv[\ti_k, \tf_k[,\quad 
\nSFR_k(t) = f_k*\frac{(t-\ti_k)}{\tau_k^2}
*\exp\left(-\frac{t-\ti_k}{\tau_k}\right),
\end{equation}
where $f_k$ is the value of \codetext{SF_peaked_mass($k$)} 
and $\tau_k$ that of \codetext{SF_peaked_timescale($k$)}.
(The star formation rate increases during $\tau_k$ after $\ti_l$, and then decreases with a
timescale of $\tau_k$, as in \citet{Sandage}.)


\paragraph{%
  $\codetext{SF_type($k$)} = \codetext{"ISM_mass"}$:
  \codetext{SF_ISM_timescale($k$)}, 
  \codetext{SF_ISM_power($k$)}, \raggedallowbreak
  \codetext{SF_ISM_threshold($k$)}
  \mbox{($\approxequiv{}$ Schmidt-Kennicutt law)}%
}
\index{SF_type@\codetext{SF_type}!ISM_mass@\codetext{\dq ISM_mass\dq}}%
\index{SF_ISM_timescale@\codetext{SF_ISM_timescale}}%
\index{SF_ISM_power@\codetext{SF_ISM_power}}%
\index{SF_ISM_threshold@\codetext{SF_ISM_threshold}}%
\kern-\baselineskip
\begin{equation}
\forall~t\in\interv[\ti_k, \tf_k[,\quad 
\nSFR_k(t) = \frac{\left(\norm\MISM[t]-\sigma_k\right)^{\alpha_k}}{\tau_k},
\end{equation}
where $\norm\MISM(t)$ is the normalized mass of the interstellar medium 
at age $t$, $\tau_k$ is the value of \codetext{SF_ISM_time\-scale($k$)},
$\alpha_k$ that of \codetext{SF_ISM_power($k$)} and 
$\sigma_k$ that of \codetext{SF_ISM_threshold($k$)}.
If $\norm\MISM(t) < \sigma_k$, $\nSFR_k(t) = 0$.


\paragraph{%
  $\codetext{SF_type($k$)} = \codetext{"infall"}$:
  \codetext{SF_infall_factor($k$)}%
}
\index{SF_type@\codetext{SF_type}!infall@\codetext{\dq infall\dq}}%
\index{SF_infall_factor@\codetext{SF_infall_factor}}%
\kern-\baselineskip
\begin{equation}
\forall~t\in\interv[\ti_k, \tf_k[,\quad 
\nSFR_k(t) = f_k*\nIfR(t),
\end{equation}
where $f_k$ is the value of \codetext{SF_infall_factor($k$)}.


\paragraph{%
  $\codetext{SF_type($k$)} = \codetext{"file"}$:
  \codetext{SF_file($k$)}%
}
\index{SF_type@\codetext{SF_type}|)}%
\index{SF_type@\codetext{SF_type}!file@\codetext{\dq file\dq}}%
\index{SF_file@\codetext{SF_file}}%
For all $t \in \interv[\ti_k, \tf_k[$, $\nSFR_k(t)$ is interpolated 
from the values read in a file.
The name of the latter is stored in parameter \codetext{SF_file($k$)}%
\footnoteref{fn:path2}. 

The first lines of the file may be blank lines or comments beginning 
with a \mention{\codetext{!}}.
On the following lines, the first column will be the age in $\unit{Myr}$, 
and the second one, the normalized star formation rate at that time.


\paragraph{%
  \codetext{SF_stochastic($k$)}:
  \codetext{SF_stoch_fluc($k$)}, 
  \codetext{SF_stoch_timescale($k$)}%
}
\label{sec:SF_stoch}
\index{SF_stochastic@\codetext{SF_stochastic}}%
\index{SF_stoch_fluc@\codetext{SF_stoch_fluc}}%
\index{SF_stoch_timescale@\codetext{SF_stoch_timescale}}%
If the value of \codetext{SF_stochastic($k$)} is \codetext{.true.},
the star formation rate of the $k$-th episode
is stochastically modulated by a log-normal function:
\begin{equation}
\SFR_k(t) = \exp(\sigma_k*\xi-\sigma_k^2/2)*\SFR_{k, 0}(t),
\end{equation}
where 
\begin{equation}
  \sigma_k = \sqrt{\ln(1+\text{\codetext{SF_stoch_fluc($k$)}}^2)}
\end{equation}
and
$\SFR_{k, 0}$ is the star formation rate in the absence of
stochastic modulation.
This expression ensures that $\SFR_k(t) \ge 0$, that the expectation of
$\SFR_k(t)$ is $\SFR_{k, 0}(t)$
and that the standard deviation of $\SFR_k(t)$ is 
$\text{\codetext{SF_stoch_fluc($k$)}} \times\SFR_{k, 0}(t)$.
The number $\xi$ is a Gaussian random deviate with mean~$0$ and variance~$1$;
it is constant during a duration drawn from a Poissonian distribution
of mean given by\codetext{SF_stoch_timescale($k$)}%
\footnote{See~\fullref{sec:random}, to define the sequence of random numbers used for this purpose.}.


\paragraph{%
  $\codetext{SF_Z_type($k$)} = \codetext{"consistent"}$  \notInToc{(default)}%
}
\index{SF_Z_type@\codetext{SF_Z_type}|(}%
\index{SF_Z_type@\codetext{SF_Z_type}!consistent@\codetext{\dq consistent\dq}}%
The metallicity of new stars formed in episode $k$ is the same as that
of the ISM.


\paragraph{%
  $\codetext{SF_Z_type($k$)} = \codetext{"constant"}$:
  \codetext{SF_Z_const_val($k$)}%
}
\index{SF_Z_type@\codetext{SF_Z_type}!constant@\codetext{\dq constant\dq}}%
\index{SF_Z_const_val@\codetext{SF_Z_const_val}}%
The metallicity of new stars formed in episode $k$ is a constant given by the value
\codetext{SF_Z_const_val($k$)}.


\paragraph{%
  $\codetext{SF_Z_type($k$)} = \codetext{"file"}$:
  \codetext{SF_Z_file($k$)}%
}
\index{SF_Z_type@\codetext{SF_Z_type}|)}%
\index{SF_Z_type@\codetext{SF_Z_type}!file@\codetext{\dq file\dq}}%
\index{SF_Z_file@\codetext{SF_Z_file}}%
The metallicity of new stars formed in episode $k$ is interpolated
from the values read in a file.
The name of the latter is stored in parameter \codetext{SF_Z_file($k$)}%
\footnoteref{fn:path2}. 

The first lines of the file may be blank lines or comments beginning 
with a \mention{\codetext{!}}.
On the following lines, the first column will be the age in $\unit{Myr}$,
and the second one, the metallicity at that time.


\paragraph{%
  \codetext{SF_inert_frac($k$)}  \notInToc{(default: $0$)}%
}
\index{SF_inert_frac@\codetext{SF_inert_frac}}%
For the purpose of the code, 
inert objects are objects, such as brown dwarfs or planets,
which formed from the ISM with live stars but 
are not taken into account in
the IMF provided to \codefile{SSPs}\index{SSPs@\codefile{SSPs}}%
\footnote{Note that objects already included in the IMF are considered
  as live stars even if their mass is in fact lower than the upper mass for a brown 
  dwarf.}. 
Contrary to live stars, inert objects are assumed to be dark and 
unevolving since their birth: they just lock mass.
The formation rate in episode~$k$ of inert objects
is $f_k*\SFR_k(t)$, where $f_k$ is the value of 
\codetext{SF_inert_frac($k$)}. 
This parameter changes the mass-to-light ratio and the chemical evolution.
(The formation rate of live stars is $(1-f_k)*\SFR_k(t)$.)
%

\subsubsection{Outflow parameters}
\label{sec:outflow_param}
\begin{RENVOI}
  \renvoi\fullref{sec:zones}, and \fullref{app:outflow_list}.
\end{RENVOI}

More than one episode of outflow%
\footnoteref{fn:max_epis}
from the galaxy into the intergalactic medium can occur. 
Each episode is characterized by its rate at age $t$, $\OfR_k(t)$.
Except as mentioned in \fullref*{sec:outflow_radical}, the total outflow rate is
\begin{equation}
\OfR(t) = \sum_k \OfR_k(t).
\end{equation}
For each $k$, the value of $\OfR_k(t)$ is computed using the values of the 
parameters defined in this section.

If the mass expelled from the galaxy in one convolution time-step, $\OfR(t)*\Delta t$, 
is larger than the mass $\MISM(t)$ of the ISM
at that age, all the $\OfR_k(t)$ are scaled down by the same factor
so that $\OfR(t)*\Delta t = \MISM(t)$\footnoteref{fn:warn}. 


\paragraph{\codetext{outflow_begin_time($k$)}, 
  \codetext{outflow_end_time($k$)}}
\index{outflow_begin_time@\codetext{outflow_begin_time}}%
\index{outflow_end_time@\codetext{outflow_end_time}}%
Whatever the value of $\OfR_k(t)$ computed using other parameters,
\begin{equation}
\forall~t \not\in \interv[\ti_k, \tf_k[,\quad \OfR_k(t) = 0,
\end{equation}
where $\ti_k$ and $\tf_k$ are the values of \codetext{outflow_begin_time($k$)}
and \codetext{outflow_end_time($k$)}.


\paragraph{$\codetext{outflow_type($k$)} = \codetext{"none"}$ \notInToc{(default)}}
\index{outflow_type@\codetext{outflow_type}|(}%
\index{outflow_type@\codetext{outflow_type}!none@\codetext{\dq none\dq}}%
\kern-\baselineskip
\begin{equation}
\forall~t,\quad \OfR_k(t) = 0.
\end{equation}


\paragraph{$\codetext{outflow_type($k$)} = \codetext{"radical"}$}
\label{sec:outflow_radical}
\index{outflow_type@\codetext{outflow_type}!radical@\codetext{\dq radical\dq}}%
For all $t\in\interv[\ti_k, \tf_k[$, all the ISM present in the galaxy is
instantaneously expelled.

 
\paragraph{$\codetext{outflow_type($k$)} = \codetext{"instantaneous"}$: 
  \codetext{outflow_inst_mass($k$)}}
\index{outflow_type@\codetext{outflow_type}!instantaneous@\codetext{\dq instantaneous\dq}}%
\index{outflow_inst_mass@\codetext{outflow_inst_mass}}%
\kern-\baselineskip
\begin{equation}
\forall~t\in\interv[\ti_k, \tf_k[,\quad \nOfR_k(t) = f_k*\dirac(t-\ti_k),
\end{equation}
where $f_k$ is the value of \codetext{outflow_inst_mass($k$)}, 
$\ti_k$ is that of \codetext{outflow_begin_time($k$)},
and $\delta$ is the Dirac distribution.
(In practice, the outflow event occurs at the nearest convolution time.)

 
\paragraph{$\codetext{outflow_type($k$)} = \codetext{"constant"}$:
  \codetext{outflow_const_mass($k$)}}
\index{outflow_type@\codetext{outflow_type}!constant@\codetext{\dq constant\dq}}%
\index{outflow_const_mass@\codetext{outflow_const_mass}}%
\kern-\baselineskip
\begin{equation}
\forall~t \in \interv[\ti_k, \tf_k[,\quad \nOfR_k(t) = \frac{f_k}{\tf_k-\ti_k},
\end{equation}
where $f_k$ is the value of \codetext{outflow_const_mass($k$)}.


\paragraph{$\codetext{outflow_type($k$)} = \codetext{"SF"}$:
  \codetext{outflow_SF_factor($k$)},
  \codetext{outflow_SF_power($k$)}}
\index{outflow_type@\codetext{outflow_type}!SF@\codetext{\dq SF\dq}}%
\index{outflow_SF_factor@\codetext{outflow_SF_factor}}%
\index{outflow_SF_power@\codetext{outflow_SF_power}}%
\kern-\baselineskip
\begin{equation}
\forall~t \in \interv[\ti_k, \tf_k[,\quad \OfR_k(t) = 
f_k*\frac{\SFR(t)}{\bigl(\norm\MISM[t]\bigr)^{\alpha_k}},
\end{equation}
where $f_k$ is the value of \codetext{outflow_SF_factor($k$)}
and $\alpha_k$ that of \codetext{outflow_SF_power($k$)}
($\alpha_k = 0$ by default%
\footnote{So, $\OfR_k(t) = f_k*\SFR(t)$, and this, with the convention $0^0=1$,
  even if $\MISM(t) = 0$.
}, 
as in most simulations \citep{DallaVechia+Schaye},
but \citet{Sharma+Nath} recommend to take $\alpha_k \approx 1$).
This rate may be appropriate if outflows in the IGM are due not only to
supernovae but also to the stellar winds of high-mass stars.


\paragraph{$\codetext{outflow_type($k$)} = \codetext{"SN"}$: 
  \codetext{outflow_SN_mass($k$)}, \codetext{outflow_SN_power($k$)}~/ \raggedallowbreak
  $\codetext{outflow_type($k$)} = \codetext{"CCSN"}$: 
  \codetext{outflow_CCSN_mass($k$)}, \codetext{outflow_CCSN_power($k$)}~/ \raggedallowbreak
  $\codetext{outflow_type($k$)} = \codetext{"SNIa"}$: \codetext{outflow_SNIa_mass($k$)}, \codetext{outflow_SNIa_power($k$)}}
\index{outflow_type@\codetext{outflow_type}!SN@\codetext{\dq SN\dq}}%
\index{outflow_SN_mass@\codetext{outflow_SN_mass}}%
\index{outflow_SN_power@\codetext{outflow_SN_power}}%
If $\codetext{outflow_type($k$)} = \codetext{"SN"}$, then
\begin{equation}
\label{eq:outflow_SN}
\forall~t \in \interv[\ti_k, \tf_k[,\quad \OfR_k(t) = 
f_k*\frac{\dot n_{\SN}(t)}{\bigl(\norm\MISM[t]\bigr)^{\alpha_k}},
\end{equation}
where $\dot n_{\SN}$ is the number rate of supernovae (both core-collapse and
type~Ia), $f_k = \codetext{outflow_SN_mass($k$)}$ is the 
\emph{unnormalized} mass of ISM expelled from the galaxy by one supernova
and $\alpha_k$ is the value of \codetext{outflow_SN_power($k$)}.

With \mention{\codetext{CCSN}} (\resp\ \mention{\codetext{SNIa}}) instead of \mention{\codetext{SN}}
in the value of \codetext{outflow_type} and the names of parameters \codetext{outflow_SN_mass}
and \codetext{outflow_SN_power}, the number rate of core-collapse
(\resp\ type~Ia) supernovae replaces the number rate of all supernova types in \fullref*{eq:outflow_SN}.


\paragraph{$\codetext{outflow_type($k$)} = \codetext{"ejecta"}$:
  \codetext{outflow_ejec_threshold($k$)}, \raggedallowbreak \codetext{outflow_ejec_factor($k$)}}
\index{outflow_type@\codetext{outflow_type}!ejecta@\codetext{\dq ejecta\dq}}%
\index{outflow_ejec_threshold@\codetext{outflow_ejec_threshold}}%
\index{outflow_ejec_factor@\codetext{outflow_ejec_factor}}%
\kern-\baselineskip
\begin{equation}
  \label{eq:outflow_ejec}
  \forall~t \in \interv[\ti_k, \tf_k[,\quad \OfR_k(t) = 
  f_k*\max\Biggl(0, 1-\frac{\MISM(t)}{
      \sigma_k*\int_{t'=0}^t \SFR(t')*\df t'}\Biggr)
  * \dot M_{\txt{ej}}(t),
\end{equation}
where $f_k$ is the value of \codetext{outflow_ejec_factor($k$)},
$\sigma_k$ that of \codetext{outflow_ejec_threshold($k$)} and
$\dot M_{\txt{ej}}$ is the mass ejection rate of stars in the ISM.

Note that, although the outflow rate is related to the mass ejection rate of stars in the ISM,
the outflow affects all the matter in the ISM, not only stellar ejecta.

In early-type galaxies, old low-mass stars eject enough matter in the ISM to 
feed a residual star formation at late ages, and this with a rate nearly 
independent from the gas-to-stars conversion factor if the mass of the ISM 
decreases (see sec.~3.2.1 in \citet{Pegase.1} and chap.~3, sec.~4 in \citet{these_Fioc_doc}). 
Conversely, if stellar ejecta are not used to form stars, they accumulate in 
the galaxy and cool down.
As neither substantial star formation nor large amounts of cold gas are 
observed in early-type galaxies, most of the late stellar ejecta must be 
expelled in the IGM through galactic winds (or be heaten up)%
\footnote{Another possibility is that only low-mass stars are formed 
  \citep{Fabian}.}.
The purely phenomenological modeling implemented through 
\fullref*{eq:outflow_ejec}, inspired from \citet{Larson}%
\footnote{We however use the mass of stars ever formed at $t$, 
  $\int_{t'=0}^t \SFR(t')*\df t'$, instead of the mass in stars.},
is aimed to rid the galaxy from its ISM, 
and thus to starve star formation, when the ISM content drops below some 
threshold (typically $\la 0.1$):
the lower the threshold, the later the age when \mention{galactic winds} occur 
and the larger the mean stellar metallicity.
A more recent paper by \citet{GW:AGB} 
also proposes that the ejecta of dying 
low-mass stars prevent residual star formation in quiescent galaxies.

 
\paragraph{$\codetext{outflow_type($k$)} = \codetext{"file"}$:
  \codetext{outflow_file($k$)}}
\index{outflow_type@\codetext{outflow_type}|)}%
\index{outflow_type@\codetext{outflow_type}!file@\codetext{\dq file\dq}}%
\index{outflow_file@\codetext{outflow_file}}%
For all $t \in \interv[\ti_k, \tf_k[$, $\nOfR_k(t)$ is interpolated 
from the values read in a file.
The name of the latter is stored in parameter \codetext{outflow_file($k$)}%
\footnoteref{fn:path2}.

The first lines of the file may be blank lines or comments beginning 
with a \mention{\codetext{!}}.
On the following lines, the first column will be the age in $\unit{Myr}$,
and the second one, the normalized outflow rate at that time.
%

\subsubsection{Dust evolution parameters}
\label{sec:dust_evol_param}
\begin{RENVOI}
  \renvoi \fullref{app:dust_evol_list}.
\end{RENVOI}

The meaning of the code parameters used to model the evolution of dust
is given in \fullref{tab:dust_evol_meaning}.
\paragraph{$\codetext{dust_evolution} = \codetext{"basic"}$: 
  \codetext{ISM_carb_deplet}, 
  \codetext{ISM_sil_deplet}}
If $\codetext{dust_evolution} = \codetext{"basic"}$, the dust evolution model described in 
\fullref{sec:dust_evol_basic}, is used.%
\index{dust_evolution@\codetext{dust_evolution}|(}%
\index{dust_evolution@\codetext{dust_evolution}!basic@\codetext{\dq basic\dq}}%
\index{ISM_carb_deplet@\codetext{ISM_carb_deplet}}%
\index{ISM_sil_deplet@\codetext{ISM_sil_deplet}}%
\index{O_sil_ratio@\codetext{O_sil_ratio}}%
\paragraph{$\codetext{dust_evolution} = \codetext{"Dwek"}$:
  \codetext{HMW_carb_deplet},
  \codetext{LMW_carb_deplet}, \raggedallowbreak
  \codetext{CCSN_carb_deplet},
  \codetext{SNIa_carb_deplet},
  \codetext{HMW_sil_deplet}, \raggedallowbreak
  \mbox{\codetext{LMW_sil_deplet}},
  \codetext{CCSN_sil_deplet},
  \codetext{SNIa_sil_deplet}, \raggedallowbreak
  \codetext{SN_swept_mass},
  \codetext{carb_accr_timescale},
  \codetext{sil_accr_timescale}}
If $\codetext{dust_evolution} = \codetext{"Dwek"}$, the dust evolution model described in 
\fullref{sec:dust_evol_Dwek} is used.
\index{dust_evolution@\codetext{dust_evolution}!Dwek@\codetext{\dq Dwek\dq}}%
\index{CCSN_carb_deplet@\codetext{CCSN_carb_deplet}}%
\index{SNIa_carb_deplet@\codetext{SNIa_carb_deplet}}%
\index{HMW_carb_deplet@\codetext{HMW_carb_deplet}}%
\index{LMW_carb_deplet@\codetext{LMW_carb_deplet}}%
\index{CCSN_sil_deplet@\codetext{CCSN_sil_deplet}}%
\index{SNIa_sil_deplet@\codetext{SNIa_sil_deplet}}%
\index{HMW_sil_deplet@\codetext{HMW_sil_deplet}}%
\index{LMW_sil_deplet@\codetext{LMW_sil_deplet}}%
\index{O_sil_ratio@\codetext{O_sil_ratio}}%
\index{SN_swept_mass@\codetext{SN_swept_mass}}%
\index{carb_accr_timescale@\codetext{carb_accr_timescale}}%
\index{sil_accr_timescale@\codetext{sil_accr_timescale}}%
\paragraph{\codetext{O_sil_ratio}}
Common to both models of dust evolution.

\begin{table}[h]
  \caption{\label{tab:dust_evol_meaning}%
    Meaning of code parameters for dust evolution models\captionPoint}
  \begin{center}
    \begin{tabular}{l@{\TBstrut\hskip2\tabcolsep\hskip2cm}l@{\hskip2\tabcolsep\hskip2cm}l}
      \hline
      \multicolumn{1}{c@{\TBstrut\hskip2\tabcolsep\hskip2cm}}{Code parameter} &
      \multicolumn{1}{@{\hskip-2cm}c@{\hskip2\tabcolsep\hskip2cm}}{
        Notation used in the modeling} &
      \multicolumn{1}{@{\hskip-2cm}c@{\hskip\tabcolsep}}{Equations where the quantity appears}\\
      \hline\hline
      \multicolumn{3}{@{\TBstrut}c@{}}{{\bfseries If $\codetext{dust_evolution} = \codetext{"basic"}$:}}\\
      \codetext{ISM_carb_deplet} & $\delta_{\txt{carb}}^{\ISM}$ & \fullref{eq:M_carb_formed_anew};\\
      \codetext{ISM_sil_deplet} & $\delta_{\txt{sil}}^{\ISM}$ & \fullref{eq:M_sil_formed_anew}.\\
      \hline
      \multicolumn{3}{@{\TBstrut}c@{}}{{\bfseries If $\codetext{dust_evolution} = \codetext{"Dwek"}$:}}\\
      \codetext{HMW_carb_deplet} & $\delta_{\txt{carb}}^{\textsc{hmw}}$ & \fullref{eq:LMW_carb};\\
      \codetext{LMW_carb_deplet} & $\delta_{\txt{carb}}^{\textsc{lmw}}$ & Idem;\\
      \codetext{CCSN_carb_deplet} & $\delta_{\txt{carb}}^{\CCSN}$ & \fullref{eq:SN_carb};\\
      \codetext{SNIa_carb_deplet} & $\delta_{\txt{carb}}^{\SNIa}$ & Idem;\\
      \codetext{HMW_sil_deplet} & $\delta_{\txt{sil}}^{\textsc{hmw}}$ & \fullref{eq:LMW_sil};\\
      \codetext{LMW_sil_deplet} & $\delta_{\txt{sil}}^{\textsc{lmw}}$ & Idem;\\
      \codetext{CCSN_sil_deplet} & $\delta_{\txt{sil}}^{\CCSN}$ & \fullref{eq:SN_sil};\\
      \codetext{SNIa_sil_deplet} & $\delta_{\txt{sil}}^{\SNIa}$ & Idem;\\
      \codetext{SN_swept_mass} & $m_{\txt{swept}}$ & \fullref{eq:dust_destr};\\
      \codetext{carb_accr_timescale} & $\tau_{\txt{carb}}^{\txt{accr}}$ & \fullref{eq:carb_accr};\\
      \codetext{sil_accr_timescale} & $\tau_{\txt{sil}}^{\txt{accr}}$ & \fullref{eq:sil_accr}.\\
      \hline 
      \multicolumn{3}{@{\TBstrut}c@{}}{{\bfseries For both dust evolution models:}}\\
      \codetext{O_sil_ratio} & $\OToSil$ & \fullref{eq:M_sil_formed_anew}.\\
      \hline
    \end{tabular}
  \end{center}
\end{table}
\index{dust_evolution@\codetext{dust_evolution}|)}%
%

\subsubsection{Dust attenuation and emission parameters}
\label{sec:dust_transfer_param}
\begin{RENVOI}
  \renvoi \fullref{app:dust_transfer_list}.
  \forbiddenBreak
\end{RENVOI}
\forbiddenBreak
\paragraph{\codetext{extinction}, \codetext{extinction_SFC}, \codetext{extinction_DISM} \notInToc{(default: \codetext{.false.})}}%
\index{extinction@\codetext{extinction}}%
\index{extinction_SFC@\codetext{extinction_SFC}}%
\index{extinction_DISM@\codetext{extinction_DISM}}%
\begin{RENVOI}
  \renvoi\fullref{sec:diffuse_attenuation}.
\end{RENVOI}
The extinction due to dust grains in clouds (\resp the diffuse medium) 
is computed if and only if the value of parameter
\codetext{extinction_SFC} (\resp \codetext{extinction_DISM}) 
is \codetext{.true.}.

The values of \codetext{extinction_SFC} and \codetext{extinction_DISM}
may be given at once through the explicit assignment of 
parameter \codetext{extinction}%
\footnote{%
  \label{fn:all_SFC_DISM}%
  For instance, setting \codetext{extinction} to \codetext{.true.} or \codetext{.false.} immediately sets 
  \codetext{extinction_SFC} and \codetext{extinc\-tion_DISM} to this same value. 

  Note that the parameters \codetext{extinction}, \codetext{grains_file},
  \codetext{dust_emission}, \codetext{stoch_heating}, \codetext{self_abs_power},
  \codetext{carb_sublim_temp}, \codetext{sil_sublim_temp}, \codetext{nebular_emission}, \codetext{neb_emis_type}
  and \codetext{neb_emis_const_frac} are not used
  in the calculations. Only the \mbox{\codetext{\usertext{*}_SFC}} and 
  \mbox{\codetext{\usertext{*}_DISM}} variants of these parameters matter
  and are written in the output file or are printed by command \codetext{echo}.
}.

 
\paragraph{\codetext{grains_file}, \codetext{grains_file_SFC}, \codetext{grains_file_DISM}}
\index{grains_file@\codetext{grains_file}}%
\index{grains_file_SFC@\codetext{grains_file_SFC}}%
\index{grains_file_DISM@\codetext{grains_file_DISM}}%
\index{grains_file@\codetext{grains_file}!ZDA.txt@\codetext{\dq ZDA.txt\dq}}%
\index{grains_file@\codetext{grains_file}!LWD.txt@\codetext{\dq LWD.txt\dq}}%
\index{grains_file@\codetext{grains_file}!MRN.txt@\codetext{\dq MRN.txt\dq}}%
\label{sec:GS_param}%
\begin{RENVOI}
  \renvoi \fullref{sec:GSD}
\end{RENVOI}
\codetext{grains_file_SFC} (\resp \codetext{grains_file_DISM})%
\footnoteref{fn:all_SFC_DISM}
holds the name of the \Emph{file of grains}\index{file of grains}
providing the size distribution in clouds (\resp the diffuse medium)
of the various species of dust grains%
\footnoteref{fn:once}.
This file also points to files (one for each species) giving the optical properties of individual grains.
Currently available files of grains are
\codetext{"ZDA.txt"}, \codetext{"LWD.txt"} and \codetext{"MRN.txt"}.

If the value of \codetext{extinction_SFC} 
(\resp \codetext{extinction_DISM}) is \codetext{.false.},
assigning \codetext{grains_file_SFC} 
(\resp \codetext{grains_file_DISM}) has no effect.

The values of \codetext{grains_file_SFC} and \codetext{grains_file_DISM} 
may be given at once through the explicit assignment of parameter \codetext{grains_file}%
\footnoteref{fn:once}%
\footnoteref{fn:all_SFC_DISM}. 


\paragraph{\codetext{geometry}}
\index{geometry@\codetext{geometry}}%
\index{geometry@\codetext{geometry}!spiral@\codetext{\dq spiral\dq}}%
\index{geometry@\codetext{geometry}!spheroidal@\codetext{\dq spheroidal\dq}}%
\index{geometry@\codetext{geometry}!slab@\codetext{\dq slab\dq}}%
\index{bulge_tot_ratio@\codetext{bulge_tot_ratio}}%
\index{M_sys_spiral@\codetext{M_sys_spiral}}%
\index{expo_radius@\codetext{expo_radius}}%
\index{M_sys_spher@\codetext{M_sys_spher}}%
\index{core_radius@\codetext{core_radius}}%
\index{slab_factor@\codetext{slab_factor}}%
\label{sec:geometry_param}%
If $\codetext{geometry} = \codetext{"spiral"}$, the spatial distribution of stars and dust described in
\fullref{sec:spiral}, is used.

If $\codetext{geometry} = \codetext{"spheroidal"}$, the spatial distribution of stars and dust described in
\fullref{sec:spher}, is used.

If $\codetext{geometry} = \codetext{"slab"}$, the spatial distribution of stars and dust described in
\fullref{sec:slab}, is used.

\begin{table}[h]
  \caption{Meaning of code parameters related to the spatial distribution of stars and dust\captionPoint}
  \begin{center}
    \begin{tabular}{l@{\TBstrut\hskip2\tabcolsep\hskip2cm}l@{\hskip2\tabcolsep\hskip2cm}l}
      \hline
      \multicolumn{1}{c@{\TBstrut\hskip2\tabcolsep\hskip2cm}}{Code parameter} &
      \multicolumn{1}{@{\hskip-2cm}c@{\hskip2\tabcolsep\hskip2cm}}{
        Notation used in the modeling} &
      \multicolumn{1}{@{\hskip-2cm}c@{\hskip\tabcolsep}}{Sections or equations where the quantity appears}\\
      \hline\hline
      \multicolumn{3}{@{\TBstrut}c@{}}{{\bfseries If $\codetext{geometry} = \codetext{"spiral"}$:}}\\
      \codetext{bulge_tot_ratio} & $\bulgeToTot$ & Eq.~(17) of \AApaper;\\
      \codetext{M_sys_spiral} & $\Mref^\spir$ & Sec.~4.1 of \AApaper;\\
      \codetext{expo_radius} & $R_\DD$ & Eq.~(13) of \AApaper.\\
      \hline
      \multicolumn{3}{@{\TBstrut}c@{}}{{\bfseries If $\codetext{geometry} = \codetext{"spheroidal"}$:}}\\
      \codetext{M_sys_spher} & $\Mref^\sph$ & Sec.~4.2 of \AApaper;\\
      \codetext{core_radius} & $\Rcore$ & Eq.~(18) of \AApaper.\\
      \hline
      \multicolumn{3}{@{\TBstrut}c@{}}{{\bfseries If $\codetext{geometry} = \codetext{"slab"}$:}}\\
      \codetext{slab_factor} & $\alpha_\txt{slab}$ & \fullref{eq:slab}.\\
      \hline
    \end{tabular}
  \end{center}
\end{table}


\paragraph{\codetext{inclin_averaged},
  \codetext{inclination}}
\index{inclin_averaged@\codetext{inclin_averaged}}%
\index{inclination@\codetext{inclination}}%
\label{sec:inclin_param}%
These parameters are not relevant if $\codetext{geometry} = \codetext{"spheroidal"}$.

If $\codetext{inclin_averaged} = \codetext{.true.}$, the
attenuation is averaged over all inclinations.
Otherwise, the attenuation is computed for the inclination
specified by parameter \codetext{inclination}.
The inclination is in degrees and relative to face-on ($0^\degree$ for face-on,
$90^\degree$ for edge-on).


\paragraph{\codetext{dust_emission}, \codetext{dust_emission_SFC}, \codetext{dust_emission_DISM} \notInToc{(default: \codetext{.false.})}}
\index{dust_emission@\codetext{dust_emission}}%
\index{dust_emission_SFC@\codetext{dust_emission_SFC}}%
\index{dust_emission_DISM@\codetext{dust_emission_DISM}}%
\begin{RENVOI}
  \renvoi\fullref{sec:diffuse_emission}.
\end{RENVOI}
The emission produced by dust grains in clouds (\resp the diffuse medium) 
is computed if and only if the value of parameter \codetext{dust_emission_SFC} 
(\resp \codetext{dust_emission_DISM}) is \codetext{.true.}
\emph{and} the value of \codetext{extinction_SFC} 
(\resp \codetext{extinction_DISM}) is \codetext{.true.} too.

The values of \codetext{dust_emission_SFC} and \codetext{dust_emission_DISM} 
may be given at once through the explicit assignment of parameter \codetext{dust_emission}%
\footnoteref{fn:all_SFC_DISM}. 


\paragraph{\codetext{stoch_heating}, \codetext{stoch_heating_SFC},
  \codetext{stoch_heating_DISM} \notInToc{(default: \codetext{.true.})}}
\index{stoch_heating@\codetext{stoch_heating}}%
\index{stoch_heating@\codetext{stoch_heating_SFC}}%
\index{stoch_heating@\codetext{stoch_heating_DISM}}%
\label{sec:stoch_heat_param}%
\begin{RENVOI}
  \renvoi\fullref{sec:stoch_heat}.
\end{RENVOI}
Stochastic heating of dust grains in clouds (\resp the diffuse medium) 
is considered if and only if the value of parameter 
\codetext{stoch_heating_SFC} 
(\resp \codetext{stoch_heating_DISM}) is \codetext{.true.}.

{\emergencystretch=1em If the value of \codetext{extinction_SFC} 
(\resp \codetext{extinction_DISM}) or \codetext{dust_emission_SFC} 
(\resp \codetext{dust_emission_DISM}) is \codetext{.false.},
assigning \codetext{stoch_heating_SFC} 
(\resp \codetext{stoch_heating_DISM}) has no effect.\par}

The values of \codetext{stoch_heating_SFC} and \codetext{stoch_heating_DISM} 
may be given at once through the explicit assignment of parameter \codetext{stoch_heating}%
\footnoteref{fn:all_SFC_DISM}. 


\paragraph{\codetext{carb_sublim_temp}, \codetext{carb_sublim_temp_SFC}, 
  \codetext{carb_sublim_temp_DISM}, \raggedallowbreak
  \codetext{sil_sublim_temp}, \codetext{sil_sublim_temp_SFC}, \codetext{sil_sublim_temp_DISM}}
\index{carb_sublim_temp@\codetext{carb_sublim_temp}}%
\index{carb_sublim_temp_SFC@\codetext{carb_sublim_temp_SFC}}%
\index{carb_sublim_temp_DISM@\codetext{carb_sublim_temp_DISM}}%
\index{sil_sublim_temp@\codetext{sil_sublim_temp}}%
\index{sil_sublim_temp_SFC@\codetext{sil_sublim_temp_SFC}}%
\index{sil_sublim_temp_DISM@\codetext{sil_sublim_temp_DISM}}%
\label{sec:sublim_param}%
\begin{RENVOI}
  \renvoi\fullref{sec:stoch_heat}.
\end{RENVOI}
{\emergencystretch=1em The values of \codetext{carb_sublim_temp_SFC} (\resp \codetext{carb_sublim_temp_DISM}) and 
\codetext{sil_sublim_temp_SFC} (\resp \codetext{sil_sublim_temp_DISM})
are the sublimation temperatures of carbonaceous grains and silicates
in clouds (\resp the diffuse medium).
Note that grains do not actually sublimate in the code, but the SEDs of grains at temperatures above
the sublimation temperature are printed separately, if requested
(see parameter \codetext{sublim_output} in \fullref{sec:spectra_output}).\par}

{\emergencystretch=1em If the value of \codetext{extinction_SFC} 
(\resp \codetext{extinction_DISM}) or \codetext{dust_emission_SFC} 
(\resp \codetext{dust_emission_DISM}) is \codetext{.false.},
assigning the \codetext{\mbox{\usertext{*}_sublim}_temp_SFC} 
(\resp \codetext{\mbox{\usertext{*}_sublim}_temp_DISM}) parameters
has no effect.\par}

The values of \codetext{carb_sublim_temp_SFC} and \codetext{carb_sublim_temp_DISM} 
may be given at once through the explicit assignment of parameter \codetext{carb_sublim_temp}%
\footnoteref{fn:all_SFC_DISM}. 
The same holds, \emph{mutatis mutandis}, for \codetext{sil_sublim_temp_SFC},
\codetext{sil_sublim_temp_DISM} and \codetext{sil_sublim_temp}.


\paragraph{\codetext{self_abs_power}, \codetext{self_abs_power_SFC}, 
  \codetext{self_abs_power_DISM} \notInToc{(defaults: (0, 0, 0))}}
\index{self_abs_power@\codetext{self_abs_power}}%
\index{self_abs_power_SFC@\codetext{self_abs_power_SFC}}%
\index{self_abs_power_DISM@\codetext{self_abs_power_DISM}}%
\label{sec:self_abs_param}%
\begin{RENVOI}
  \renvoi\fullref{sec:self_abs}.
\end{RENVOI}
The value of \codetext{self_abs_power_SFC} (\resp \codetext{self_abs_DISM}) 
is the value in clouds (\resp the diffuse medium) of the parameter $\gamma$ 
used in \fullref{eq:SA}. The default value corresponds to no self-absorption.

{\emergencystretch=1em If the value of \codetext{extinction_SFC} 
(\resp \codetext{extinction_DISM}) or \codetext{dust_emission_SFC} 
(\resp \codetext{dust_emission_DISM}) is \codetext{.false.},
assigning \codetext{self_abs_power_SFC} 
(\resp \codetext{self_abs_power_DISM}) has no effect.\par}

The values of \codetext{self_abs_power_SFC} and \codetext{self_abs_power_DISM} 
may be given at once through the explicit assignment of parameter \codetext{self_abs_power}%
\footnoteref{fn:all_SFC_DISM}. 
%

\subsubsection{Parameters for star-forming clouds and nebular emission}
\label{sec:cloud_neb_param}
\begin{RENVOI}
  \renvoi \AApaper, sec.~6.
  \renvoi \fullref{app:cloud_list}.
\end{RENVOI}


\paragraph{\codetext{cloud_init_frac}, 
  \codetext{cloud_duration},
  \codetext{cloud_power}, 
  \codetext{cluster_stel_mass}}
\index{cloud_init_frac@\codetext{cloud_init_frac}}%
\index{cloud_duration@\codetext{cloud_duration}}%
\index{cloud_power@\codetext{cloud_power}}%
\index{cluster_stel_mass@\codetext{cluster_stel_mass}}%

\codetext{cloud_init_frac}, \codetext{cloud_duration} and \codetext{cloud_power}
are the values of the parameters $\varphi_0$, $\theta$ and $\beta$ 
in \fullref{eq:cloud_frac}.
The parameter
\codetext{cluster_stel_mass} is the typical initial stellar mass of a star cluster 
(quantity $M_\textsc{sc}$ in eq.~(29) of \AApaper).


\paragraph{\codetext{nebular_emission}, \codetext{nebular_emission_SFC}, \codetext{nebular_emission_DISM} \notInToc{(default: \codetext{.false.})}}%
\index{nebular_emission@\codetext{nebular_emission}}%
\index{nebular_emission_SFC@\codetext{nebular_emission_SFC}}%
\index{nebular_emission_SFC@\codetext{nebular_emission_SFC}}%
The emission of the nebular gas in clouds (\resp the diffuse medium) 
is computed if and only if the value of parameter
\codetext{nebular_emission_SFC} (\resp \codetext{nebular_emission_DISM}) 
is \codetext{.true.}.

The values of \codetext{nebular_emission_SFC} and \codetext{nebular_emission_DISM}
may be given at once through the explicit assignment of 
parameter \codetext{nebular_emission}%
\footnoteref{fn:all_SFC_DISM}. 
\subparagraph{\codetext{neb_emis_type} or \codetext{neb_emis_type_SFC} or \codetext{neb_emis_type_DISM}${} = {}$\raggedallowbreak\codetext{"automatic"}\raggedallowbreak \notInToc{(default)}}%
\index{neb_emis_type@\codetext{neb_emis_type}}%
\index{neb_emis_type_SFC@\codetext{neb_emis_type_SFC}}%
\index{neb_emis_type_DISM@\codetext{neb_emis_type_DISM}}%
If \codetext{neb_emis_type_SFC} (\resp \codetext{neb_emis_type_DISM}) is set to \codetext{"automatic"}, 
the fraction of Lyman continuum photons emitted by young stars in star-forming clouds (\resp the diffuse medium)
which are absorbed (in the same region) by gas rather than by dust
is computed as explained in \fullref{sec:cloud_neb}, and app.~A of \AApaper.

The values of \codetext{neb_emis_type_SFC} and \codetext{neb_emis_type_DISM}
may be given at once through the explicit assignment of 
parameter \codetext{neb_emis_type}%
\footnoteref{fn:all_SFC_DISM}. 

\subparagraph{\codetext{neb_emis_type} or \codetext{neb_emis_type_SFC} or \codetext{neb_emis_type_DISM}${} ={}$\raggedallowbreak \codetext{"constant"}: 
\raggedallowbreak\codetext{neb_emis_const_frac}, \codetext{neb_emis_const_frac_SFC}, \raggedallowbreak\codetext{neb_emis_const_frac_DISM}}%
\index{neb_emis_const_frac@\codetext{neb_emis_const_frac}}%
\index{neb_emis_const_frac_SFC@\codetext{neb_emis_const_frac_SFC}}%
\index{neb_emis_const_frac_DISM@\codetext{neb_emis_const_frac_DISM}}%
If \codetext{neb_emis_type_SFC} (\resp \codetext{neb_emis_type_DISM}) is set to \codetext{"constant"}, 
the fraction of Lyman continuum photons emitted by young stars in star-forming clouds (\resp the diffuse medium)
which are absorbed by the surrounding gas is assumed to be constant and is given by \codetext{neb_emis_const_frac_SFC}
(\resp \codetext{neb_emis_const_frac_DISM}).

The values of \codetext{neb_emis_const_frac_SFC} and \codetext{neb_emis_const_frac_DISM}
may be given at once through the explicit assignment of 
parameter \codetext{neb_emis_const_frac}%
\footnoteref{fn:all_SFC_DISM}. 
\subparagraph{\codetext{l10_mean_U_DISM}}%
\index{l10_mean_U_DISM@\codetext{l10_mean_U_DISM}}%
\begin{RENVOI}
  \renvoi \AApaper, sec.~6.2.
\end{RENVOI}
Decimal logarithm of the mean value of the unitless ionization parameter 
in the diffuse interstellar medium.
%

\subsubsection{Output files}
\label{sec:output_param}
\begin{RENVOI}
\renvoi \fullref{app:output_list}.
\end{RENVOI}


\paragraph{Main output file of \codefile{spectra}}
\index{spectra@\codefile{spectra}}%


\subparagraph{\codetext{spectra_output}, 
  \codetext{RF_output}, 
  \codetext{sublim_output}
  \notInToc{(defaults: \codetext{"basic"}, \codetext{.false.}, \codetext{.false.})}}
\index{spectra_output@\codetext{spectra_output}}%
\index{RF_output@\codetext{RF_output}}%
\index{sublim_output@\codetext{sublim_output}}%
\label{sec:spectra_output}%
Parameter \codetext{spectra_output} controls writing in the main output file of \codefile{spectra}\index{spectra@\codefile{spectra}}.

If $\codetext{spectra_output} = \codetext{"basic"}$ (default), only the global SED is printed in the file of spectra
(however, emission lines are distinguished from the continuum).

If $\codetext{spectra_output} = \codetext{"detailed"}$, the SEDs produced by the various galactic components (stars, ionized gas,
grain species; see \fullref{tab:output_detailed}) in the various regions (star-forming clouds and diffuse ISM)
are printed separately, in addition to the global SED.

Whether the value of \codetext{spectra_output} is \codetext{"basic"} or \codetext{"detailed"}, a series of other 
quantities \dashOpen masses of components, chemical composition, star formation and supernova 
rates\textellipsis\dashClos are also written (see \fullref{tab:output_main}).
If $\codetext{spectra_output} = \codetext{"none"}$, no file of spectra is produced.

If \codetext{RF_output} is true, the mean radiation fields in star-forming clouds and the diffuse medium
are written in the file of spectra;
if \codetext{sublim_output} is true, the SEDs of grains at temperatures above the sublimation
temperature are written separately. (See \fullref{tab:output_detailed}.)

{\emergencystretch=1em
The file of spectra may be read with the procedure \codetext{read_spectra_output} defined in
\codefile{mod_read_spectra_output.f90}, as exemplified by codes \codefile{colors} and \codefile{plot_spectra.f90}.\par}

\subparagraph{\codetext{spectra_file},
  \codetext{prefix}, \codetext{stamp_time}, \codetext{overwrite}
  \notInToc{(defaults: \none, \codetext{""}, \codetext{.false.}, \codetext{.false.})}}
\label{sec:spectra_file}%
\index{spectra_file@\codetext{spectra_file}}%
\index{prefix@\codetext{prefix}}%
\index{stamp_time@\codetext{stamp_time}}%
\index{overwrite@\codetext{overwrite}}%
\index{time-stamp}%
The value of \codetext{spectra_file} is the name of the main output file of \codefile{spectra}\index{spectra@\codefile{spectra}}
for the current scenario.
Contrary to other parameters, this value is erased at the beginning of each scenario.
If no value is provided, a default name is built from the date and time at the
beginning of the execution of \codefile{spectra}\index{spectra@\codefile{spectra}} (a \mention{time-stamp}) and from
the ordinal number of the scenario in the file of scenarios\index{file of scenarios}.

If the file of spectra already exists, a modified name is derived from the original name
and from the time-stamp, unless \codetext{overwrite} is \codetext{.true.} (in which case the old file of spectra is replaced 
by the new one).

If $\codetext{prefix} \neq \codetext{""}$, the name of the main output file is built by prepending
the value of \codetext{prefix} to either the value provided for \codetext{spectra_file}, the default name or the modified name. 

In all cases, the time-stamp is inserted between \codetext{prefix} and \codetext{spectra_file} if $\codetext{stamp_time} = \codetext{.true.}$.

Underscores are used to separate the prefix and the time-stamp from what follows.

\subparagraph{\codetext{ages_file}}
\index{ages_file@\codetext{ages_file}}%
{\emergencystretch=1em
The value of \codetext{ages_file} is the name of the file containing the ages (variable \codetext{output_age} in code \codefile{spectra}\index{spectra@\codefile{spectra}})%
\footnote{Actually, \codefile{spectra}\index{spectra@\codefile{spectra}} outputs the spectrum and other
  evolving quantities at the nearest age (\codetext{convol_time})
  for which the state of the galaxy was computed.
  These values are identical if \codetext{output_age} is a multiple of
  the convolution time-step (variable \codetext{time_step}) used in \codefile{spectra}\index{spectra@\codefile{spectra}}
  to convolve the star formation history with the properties of SSPs.}
at which quantities are printed in the file of spectra\index{file of spectra} (one age in $\unit{Myr}$ per line).
This file should be in \codefile{ages_dir/}.\par}


\paragraph{Output files related to grain properties}
\label{sec:output_grain}


\subparagraph{\codetext{grain_temp_output}, 
  \codetext{grain_temp_file}, 
  \codetext{grain_SED_output}, \raggedallowbreak
  \codetext{grain_SED_file}}
\label{sec:grain_files}%
\index{grain_temp_output@\codetext{grain_temp_output}}%
\index{grain_temp_file@\codetext{grain_temp_file}}%
\index{grain_SED_output@\codetext{grain_SED_output}}%
\index{grain_SED_file@\codetext{grain_SED_file}}%
By default, $\codetext{grain_temp_output} = \codetext{.false.}$.
If set to \codetext{.true.}, the temperature probability distributions of individual dust
grains are written in a file of grain temperatures.

The name of this file is given by parameter \codetext{grain_temp_file} and is
processed in the same way as \codetext{spectra_file}.
This name is erased at the beginning of each scenario;
if not provided, a name is created from the processed value of \codetext{spectra_file}.

{\emergencystretch=1em
The output file may be read with the procedure \codetext{read_grain_temp} defined in
\codefile{mod_read_grain_temp.f90}, as exemplified by code \codefile{plot_grain_temp.f90}.\par}
\separate
If $\codetext{grain_SED_output} = \codetext{.true.}$ (default: \codetext{.false.}), 
the spectral energy distributions of individual dust
grains are written in a file of grain SEDs.
The name of this file is given by parameter \codetext{grain_SED_file} (same remarks as for
\codetext{grain_temp_file}).
The output file may be read with the procedure \codetext{read_grain_SED} defined in
\codefile{mod_read_grain_SED.f90}, as exemplified by code \codefile{plot_grain_SED.f90}.
\separate
The parameters \paramname{prefix}, \paramname{stamp_time} and \paramname{overwrite}
also apply to the files of grain temperatures and grain SEDs.%
\index{prefix@\codetext{prefix}}%
\index{stamp_time@\codetext{stamp_time}}%
\index{overwrite@\codetext{overwrite}}%
\index{time-stamp}%
The output files produced if \codetext{grain_temp_output} or 
\codetext{grain_SED_output} are \codetext{.true.} may be huge.
The values of the parameters \codetext{output_grain_*} defined hereafter may be changed to reduce 
their size.


\subparagraph{\codetext{grain_output_SFC}, 
  \codetext{grain_output_DISM}}
\index{grain_output_SFC@\codetext{grain_output_SFC}}%
\index{grain_output_DISM@\codetext{grain_output_DISM}}%
The temperature probability and spectral energy distributions of dust grains in star-forming clouds
are written only if $\codetext{grain_output_SFC} = \codetext{.true.}$ (default).

Idem for dust grains in the diffuse medium if $\codetext{grain_output_DISM} = \codetext{.true.}$.


\subparagraph{\codetext{grain_output_min_age},
  \codetext{grain_output_max_age}}
\index{grain_output_min_age@\codetext{grain_output_min_age}}%
\index{grain_output_max_age@\codetext{grain_output_max_age}}%
{\emergencystretch=1em
Grain temperature and spectral energy distributions are written
for all ages in the interval 
$\interv[$\codetext{grain_output_min_age}$,\allowbreak $\codetext{grain_output_max_age}$]$.
(By default, 
at all the ages defined by \codetext{ages_file}.)\par}


\subparagraph{\codetext{grain_output_min_size}, 
  \codetext{grain_output_max_size}}
\index{grain_output_min_size@\codetext{grain_output_min_size}}%
\index{grain_output_max_size@\codetext{grain_output_max_size}}%
{\emergencystretch=1em
Grain temperature and spectral energy distributions are written
for all grain radii in the interval $\interv[$\codetext{grain_output_min_size}$,\allowbreak$\codetext{grain_output_max_size}$]$.
(By default, 
for all available radii.)\par}
%

\subsection{Random numbers: 
  \codetext{seed}, 
  \codetext{initialize_seed}}
\label{sec:random}%
\index{seed@\codetext{seed}}%
\index{initialize_seed@\codetext{initialize_seed}}%
Random numbers are used to simulate stochastic star formation (see \fullref{sec:SF_stoch});
they are generated from integer seeds $s_1$, \ldots, $s_n$
(beware, the number~$n$ of seeds depends on the compiler!).
To set the seeds to specific values, 
assign the parameters $\codetext{seed($i$)}$ for $i \in \IE[1, n]$.

The seeds generating the sequence of random numbers used for a given scenario
are written in the main output file of \codefile{spectra}\index{spectra@\codefile{spectra}};
search for the line starting with
\Mention[.]{\codetext{seed(1:$n$) = [}} 

To generate the same sequence of random numbers for some other scenario, 
enter this line (and, if necessary, following lines up to the closing \mention{]} included)
in the section of the file of scenarios\index{file of scenarios} describing this other scenario. 
(You may need to break the line; 
see~\fullref{app:syntax}, item \mention{Line length and continuation character}.)

To set the seeds to some random values determined from the computer's clock, use
instead the command \codetext{initialize_seed}.
%

\subsection{End of file and other statements}
\label{sec:other_statements}

 
\subsubsection{End of file}
\index{end of file}%
When the end of the file of scenarios\index{file of scenarios} (the end of the most-outer file in the case of included files; 
see \fullref{sec:included}) is reached,
control is returned to \codefile{spectra}\index{spectra@\codefile{spectra}} and the last scenario is run (normally).


\subsubsection{\codetext{end}}
\index{end@\codetext{end}}%
The \codetext{end} command has the same effect as the end of the file of scenarios\index{file of scenarios}. 
Nothing will be read thereafter. 
Practically, you may write \mention{\codetext{end}} after the last scenario you want to run if this is not the last one
in the file of scenarios\index{file of scenarios}.


\subsubsection{\codetext{return}}
\index{return@\codetext{return}}%
The \codetext{return} command returns control to \codefile{spectra}\index{spectra@\codefile{spectra}};
the most recently read scenario is then run. 
Once this one has been computed, \codefile{spectra}\index{spectra@\codefile{spectra}} reads the next scenario in the file of 
scenarios\index{file of scenarios}.
The string \mention{\codetext{return}} must be written after each scenario (except, possibly, the last one since an
end of file or an \codetext{end} statement will have the same effect).

 
\subsubsection{\codetext{stop}}
\index{stop@\codetext{stop}}%
The \codetext{stop} command interrupts the execution of \codefile{spectra}\index{spectra@\codefile{spectra}}.
\cache{\codetext{end} is not exactly equivalent to \codetext{return~; stop}}

 
\subsubsection{\codetext{echo}}
\index{echo@\codetext{echo}}%
If the \codetext{echo} command is encountered while reading the file of
scenarios\index{file of scenarios} or a file included in it with \codetext{include},
the name of this file, the line number where the statement
\codetext{echo} appears
and the values of all the relevant parameters are printed on the screen and in
the log file\index{log file} (see \fullref{sec:run_spectra}).
The text produced by all the \codetext{add_text} commands is also printed (see \fullref{sec:add_text}).


\subsubsection{\codetext{include}}
\index{include@\codetext{include}}%
\label{sec:included}%
The contents of an external file (e.g.\ \userfile{included_file})
may be inserted in the file of scenarios\index{file of scenarios} by typing
\mention{\codetext{include "\usertext{\optional{path/}included_file}"}}
(or \mention{\codetext{include '\usertext{\optional{path/}included_file}'}}),
where the optional path \usertext{path/} is required only if
the included file is not in \codefile{scenarios_dir/}.
An included file may itself contain \codetext{include} statements.


\subsubsection{\codetext{verbosity}}
\label{sec:verbosity}%
\index{verbosity@\codetext{verbosity}}%
If the value of the parameter \codetext{verbosity} is $\ge 0$,
all warnings are written on the screen.
If $\codetext{verbosity} \ge 1$, major steps in the execution of
\codefile{spectra}\index{spectra@\codefile{spectra}} are also shown.

 
\subsubsection{\codetext{check_only}}
\index{check_only@\codetext{check_only}}%
By default, the value of the
parameter \codetext{check_only} is \codetext{.false.}.
If set to \codetext{.true.}, the scenarios are not run when
the statements \codetext{return} and \codetext{end}
or the end of the most outer file are encountered
(see \fullref{sec:other_statements}), and this until
\mention{\codetext{check_only~= .false.}} is met.

Note that setting \codetext{check_only} to \codetext{.true.} does not interrupt the reading
of input parameters, so this may be used to check the syntax of the file of scenarios.

 
\subsubsection{\codetext{add_text}, \codetext{erase_text}}
\index{add_text@\codetext{add_text}}%
\index{erase_text@\codetext{erase_text}}%
\label{sec:add_text}%
Lines of text containing additional informations (the purpose of the scenario, for instance) may be written 
at the beginning of the output files of \codefile{spectra} with these commands.
The line \mention{\codetext{add_text "\placeholder{string}"}} (or
\mention{\codetext{add_text '\placeholder{string}'}}),
where \placeholder{string} denotes a string of characters, \emph{appends} 
the line \placeholder{string} to this text.
To erase \emph{all} these lines, use the command \codetext{erase_text}.

 
\subsubsection{\texorpdfstring{\codetext{reset_\placeholder{*}}}{\codetext{reset_*}} commands}
\label{sec:reset}%


\paragraph{\codetext{reset_cosmo}}
\index{reset_cosmo@\codetext{reset_cosmo}}%
This command resets all the cosmological parameters (listed in \fullref{sec:cosmo_param}) 
to their default values if they exist, or to \mention{undefined} otherwise.

 
\paragraph{\codetext{reset_reserv_infall}}
\index{reset_reserv_infall@\codetext{reset_reserv_infall}}%
This command resets all the parameters related to reservoirs and infall episodes
(listed in \fullref{sec:reserv_infall_param}) to their default values
if they exist, or to \mention{undefined} otherwise.

 
\paragraph{\codetext{reset_SF}}
\index{reset_SF@\codetext{reset_SF}}%
This command resets all the parameters related to star formation episodes
(listed in \fullref{sec:SF_param}) to their default values if they exist,
or to \mention{undefined} otherwise.

 
\paragraph{\codetext{reset_outflow}}
\index{reset_outflow@\codetext{reset_outflow}}%
This command resets all the parameters related to outflow episodes
(listed in \fullref{sec:outflow_param}) to their default values if they exist,
or to \mention{undefined} otherwise.

 
\paragraph{\codetext{reset_dust_evol}}
\index{reset_dust_evol@\codetext{reset_dust_evol}}%
This command resets all the parameters related to dust evolution 
(listed in \fullref{sec:dust_evol_param}) to their default values
if they exist, or to \mention{undefined} otherwise.

 
\paragraph{\codetext{reset_dust_transfer}}
\index{reset_dust_transfer@\codetext{reset_dust_transfer}}%
This command resets all the parameters related to dust extinction and emission
(listed in \fullref{sec:dust_transfer_param}) to their default values
if they exist, or to \mention{undefined} otherwise.


\paragraph{\codetext{reset_cloud_neb}}
\index{reset_cloud_neb@\codetext{reset_cloud_neb}}%
This command resets all the parameters related to clouds and nebular emission
(listed in \fullref{sec:cloud_neb_param}) to their default values
if they exist, or to \mention{undefined} otherwise.


\paragraph{\codetext{reset_output}}
\index{reset_output@\codetext{reset_output}}%
This command resets all the parameters related to output files
(listed in \fullref{sec:output_param}) to their default values
if they exist, or to \mention{undefined} otherwise.
It also executes the command \codetext{erase_text} (see \fullref{sec:add_text}).


\paragraph{\codetext{reset_others}}
\index{reset_others@\codetext{reset_others}}%
This command resets all the parameters listed in \fullref{sec:SSPs_param}, 
in \fullref{sec:chemical_param}, and
in \fullref{sec:verbosity}, 
to their default values if they exist, or to \mention{undefined} otherwise.


\paragraph{\codetext{reset_all}}
\index{reset_all@\codetext{reset_all}}%
The \codetext{reset_all} statement executes all the \codetext{reset_\usertext{*}} statements listed above.
%
\section{Outputs}


\subsection{Outputs of \codefile{SSPs}}
\index{SSPs@\codefile{SSPs}}%
The output files\index{files of SSPs}\index{set of SSPs}
of \codefile{SSPs}\index{SSPs@\codefile{SSPs}} are not written to be read by the user but to be
processed by code \codefile{spectra}\index{spectra@\codefile{spectra}}.
See \fullref{fn:SSPs_outputs}.


\subsection{Outputs of \codefile{spectra}}
\index{spectra@\codefile{spectra}}%


\subsubsection{Main output file}
\label{sec:spectra_main_output}%

\paragraph{Reading procedure}
\label{sec:spectra_main_output_proc}%
The data written by \codefile{spectra}\index{spectra@\codefile{spectra}} in a file of spectra may be read using
the Fortran subroutine \codetext{read_spectra_\-out\-put}%
\footnote{%
  This subroutine is also called in \codefile{plot_spectra.f90} and
  \codefile{colors.f90}.
}.
To read a file \userfile{file_name} from within a Fortran~95 program, do the following:
\begin{enumerate}

\item
  \label{enum:step_use}
  Type the line
  \Mention{\codetext{use mod_read_spectra_output}}
  right after the \codetext{program} statement
  to load the module \codetext{mod_read_\-spectra_output} defined in
  \codefile{mod_read_spectra_output.f90}.
  This module
  provides
  the subroutine \codetext{read_spectra_output}
  and a data structure, \codetext{struct_spectra_output}%
  \index{struct_spectra_output@\codetext{struct_spectra_output}}.
  Tables~\ref{tab:output_sizes} and~\ref{tab:output_constants}
  to~\ref{tab:output_warn} list all the fields in this structure;

\item
  \label{enum:step_declare}
  Declare a variable, say \usertext{data}, which will contain all the
  data in the file of spectra\index{file of spectra}.
  To do this, insert the line
  \Mention{\codetext{type(struct_spectra_output) :: \usertext{data}}}%
  \index{struct_spectra_output@\codetext{struct_spectra_output}}
  among the declarations of variables;

\item
  \label{enum:step_read}
  Read the data in the executable part of the program with the line
  \Mention[.]{\codetext{call read_spectra_output(\optional{file_name = }"\usertext{file_name}",
      \optional{file = }\usertext{data})}}
  This statement also allocates all the array fields of \usertext{data}
  to the right size;

\item
  \label{enum:step_access}
  The value of any component \usertext{field}
  (see Tables~\ref{tab:output_sizes} and~\ref{tab:output_constants}
  to~\ref{tab:output_warn} for a complete list)
  of \usertext{data} is given by
  \CodeText[.]{\usertext{data} \% \usertext{field}}
\end{enumerate}
To compile this program, link it (with the appropriate paths)
to \codefile{util_dir/mod_types.f90}, \codefile{util_dir/mod_dir_access.f90}, 
\codefile{util_dir/mod_file_access.f90}, \codefile{util_dir/mod_strings.f90},
\codefile{source_dir/mod_directories.f90} and \codefile{source_dir/mod_read_spectra_output.f90}.

Here is a basic Fortran example showing how to read, with the program \codefile{source_dir/example_read_spectra.f90}
provided with the code, the files \userfile{example_spectra1.txt} and \userfile{example_spectra2.txt} computed in \fullref{sec:run_spectra},
and how to write to the screen the corresponding continuous spectrum
at all ages and wavelengths:
\begin{lstlisting}[name=output, frame=trlb, numbers=none, xleftmargin=5pt, xrightmargin=5.5em]
!/*\string#*/ With "example_read_spectra.f90" in "source_dir/", type for instance
!/*\string#*/ ''gfortran -o example_read_spectra ../util_dir/mod_types.f90 \
/*\ \ \ \ \ \ \ \ */../util_dir/mod_dir_access.f90 ../util_dir/mod_file_access.f90 \
/*\ \ \ \ \ \ \ \ */../util_dir/mod_strings.f90 ../source_dir/mod_directories.f90 \
/*\ \ \ \ \ \ \ \ */../source_dir/mod_read_spectra_output.f90 \
/*\ \ \ \ \ \ \ \ */../source_dir/example_read_spectra.f90''
!/*\string#*/ (without quotation marks!) to compile it from within "bin_dir/" 
!/*\string#*/ with `gfortran`.

program example_read_spectra

/*\ */use mod_read_spectra_output/*%
\hfill\rlap{\quad$\rightarrow$~\textrm{Step~\ref{enum:step_use}}}*/
/*\ */implicit none/*%
\hfill\rlap{\quad\phantom{$\rightarrow$}~\textrm{above.}}*/
/*\ */type(struct_spectra_output), dimension(2) :: data/*%
\hfill\rlap{\quad$\rightarrow$~\textrm{Step~\ref{enum:step_declare}.}}%
\index{struct_spectra_output@\codetext{struct_spectra_output}}*/
/*\ */integer :: i, j, k

/*\ */call read_spectra_output("example_spectra1.txt", data(1))/*%
\hfill\rlap{\quad$\rightarrow$~\textrm{Step~\ref{enum:step_read}.}}*/
/*\ */call read_spectra_output("example_spectra2.txt", data(2))/*%
\hfill\rlap{\quad\phantom{$\rightarrow$~}\textrm{Idem.}}*/
/*\ */do j = 1, data(1) % dim_output_age/*%
\hfill\rlap{\quad$\rightarrow$~\textrm{Step~\ref{enum:step_access}.}}*/
/*\ \ \ \ */write(*,*) "Wavelength, monochromatic luminosities in &
/*\ \ \ \ \ \ \ \ \ */&''example_spectra1.txt'' and ''example_spectra2.txt'' &
/*\ \ \ \ \ \ \ \ \ */&at age = ", data(1) % output_age(j), ":"/*%
\hfill\rlap{\quad\phantom{$\rightarrow$~}\textrm{Idem.}}*/
/*\ \ \ \ */do k = 1, data(1) % dim_cont/*%
\hfill\rlap{\quad\phantom{$\rightarrow$~}\textrm{Idem.}}*/
/*\ \ \ \ \ \ \ */write(*,*) data(1) % lambda_cont(k), (data(i) % lum_cont(j, k), i = 1, 2)/*%
\hfill\rlap{\quad\phantom{$\rightarrow$~}\textrm{Idem.}}*/
/*\ \ \ \ */enddo
/*\ */enddo

end program example_read_spectra
\end{lstlisting}

\paragraph{Age-independent quantities}
\begin{longtable}{l>{\raggedright\Tstrut}l<{\Bstrut}}
  \caption{Fields defined in structure \codetext{struct_spectra_output}:
    array sizes\captionPoint}%
  \index{struct_spectra_output@\codefile{struct_spectra_output}}%
  \label{tab:output_sizes}%
  \tabularnewline*%
  \nobreakhline
  \multicolumn{1}{>{\TBstrut}c}{Field} & \multicolumn{1}{c}{Meaning}
  \tabularnewline* 
  \nobreakhline\nobreakhline
  \codetext{\usertext{data} \% dim_output_age}
  & \UCase number of galactic ages for which quantities in \fullref{tab:output_main}, are printed
  \tabularnewline*
  \nobreakhline
  \codetext{\usertext{data} \% dim_elem}
  & \UCase number of elements followed during chemical evolution
  \tabularnewline*
  \nobreakhline
  \codetext{\usertext{data} \% dim_species_SFC}
  & \begin{tabular}[t]{@{\Tstrut}l@{\Bstrut}}%
    \UCase number of dust species (\ie graphites, silicates, \\
    various kinds
    of PAHs\textellipsis)\ in star-forming clouds
  \end{tabular}%
  \tabularnewline*
  \nobreakhline
  \codetext{\usertext{data} \% dim_species_DISM}
  & \begin{tabular}[t]{@{\Tstrut}l@{\Bstrut}}%
    \UCase number of dust species in the diffuse ISM
  \end{tabular}%
  \tabularnewline*
  \nobreakhline
  \codetext{\usertext{data} \% dim_cont}
  & \UCase number of continuum wavelengths
  \tabularnewline*
  \nobreakhline
  \codetext{\usertext{data} \% dim_line} 
  & \UCase number of emission lines
  \tabularnewline*
  \nobreakhline
\end{longtable}

%
\begin{longtable}{l>{\Tstrut}l<{\Bstrut}l}
  \caption{Meaning of indices used in Tables~\ref{tab:output_constants}
    to~\ref{tab:output_warn}\captionPoint}
  \label{tab:output_indices}
  \\*
  \nobreakhline
  \multicolumn{1}{>{\TBstrut}c}{Index} & \multicolumn{1}{c}{Quantity referred to} & \multicolumn{1}{c}{Index range}
  \\*
  \nobreakhline\nobreakhline
  $\ielem$
  &  \UCase element \codetext{\usertext{data} \% elem_id(\ielem)}
  & $\IE[1, \text{\codetext{\usertext{data} \% dim_elem}}]$
  \\*
  \nobreakhline
  $\iage$
  & \UCase age \codetext{\usertext{data} \% output_age(\iage)}
  & $\IE[1, \text{\codetext{\usertext{data} \% dim_output_age}}]$
  \\
  \hline
  $\icont$
  & \UCase continuum wavelength
  \codetext{\usertext{data} \% lambda_cont(\icont)}
  & $\IE[1, \text{\codetext{\usertext{data} \% dim_cont}}]$
  \\*
  \nobreakhline
  $\iline$
  & \UCase emission line
  \codetext{\usertext{data} \% line_id(\iline)}
  & $\IE[1, \text{\codetext{\usertext{data} \% dim_line}}]$
  \\*
  \nobreakhline
  $\ispecies$
  & \begin{tabular}[t]{@{}l@{}}%
    \UCase grain species \codetext{\usertext{data} \% species_id_SFC(\ispecies)}\\
    or \codetext{\usertext{data} \% species_id_DISM(\ispecies)}
  \end{tabular}
  & \begin{tabular}[t]{@{}l@{}}%
    $\IE[1, \text{\codetext{\usertext{data} \% dim_species_SFC}}]$\\
    or $\IE[1, \text{\codetext{\usertext{data} \% dim_species_DISM}}]$\Bstrut
  \end{tabular}
  \\*
  \nobreakhline
\end{longtable}
%
%
\saveFootnoteNumber
\begin{longtable}{l>{\Tstrut\raggedright}p{9.5cm}<{\Bstrut}l}
  \caption{Fields defined in structure \codetext{struct_spectra_output}:
    constant quantities\captionPoint}%
  \index{struct_spectra_output@\codefile{struct_spectra_output}}%
  \label{tab:output_constants}
  \\*
  \nobreakhline
  \multicolumn{1}{>{\TBstrut}c}{Field} & \multicolumn{1}{c}{Meaning} & \multicolumn{1}{c}{Unit}
  \\*
  \nobreakhline\nobreakhline
  \endfirsthead
  \multicolumn{3}{c}{$\downarrow$}
  \\*
  \nobreakhline
  \multicolumn{1}{>{\TBstrut}c}{Field} & \multicolumn{1}{c}{Meaning} & \multicolumn{1}{c}{Unit}
  \\*
  \nobreakhline\nobreakhline
  \endhead
  \multicolumn{3}{c}{$\downarrow$} \\*
  \endfoot
  \nobreakhline
  \endlastfoot
  \codetext{\usertext{data} \% version_id}
  & \UCase identifier of the code version
  & \none
  \\*
  \nobreakhline
  \codetext{\usertext{data} \% version_date}
  & \UCase date of the code version
  & \none
  \\*
  \nobreakhline
  \codetext{\usertext{data} \% spectra_output}
  & \UCase value of the scenario parameter \codetext{spectra_output}\index{spectra_output@\codetext{spectra_output}} (see \fullref{sec:spectra_output})
  & \none
  \\
  \hline
  \codetext{\usertext{data} \% RF_output}
  & \UCase value of the scenario parameter \codetext{RF_output}\index{RF_output@\codetext{RF_output}} (see \fullref{sec:spectra_output})
  & \none
  \\*
  \nobreakhline
  \codetext{\usertext{data} \% sublim_output}
  & \UCase value of the scenario parameter \codetext{sublim_output}\index{sublim_output@\codetext{sublim_output}} (see \fullref{sec:spectra_output})
  & \none
  \\
  \hline
  \codetext{\usertext{data} \% time_step} 
  & \UCase time-step used in the evolution procedure of \codefile{spectra}\index{spectra@\codefile{spectra}} & Myr
  \\
  \hline
  \codetext{\usertext{data} \% elem_id(\ielem)}
  & \UCase identifier of the $\ielem$-th chemical element%
  \footnotemark
  \vadjust{\footnotetext{In the current implementation: 
      1$\,\to\,$O; 
      2$\,\to\,$C; 
      3$\,\to\,$Fe;
      4$\,\to\,$He;
      5$\,\to\,$N; 
      6$\,\to\,$Ne;
      7$\,\to\,$Mg;
      8$\,\to\,$Si;
      9$\,\to\,$S; 
      10$\,\to\,$Ca.}%
  }%
  & \none
  \\
  \hline
  \codetext{\usertext{data} \% species_id_SFC(\ispecies)}
  & \UCase identifier of the $\ispecies$-th dust species%
  \footnotemark\vadjust{\footnotetext{\label{fn:grain_species}%
      In the current implementation: 
      1$\,\to\,$graphites; 
      2$\,\to\,$neutral PAHs; 
      3$\,\to\,$ionized PAHs; 
      4$\,\to\,$silicates.}%
  } in star-forming clouds
  & \none
  \\
  \hline
  \codetext{\usertext{data} \% species_id_DISM(\ispecies)}
  & \UCase identifier of the $\ispecies$-th dust species%
  \footnoteref{fn:grain_species}\ in the diffuse ISM
  & \none
  \\*
  \nobreakhline
  \codetext{\usertext{data} \% lambda_cont(\icont)}
  & \UCase value of the $\icont$-th continuum wavelength
  & $\AA$
  \\
  \hline
  \codetext{\usertext{data} \% line_id(\iline)}
  & \UCase identifier\footnotemark\ of the $\iline$-th emission line
  & \none
  \\*
  \nobreakhline
  \codetext{\usertext{data} \% lambda_line(\iline)}
  & \UCase wavelength of the $\iline$-th emission line
  & $\AA$
  \\*
\end{longtable}
\restoreFootnoteNumber%
\stepcounter{footnote}%
\stepcounter{footnote}%
\stepcounter{footnote}%
\footnotetext{Warning: each value of \codetext{line_id} must refer to a single emission line.
  All emission lines are listed with their identifier and their wavelength in \codefile{Cloudy_dir/list_neb_lines.txt}.}%

The grid of wavelengths for the continuum spectra of galaxies merges
the wavelengths of the library of stellar spectra (see \fullref{sec:stel_lib})
and those of optical properties of grains (see \fullref{sec:GSD}).

\paragraph{Age-dependent quantities}
\saveFootnoteNumber
\begin{longtable}{>{\raggedright}p{5cm}>{\raggedright\Tstrut}p{8cm}<{\Bstrut}l}
  \caption{Fields defined in structure \codetext{struct_spectra_output}:
    main variable quantities\captionPoint}%
  \index{struct_spectra_output@\codefile{struct_spectra_output}}%
  \label{tab:output_main}
  \\*
  \nobreakhline
  \multicolumn{1}{>{\TBstrut}c}{Field} & \multicolumn{1}{c}{Meaning} & \multicolumn{1}{c}{Unit}
  \\*
  \nobreakhline\nobreakhline
  \endfirsthead
  \multicolumn{3}{c}{$\downarrow$}
  \\*
  \nobreakhline
  \multicolumn{1}{>{\TBstrut}c}{Field} & \multicolumn{1}{c}{Meaning} & \multicolumn{1}{c}{Unit}
  \\*
  \nobreakhline\nobreakhline
  \endhead
  \multicolumn{3}{c}{$\downarrow$} \\
  \endfoot
  \nobreakhline
  \endlastfoot
  \codetext{\usertext{data} \% output_age(\iage)} 
  & \UCase Galactic age for which output of the spectrum is requested
  & Myr
  \\*
  \nobreakhline
  \codetext{\usertext{data} \% convol_time(\iage)} 
  & \UCase age for which the spectrum is given
  (differs from \codetext{\usertext{data} \% output_age(\iage)} by at most \codetext{\usertext{data} \% time_step})
  & Myr
  \\*
  \nobreakhline
  \codetext{\usertext{data} \% cosmic_time(\iage)} 
  & \UCase cosmic time & Myr
  \\*
  \nobreakhline
  \codetext{\usertext{data} \% redshift(\iage)} 
  & \UCase redshift & \none
  \\*
  \hline
  \codetext{\usertext{data} \% galaxy_mass(\iage)} 
  & \UCase normalized mass of the galaxy
  & \none
  \\
  \hline
  \codetext{\usertext{data} \% live_stars_mass(\iage)} 
  & \UCase normalized mass of stars still alive
  & \none
  \\
  \hline
  \codetext{\usertext{data} \% WD_mass(\iage)} 
  & \UCase normalized mass of white dwarfs
  & \none
  \\
  \hline
  \codetext{\usertext{data} \% BHNS_mass(\iage)} 
  & \UCase normalized mass of black holes and neutron stars
  & \none
  \\
  \hline
  \codetext{\usertext{data} \% inert_mass(\iage)} 
  & \UCase normalized mass of inert objects (brown dwarfs, etc.)
  & \none
  \\
  \hline
  \codetext{\usertext{data} \% ISM_mass(\iage)} 
  & \UCase normalized mass of the ISM
  & \none
  \\
  \hline
  \codetext{\usertext{data} \% ISM_Z(\iage)} 
  & \UCase metallicity of the ISM
  & \none
  \\
  \hline
  \codetext{\usertext{data} \% stel_Z_mass_avrg(\iage)} 
  & \UCase mean birth-metallicity of stars, averaged over the masses of stars still alive
  & \none
  \\
  \hline
  \codetext{\usertext{data} \% stel_Z_bol_avrg(\iage)} 
  & \UCase mean birth-metallicity of stars, averaged over the bolometric luminosities
  of stars still alive
  & \none
  \\
  \hline
  \codetext{\usertext{data} \% carb_abund(\iage)} 
  & \UCase mass fraction of carbonaceous dust relative to the ISM mass
  & \none
  \\
  \hline
  \codetext{\usertext{data} \% sil_abund(\iage)} 
  & \UCase mass fraction of silicate dust relative to the ISM mass
  & \none
  \\
  \hline
  \codetext{\usertext{data} \% ISM_abund(\iage, \ielem)} &
  ISM abundance of the $\ielem$-th element, in mass fraction, relative to all
  the elements in the ISM%
  \footnotemark
  \vadjust{\footnotetext{In \codefile{spectra.f90}, \codetext{\usertext{data} \% ISM_abund(\iage, $0$)}
      also exists and is the same as \codetext{\usertext{data} \% ISM_Z(\iage)}.}%
  }%
  & \none
  \\
  \hline
  \codetext{\usertext{data} \% L_bol(\iage)} 
  & \UCase normalized bolometric luminosity
  (radiant power) of the galaxy
  & $\Lsol/\Msol$
  \\
  \hline
  \codetext{\usertext{data} \% tau_V(\iage)}
  & \UCase optical depth in the $V$-band of dust in the diffuse ISM.
  Computed from side to side through the center of the galaxy
  (along the rotation axis for axisymmetric galaxies)
  & \none
  \\
  \hline
  \codetext{\usertext{data} \% dust_bol_ratio(\iage)} 
  & \UCase ratio of the dust luminosity to the bolometric one
  & \none
  \\
  \hline
  \codetext{\usertext{data} \% SF_rate(\iage)} 
  & \UCase normalized star formation (mass) rate
  & Myr$^{-1}$
  \\
  \hline
  \codetext{\usertext{data} \% Lyman_cont_rate(\iage)}
  & \UCase normalized number rate of Lyman continuum photons emitted by stars
  & $\scd^{-1}*\Msol^{-1}$
  \\
  \hline
  \codetext{\usertext{data} \% CCSN_rate(\iage)} 
  & \UCase normalized number rate of core collapse supernovae
  & $\text{Myr}^{-1}*\Msol^{-1}$
  \\
  \hline
  \codetext{\usertext{data} \% SNIa_rate(\iage)} 
  & \UCase idem for type~Ia supernovae
  & $\text{Myr}^{-1}*\Msol^{-1}$
  \\
  \hline
  \codetext{\usertext{data} \% stel_age_mass_avrg(\iage)} 
  & \UCase mean stellar age, averaged over the masses of stars still alive
  & Myr
  \\
  \hline
  \codetext{\usertext{data} \% stel_age_bol_avrg(\iage)} 
  & \UCase mean stellar age, averaged over the bolometric luminosities of stars
  still alive
  & Myr
  \\
  \hline
  \codetext{\usertext{data} \% Lyman_cont_gas_abs(\iage)} 
  & \UCase fraction (in number)
  of Lyman continuum photons emitted by stars which are absorbed by gas
  & \none
  \\
  \hline
  \codetext{\usertext{data} \% Lyman_cont_dust_abs(\iage)} 
  & \UCase fraction (in number)
  of Lyman continuum photons emitted by stars which are absorbed by dust
  & \none
  \\
  \hline
  \codetext{\usertext{data} \% ejec_rate_tot(\iage)} 
  & \UCase normalized mass ejection rate  by stars of matter into the ISM of the galaxy
  & Myr$^{-1}$
  \\
  \hline
  \codetext{\usertext{data} \% infall_rate(\iage)} 
  & \UCase normalized mass infall rate from all reservoirs onto the galaxy
  & Myr$^{-1}$
  \\
  \hline
  \codetext{\usertext{data} \% outflow_rate(\iage)} 
  & \UCase normalized mass outflow rate from the ISM of the galaxy into the IGM
  & Myr$^{-1}$
  \\
  \hline
  \codetext{\usertext{data} \% ejec_cumul_mass(\iage)} 
  & \UCase cumulative normalized mass of stellar ejecta since the beginning
  & \none
  \\
  \hline
  \codetext{\usertext{data} \% SF_live_cumul_mass(\iage)}
  & \UCase cumulative normalized mass of live stars formed since the beginning%
  \footnotemark
  \vadjust{\footnotetext{Contrary to \codetext{\usertext{data} \% live_stars_mass},
      the quantity \codetext{\usertext{data} \% SF_live_cumul_mass} accounts for all the mass in stars considered
      as alive at birth:
      beside the mass in stars still alive at age \codetext{\usertext{data} \% output_age($\iage$)},
      it includes the mass of stellar ejecta and compact stellar remnants, but not that in
      objects inert from the outset.}%
  }%
  & \none
  \\
  \hline
  \codetext{\usertext{data} \% infall_cumul_mass(\iage)} 
  & \UCase cumulative normalized mass of matter fallen from all reservoirs onto the galaxy
  since the beginning
  & \none
  \\
  \hline
  \codetext{\usertext{data} \% outflow_cumul_mass(\iage)} 
  & \UCase cumulative normalized mass of matter expelled by the galaxy into the IGM
  since the beginning
  & \none
  \\
  \hline
  \codetext{\usertext{data} \% L_dust_SFC(\iage)} 
  & \UCase normalized bolometric luminosity
  of dust grains in star-forming clouds
  & $\erg*\scd^{-1}*\Msol^{-1}$
  \\
  \hline
  \codetext{\usertext{data} \% L_dust_DISM(\iage)} 
  & \UCase idem for dust grains in the diffuse ISM
  & $\erg*\scd^{-1}*\Msol^{-1}$
  \\*
  \hline
  \codetext{\usertext{data} \% lum_cont(\iage, \icont)}
  & \UCase normalized attenuated (see \fullref{sec:inclin_param}) monochromatic continuum luminosity (spectral power) of the galaxy
  & $\erg*\scd^{-1}*\AA^{-1}*\Msol^{-1}$
  \\*
  \hline
  \codetext{\usertext{data} \% L_line(\iage, \iline)}
  & \UCase normalized attenuated in-line luminosity of the galaxy
  & $\erg*\scd^{-1}*\Msol^{-1}$
  \\*
\end{longtable}%
\restoreFootnoteNumber
\stepcounter{footnote}%
\stepcounter{footnote}%
%
%
\saveFootnoteNumber
\begin{longtable}{>{\raggedright}p{5cm}>{\raggedright\Tstrut}p{8cm}<{\Bstrut}l}
  \caption{Fields defined in structure \codetext{struct_spectra_output}
    if and only if field $\codetext{\usertext{data} \% \raggedallowbreak spectra_output} = \raggedallowbreak \codetext{"detailed"}$\captionPoint}%
  \index{struct_spectra_output@\codefile{struct_spectra_output}}%
  \index{spectra_output@\codetext{spectra_output}}%
  \label{tab:output_detailed}
  \\*
  \nobreakhline
  \multicolumn{1}{>{\TBstrut}c}{Field} & \multicolumn{1}{c}{Meaning} & \multicolumn{1}{c}{Unit}
  \\*
  \nobreakhline\nobreakhline
  \endfirsthead
  \multicolumn{3}{c}{$\downarrow$}
  \\*
  \nobreakhline
  \multicolumn{1}{>{\TBstrut}c}{Field} & \multicolumn{1}{c}{Meaning} & \multicolumn{1}{c}{Unit}
  \\*
  \nobreakhline\nobreakhline
  \endhead
  \multicolumn{3}{c}{$\downarrow$} \\
  \endfoot
  \nobreakhline
  \endlastfoot
  \codetext{\usertext{data} \% lum_stel_SFC_unatt(\iage,~\icont)}
  & \UCase normalized unattenuated monochromatic continuum luminosity of stars in 
  star-forming clouds
  & $\erg*\scd^{-1}*\AA^{-1}*\Msol^{-1}$
  \\*
  \nobreakhline
  \codetext{\usertext{data} \% lum_stel_DISM_unatt(\iage,~\icont)}
  & \UCase idem for stars in the diffuse medium
  & $\erg*\scd^{-1}*\AA^{-1}*\Msol^{-1}$
  \\*
  \nobreakhline
  \codetext{\usertext{data} \% lum_neb_cont_SFC_unatt(\iage,~\icont)}
  & \UCase normalized unattenuated monochromatic luminosity of the nebular continuum
  produced by ionized gas in star-forming clouds
  & $\erg*\scd^{-1}*\AA^{-1}*\Msol^{-1}$
  \\*
  \nobreakhline
  \codetext{\usertext{data} \% lum_neb_cont_DISM_unatt(\iage,~\icont)}
  & \UCase idem for ionized gas in the diffuse medium
  & $\erg*\scd^{-1}*\AA^{-1}*\Msol^{-1}$
  \\*
  \nobreakhline
  \codetext{\usertext{data}~\%~lum_species_SFC(\ispecies,~\iage,~\icont)}
  & \UCase normalized monochromatic continuum luminosity emitted
  in star-forming clouds by the $\ispecies$-th dust species
  & $\erg*\scd^{-1}*\AA^{-1}*\Msol^{-1}$
  \\
  \hline
  \codetext{\usertext{data}~\%~lum_species_DISM(\ispecies,~\iage,~\icont)}
  & \UCase idem for the $\ispecies$-th dust species in the diffuse medium
  & $\erg*\scd^{-1}*\AA^{-1}*\Msol^{-1}$
  \\
  \hline
  \codetext{\usertext{data} \% L_line_SFC_unatt(\iage,~\iline)}
  & \UCase normalized unattenuated luminosity of emission lines produced by ionized gas
  in star-forming clouds
  & $\erg*\scd^{-1}*\Msol^{-1}$
  \\
  \hline
  \codetext{\usertext{data} \% L_line_DISM_unatt(\iage,~\iline)}
  & \UCase idem for ionized gas in the diffuse medium
  & $\erg*\scd^{-1}*\Msol^{-1}$
  \\
  \hline
    \multicolumn{3}{>{\TBstrut}c}{\textbf{Quantities printed only if $\codetext{\usertext{data} \% RF_output} = \codetext{.true.}$:}} 
    \\*
  \codetext{\usertext{data} \% RF_cont_SFC(\iage,~\icont)}
  & \UCase mean monochromatic radiation field\footnotemark\ in star-forming clouds
  & $\unit{erg*cm^{-3}*\micron^{-1}}$
  \\
  \hdottedline
  \codetext{\usertext{data} \% RF_cont_DISM(\iage,~\icont)}
  & \UCase idem in the diffuse medium
  & $\unit{erg*cm^{-3}*\micron^{-1}}$
  \\
  \hdottedline
  \codetext{\usertext{data} \% RF_line_SFC(\iage,~\iline)}
  & \UCase mean in-line radiation field\footnotemark\ in star-forming clouds
  & $\unit{erg*cm^{-3}}$
  \\
  \hdottedline
  \codetext{\usertext{data} \% RF_line_DISM(\iage,~\iline)}
  & \UCase idem in the diffuse medium
  & $\unit{erg*cm^{-3}}$
  \\
  \hline
  \multicolumn{3}{>{\TBstrut}c}{\textbf{Quantities printed only if $\codetext{\usertext{data} \% sublim_output} = \codetext{.true.}$:}}
  \\*
  \codetext{\usertext{data}~\%~sublim_lum_species_SFC(\ispecies,~\iage,~\icont)}
  & \UCase normalized monochromatic continuum luminosity emitted
  in star-forming clouds by dust grains of the $\ispecies$-th dust species
  at temperatures above the sublimation temperature
  & $\erg*\scd^{-1}*\AA^{-1}*\Msol^{-1}$
  \\*
  \hdottedline
  \codetext{\usertext{data}~\%~sublim_lum_species_DISM(\ispecies,~\iage,~\icont)}
  & \UCase idem for dust grains in the diffuse medium
  & $\erg*\scd^{-1}*\AA^{-1}*\Msol^{-1}$
  \\*
\end{longtable}
\restoreFootnoteNumber
\stepcounter{footnote}%
\footnotetext{\label{fn:RF_cont}%
  This is the electromagnetic energy density per unit wavelength in the continuum,
  averaged in the medium,
  \ie\ the quantity $\langle u_\lambda\rangle$ given
  by \fullref{eq:mean_ISRF}.}%
\stepcounter{footnote}%
\footnotetext{\label{fn:RF_line}%
  This is the total electromagnetic energy density in the emission line,
  averaged in the medium,
  \ie\ $\int_{\text{line \diese}\ell}{\langle u_\lambda\rangle *\df\lambda}$, where $\langle u_\lambda\rangle$ is
  given by \fullref{eq:mean_ISRF}.}%
%
%
\pagebreak
\paragraph{Warnings}
\begin{longtable}{>{\raggedright}p{5cm}>{\raggedright\Tstrut}p{8cm}<{\Bstrut}l}
  \caption{Fields defined in structure \codetext{struct_spectra_output}:
    warnings\captionPoint}%
  \index{struct_spectra_output@\codefile{struct_spectra_output}}%
  \index{warnings}%
  \label{tab:output_warn}
  \\*
  \nobreakhline
  \multicolumn{1}{>{\TBstrut}c}{Field} & \multicolumn{1}{c}{Meaning} & \multicolumn{1}{c}{Unit}
  \\*
  \nobreakhline\nobreakhline
  \endfirsthead
  \multicolumn{3}{c}{$\downarrow$}
  \\*
  \nobreakhline
  \multicolumn{1}{>{\TBstrut}c}{Field} & \multicolumn{1}{c}{Meaning} & \multicolumn{1}{c}{Unit}
  \\*
  \nobreakhline\nobreakhline
  \endhead
  \multicolumn{3}{c}{$\downarrow$} \\
  \endfoot
  \nobreakhline
  \endlastfoot
  \codetext{\usertext{data} \% reserv_warn_present}
  & \UCase flag set to \codetext{.true.} if the sum of the initial masses of the
  reservoirs exceeds the total mass of the system
  & \none
  \\*
  \nobreakhline
  \codetext{\usertext{data} \% SF_warn_present}
  & \UCase flag set to \codetext{.true.} if the total star formation rate
  exceeds the maximal possible value at some age
  & \none
  \\*
  \hdottedline
  \codetext{\usertext{data} \% SF_warn_age}
  & \UCase age at which the total star formation rate exceeds the maximal possible value
  & Myr
  \\*
  \nobreakhline
  \codetext{\usertext{data} \% dim_infall_warn}
  & \UCase number of reservoirs for which the infall rate exceeds
  the maximal possible value at some age
  & \none
  \\*
  \multicolumn{3}{l}{The index $\irsv$ below is in $\IE[1, \codetext{\usertext{data} \% dim_infall_warn}]$.}
  \\
  \hdottedline
  \codetext{\usertext{data} \% infall_warn_present($\irsv$)}
  & \UCase index of the reservoir for which the infall
  rate exceeds the maximal possible value at some age
  & \none
  \\
  \hdottedline
  \codetext{\usertext{data} \% infall_warn_age($\irsv$)}
  & \UCase age at which the infall
  rate exceeds the maximal possible value in this reservoir
  & Myr
  \\
  \hline
  \codetext{\usertext{data} \% outflow_warn_present}
  & \UCase flag set to \codetext{.true.}
  if the total outflow rate exceeds the maximal possible value
  at some age
  & \none
  \\
  \hdottedline
  \codetext{\usertext{data} \% outflow_warn_age}
  & \UCase age at which the total outflow
  rate possibly exceeds the maximal possible value
  & Myr
  \\
  \hline
  \codetext{\usertext{data} \% opt_depth_warn_present(\iage)}
  & \UCase flag set to \codetext{.true.}
  if the optical depth is larger than the maximal one
  used in radiative transfer precomputations at age
  \codetext{\usertext{data} \% output_age($\iage$)}
  & \none
  \\*
  \hdottedline
  \codetext{\usertext{data} \% opt_depth_warn_min_lambda(\iage)}
  & \UCase lower wavelength for which this occurs
  & $\AA$
  \\*
  \hdottedline
  \codetext{\usertext{data} \% opt_depth_warn_max_lambda(\iage)}
  & \UCase upper wavelength for which this occurs
  & $\AA$
  \\*
\end{longtable}


\subsubsection{Outputs for grains}
\paragraph{Reading a file of grain temperatures}
The data written in a file of grain temperatures\index{file of grain temperatures} may be read with the subroutine
\codetext{read_grain_temp} of module \codefile{source_dir/mod_read_grain_temp.f90}.
To do this, create a Fortran~95 code loading this module and declare a structure
of type \codetext{struct_grain_temp}%
\index{struct_grain_temp@\codetext{struct_grain_temp}};
its fields are described in \fullref{tab:struct_grain_temp}.
Detailed explanations on the creation and the compilation of this code are provided in the header of
\codefile{mod_read_grain_temp.f90}
(the procedure is similar to the one used for the main output file of \codefile{spectra}\index{spectra@\codefile{spectra}};
see \fullref{sec:spectra_main_output_proc}).
See \codefile{source_dir/plot_grain_temp.f90} for an example of use.

\begin{longtable}{l>{\raggedright\Tstrut}p{9cm}<{\Bstrut}l}
  \caption{Fields defined for a structure \usertext{data} of type \codetext{struct_grain_temp}\captionPoint}%
  \index{struct_grain_temp@\codetext{struct_grain_temp}}
  \label{tab:struct_grain_temp}\\*
  \\*
  \nobreakhline
  \multicolumn{1}{>{\TBstrut}c}{Field} & \multicolumn{1}{c}{Meaning} & \multicolumn{1}{c}{Unit}
  \\*
  \nobreakhline\nobreakhline
  \endfirsthead
  \multicolumn{3}{c}{$\downarrow$}
  \\*
  \nobreakhline
  \multicolumn{1}{>{\TBstrut}c}{Field} & \multicolumn{1}{c}{Meaning} & \multicolumn{1}{c}{Unit}
  \\*
  \nobreakhline\nobreakhline
  \endhead
  \multicolumn{3}{c}{$\downarrow$} \\
  \endfoot
  \nobreakhline
  \endlastfoot
  \codetext{\usertext{data} \% version_id} 
  & \UCase  identifier of the code version & \none
  \\* \hline
  \codetext{\usertext{data} \% version_date} 
  & \UCase  date of the code version  & \none
  \\ \hline
  \codetext{\usertext{data} \% dim_age_grains} 
  & \UCase number of ages & \none
  \\*
  \multicolumn{3}{>{\TBstrut}l}{The index $\iage$ below is in $\IE[1, \codetext{\usertext{data} \% dim_age_grains}]$.}
  \\ \hline
  \codetext{\usertext{data} \% dim_region} 
  & \UCase number of regions (star-forming clouds, diffuse ISM) & \none
  \\*
  \multicolumn{3}{>{\TBstrut}l}{The index $\iregion$ below is in $\IE[1, \codetext{\usertext{data} \% dim_region}]$.}
  \\ \hline
  \codetext{\usertext{data} \% age_grains($\iage$)} 
  & \UCase age & $\unit{Myr}$
  \\*
  \multicolumn{3}{>{\TBstrut}l}{These ages are a subset of the \codetext{output_age(1:dim_output_age)} of \fullref{tab:output_main}.}
  \\ \hline
  \codetext{\usertext{data} \% region($\iregion$)} 
  & \UCase data for the $\iregion$-th region & \none
  \\[1ex]
  \hdottedline
  \multicolumn{3}{>{\TBstrut}c}{\bfseries Subfields of \codetext{\usertext{data} \% region($\iregion$)}
    (denoted by \mention{\usertext{\abrege{region}}} below):}
  \\*
  \codetext{\usertext{\abrege{region}} \% id} 
  & \UCase identifier of the region: \codetext{"SFC"} for
  star-forming clouds, \codetext{"DISM"} for the diffuse ISM & \none
  \\*
  \codetext{\usertext{\abrege{region}} \% dim_species} 
  & \UCase number of grain species (graphite,
  silicate, etc.) for this region & \none
  \\*
  \multicolumn{3}{>{\TBstrut}l}{The index $\ispecies$ below is in $\IE[1, \codetext{\usertext{\abrege{region}} \% dim_species}]$.}
  \\*
  \codetext{\usertext{\abrege{region}} \% stoch_heating} 
  & \UCase boolean set to \codetext{.true.}
  if the probability distribution of grain temperatures is computed (so,
  grains are stochastically heated), and to \codetext{.false.} if the equilibrium
  temperature is used & \none
  \\*
  \codetext{\usertext{\abrege{region}} \% species($\ispecies$)} 
  & \UCase data for the $\ispecies$-th grain species in the $\iregion$-th region & \none
  \\[1ex]
  \hdottedline
  \multicolumn{3}{>{\TBstrut}l}{\bfseries Subfields of \codetext{\usertext{data} \% region($\iregion$) \% species($\ispecies$)}
    (denoted by \mention{\usertext{\abrege{species}}} below):}
  \\*
  \codetext{\usertext{\abrege{species}} \% id} 
  & \UCase identifier of the grain species & \none
  \\*
  \codetext{\usertext{\abrege{species}} \% dim_radius} 
  & \UCase number of grain radii & \none
  \\*
  \multicolumn{3}{>{\TBstrut}l}{The index $\irad$ below is in $\IE[1, \codetext{\usertext{\abrege{species}} \% dim_radius}]$.}
  \\*
  \codetext{\usertext{\abrege{species}} \% radius($\irad$)} 
  & \UCase radius of the grain with the $\irad$-th radius & $\micron$
  \\*
  \codetext{\usertext{\abrege{species}} \% state($\iage$, $\irad$)} 
  & \UCase data for this grain at the $\iage$-th time & \none
  \\[1ex]
  \hdottedline
  \multicolumn{3}{>{\TBstrut}l}{\bfseries Subfields of \codetext{\usertext{data} \% region($\iregion$) \% species($\ispecies$) \% state($\iage$, $\irad$)} (denoted by \mention{\usertext{\abrege{state}}} below):}
  \\*
  \codetext{\usertext{\abrege{state}} \% temp_eq} 
  & \UCase equilibrium temperature of the grain & $\unit{K}$
  \\*
  \codetext{\usertext{\abrege{state}} \% dim_temp} 
  & \UCase number of grain temperatures & \none
  \\*
  \multicolumn{3}{>{\TBstrut}l}{The index $\itemp$ below is in $\IE[1, \codetext{\usertext{\abrege{state}} \% dim_temp}]$.}
  \\*
  \codetext{\usertext{\abrege{state}} \% temp($\itemp$)} 
  & \UCase temperature $T$ of the grain & $\unit{K}$
  \\*
  \codetext{\usertext{\abrege{state}} \% prob($\itemp$)} 
  & \UCase value of $\df P/\df(\log_{10} T)$ for this temperature,
  where $P(x)$ is the probability that $T < x$ & \none
  \\*
\end{longtable}
%
%
\paragraph{Reading a file of grain SEDs}
The data written in a file of grain SEDs\index{file of grain SEDs} may be read with the subroutine
\codetext{read_grain_SED} of module \codefile{source_dir/mod_read_grain_SED.f90}.
To do this, create a Fortran~95 code loading this module and declare a structure
of type \codetext{struct_grain_SED}%
\index{struct_grain_SED@\codetext{struct_grain_SED}};
its fields are described in \fullref{tab:struct_grain_SED}.
Detailed explanations on the creation and the compilation of this code are provided in the header of
\codefile{mod_read_grain_SED.f90}
(the procedure is similar to the one used for the main output file of \codefile{spectra}\index{spectra@\codefile{spectra}};
see \fullref{sec:spectra_main_output_proc}).
See \codefile{source_dir/plot_grain_SED.f90} for an example of use.
\begin{longtable}{l>{\raggedright\Tstrut}p{7cm}<{\Bstrut}l}
  \caption{Fields defined for a structure \usertext{data} of type \codetext{struct_grain_SED}\captionPoint}%
  \index{struct_grain_SED@\codetext{struct_grain_SED}}
  \label{tab:struct_grain_SED} 
  \\*
  \nobreakhline
  \multicolumn{1}{>{\TBstrut}c}{Field} & \multicolumn{1}{c}{Meaning} & \multicolumn{1}{c}{Unit}
  \\*
  \nobreakhline\nobreakhline
  \endfirsthead
  \multicolumn{3}{c}{$\downarrow$}
  \\*
  \nobreakhline
  \multicolumn{1}{>{\TBstrut}c}{Field} & \multicolumn{1}{c}{Meaning} & \multicolumn{1}{c}{Unit}
  \\*
  \nobreakhline\nobreakhline
  \endhead
  \multicolumn{3}{c}{$\downarrow$} \\
  \endfoot
  \nobreakhline
  \endlastfoot
  \codetext{\usertext{data} \% version_id} 
  & \UCase  identifier of the code version & \none
  \\* \hline
  \codetext{\usertext{data} \% version_date} 
  & \UCase  date of the code version & \none
  \\ \hline
  \codetext{\usertext{data} \% dim_age_grains} 
  & \UCase  number of ages & \none
  \\*
  \multicolumn{3}{>{\TBstrut}l}{The index $\iage$ below is in $\IE[1, \codetext{\usertext{data} \% dim_age_grains}]$.}
  \\ \hline
  \codetext{\usertext{data} \% dim_region} 
  & \UCase  number of regions (star-forming cloud, diffuse ISM) & \none
  \\*
  \multicolumn{3}{>{\TBstrut}l}{The index $\iregion$ below is in $\IE[1, \codetext{\usertext{data} \% dim_region}]$.}
  \\ \hline
  \codetext{\usertext{data} \% age_grains($\iage$)} 
  &  \UCase age & $\unit{Myr}$
  \\*
  \multicolumn{3}{>{\TBstrut}l}{These ages are a subset of the \codetext{output_age(1:dim_output_age)} of \fullref{tab:output_main}.}
  \\ \hline
  \codetext{\usertext{data} \% region($\iregion$)} 
  &  \UCase data for the $\iregion$-th region & \none
  \\[1ex]
  \hdottedline
  \multicolumn{3}{>{\TBstrut}c}{\bfseries Subfields of \codetext{\usertext{data} \% region($\iregion$)}
    (denoted by \mention{\usertext{\abrege{region}}} below):}
  \\*
  \codetext{\usertext{\abrege{region}} \% id} 
  &  \UCase identifier of the region: \codetext{"SFC"} for
  star-forming clouds, \codetext{"DISM"} for the diffuse ISM & \none
  \\*
  \codetext{\usertext{\abrege{region}} \% dim_species} 
  & \UCase  number of grain species (graphites,
  silicates, etc.) for this region & \none
  \\*
  \multicolumn{3}{>{\TBstrut}l}{The index $\ispecies$ below is in $\IE[1, \codetext{\usertext{\abrege{region}} \% dim_species}]$.}
  \\*
  \codetext{\usertext{\abrege{region}} \% stoch_heating} 
  &  \UCase boolean set to \codetext{.true.}
  if SEDs are computed both for stochastically heated grains (so, with a
  temperature distribution) and at the equilibrium temperature, and to
  \codetext{.false.} if SEDs are computed only at the equilibrium temperature
  & \none
  \\*
  \codetext{\usertext{\abrege{region}} \% species($\ispecies$)} 
  &  \UCase data for the $\ispecies$-th grain species
  in the  $\iregion$-th region & \none
  \\[1ex]
  \hdottedline
  \multicolumn{3}{>{\TBstrut}c}{\bfseries Subfields of \codetext{\usertext{data} \% region($\iregion$) \% species($\ispecies$)} (denoted by \mention{\usertext{\abrege{species}}} below):}
  \\*
  \codetext{\usertext{\abrege{species}} \% id} 
  & \UCase  identifier of the grain species & \none
  \\*
  \codetext{\usertext{\abrege{species}} \% dim_lambda} 
  & \UCase  number of wavelengths & \none
  \\*
  \multicolumn{3}{>{\TBstrut}l}{The index $\icont$ below is in $\IE[1, \codetext{\usertext{\abrege{species}} \% dim_lambda}]$.}
  \\*
  \codetext{\usertext{\abrege{species}} \% lambda($\icont$)} 
  &  \UCase wavelength & $\micron$
  \\*
  \multicolumn{3}{>{\TBstrut}l}{These wavelengths are a subset of the \codetext{lambda_cont(1:dim_cont)} of \fullref{tab:output_constants}. Beware the unit!}
  \\*
  \codetext{\usertext{\abrege{species}} \% dim_radius} 
  &  \UCase number of grain radii & \none
  \\*
  \multicolumn{3}{>{\TBstrut}l}{The index $\irad$ below is in $\IE[1, \codetext{\usertext{\abrege{species}} \% dim_radius}]$.}
  \\*
  \codetext{\usertext{\abrege{species}} \% radius($\irad$)} 
  &  \UCase radius of the
  $\irad$-th grain& $\micron$
  \\*
  \codetext{\usertext{\abrege{species}} \% lum_stoch($\iage$, $\irad$, $\icont$)} 
  & \UCase   monochromatic luminosity for a stochastically heated grain
  (void if \codetext{\usertext{\abrege{region}} \% stoch_heating} is
  \codetext{.false.}) & $\unit{erg*s^{-1}*\micron^{-1}}$
  \\*
  \codetext{\usertext{\abrege{species}} \% lum_eq($\iage$, $\irad$, $\icont$)} 
  & \UCase monochromatic luminosity for a grain at equilibrium temperature
  & $\unit{erg*s^{-1}*\micron^{-1}}$
  \\*
\end{longtable}
%
\subsection{Outputs of \codefile{calib} and \codefile{colors}}
\index{calib@\codefile{calib}}%
\index{colors@\codefile{colors}}%

Code \codefile{colors}\index{colors@\codefile{colors}} takes as input a file of spectra\index{file of spectra} produced by
\codefile{spectra}\index{spectra@\codefile{spectra}}.
Beside other quantities, it computes the in-band luminosities and magnitudes
of the evolving galaxy SED through a series of filter passbands.

\codefile{colors}\index{colors@\codefile{colors}} also needs to read the file \codefile{calib.txt}%
\index{calib.txt@\codefile{calib.txt}},
which
is created by \codefile{calib}\index{calib@\codefile{calib}} and contains various filter-related properties,
in particular the in-band fluxes of some reference stars.

Before we turn to the outputs of \codefile{colors}\index{colors@\codefile{colors}}, we first have
to look at how, depending on the filter, in-band quantities are defined.
\subsubsection{Preliminary: filters, in-band quantities and magnitudes}
\paragraph{Filters}
\label{sec:filters}
For any filter~$f$ used by \codefile{calib}\index{calib@\codefile{calib}} or \codefile{colors}\index{colors@\codefile{colors}}, a \Emph{file of filter}%
\index{file of filter} containing the following informations must exist:
\begin{description}

\item[First (non-comment) line]
  \Emph{filter's identifier} (string of characters) used in \codefile{calib.txt}%
  \index{calib.txt@\codefile{calib.txt}}
  and \codefile{colors}\index{colors@\codefile{colors}};

\item[Second line]
  \Emph{type of transmission} (string): either \codetext{"energy"} or \codetext{"nb_phot"}.%
  \index{type of transmission|(}
  \begin{description}

  \item[\codetext{"energy"}]%
    \index{type of transmission!energy@\codetext{\dq energy\dq}}
    the transmission is the value of the system's (filter~+ detector) \emph{energy} response function,
    denoted by \mention{$T_{f, \lambda}$} hereafter (quantity \mention{$S'(\lambda)$} in \citet{BM2012}, app.~A2);

  \item[\codetext{"nb_phot"}]
    \index{type of transmission!nb_phot@\codetext{\dq nb_phot\dq}}
    the transmission is the value of the \emph{photon} response function
    (quantity \mention{$S(\lambda)$} in \citet{BM2012}, with $S(\lambda) \propto T_{f, \lambda}/\lambda$);
  \end{description}

\item[Third line] \emergencystretch=1em\relax
  \Emph{type of calibration} (string): either \codetext{"Vega"}, \codetext{"AB"}, \codetext{"TG"}, \codetext{"HST"},
  \codetext{"FOCA"}, \codetext{"IRAS"} or \codetext{"D4000"}.%
  \index{type of calibration|(}%
  \index{type of calibration!Vega@\codetext{\dq Vega\dq}}%
  \index{type of calibration!AB@\codetext{\dq AB\dq}}%
  \index{type of calibration!TG@\codetext{\dq TG\dq}}%
  \index{type of calibration!HST@\codetext{\dq HST\dq}}%
  \index{type of calibration!FOCA@\codetext{\dq FOCA\dq}}%
  \index{type of calibration!IRAS@\codetext{\dq IRAS\dq}}%
  \index{type of calibration!D4000@\codetext{\dq D4000\dq}}%

  For type \codetext{"Vega"}, the zero-point offset (real number) \emph{may} be given on the same line (set to $0$ if absent).
  For type \codetext{"IRAS"}, the nominal wavelength in~$\AA$ (real number) \emph{must} be given on the same line;

\item[Following lines]
  \Emph{passband of the filter}, with, on each line,
  \begin{itemize}

  \item
    the wavelength~$\lambda$ in~$\AA$,

  \item
    the transmission at this wavelength.
  \end{itemize}
\end{description}

The names of the files of filters%
\index{file of filter} are listed in \codefile{calib_dir/list_filters.txt}%
\index{list_filters.txt@\codefile{list_filters.txt}} (one per line).
To add another filter,
create a file of filter as explained in this section, and insert its name in
\codefile{list_filters.txt}.
Code \codefile{calib} needs to be run before the filter may be used by
\codefile{colors}.

\paragraph{In-band fluxes and luminosities}
\label{sec:in-band}
The \Emph{in-band flux} $\Ff$ through filter $f$ of the light emitted by some object  (denoted by \mention{$\star$} hereafter) is defined as follows,
depending on the type of calibration:%
\index{type of calibration|(}%
\begin{description}

\item[\codetext{"Vega"}, \codetext{"AB"}, \codetext{"TG"}, \codetext{"HST"} or \codetext{"FOCA"}]
  \index{type of calibration!Vega@\codetext{\dq Vega\dq}}
  \index{type of calibration!AB@\codetext{\dq AB\dq}}
  \index{type of calibration!TG@\codetext{\dq TG\dq}}
  \index{type of calibration!HST@\codetext{\dq HST\dq}}
  \index{type of calibration!FOCA@\codetext{\dq FOCA\dq}}
  \begin{equation}
    \label{eq:std_Flf}
    \Ff(\star) \egdef 
    \frac{1}{N_f}*\int{F_\lambda(\star)*T_{f, \lambda}*\df\lambda},
  \end{equation}
  where
  \begin{equation}
    N_f \egdef \int{T_{f, \lambda}*\df\lambda}
  \end{equation}
  is the \Emph{normalisation} of the filter,
  and $F_\lambda(\star)$ is the value, at wavelength~$\lambda$ and per unit wavelength,
  of the monochromatic flux (spectral irradiance) of the object;

\item[\codetext{"D4000"}]
  \index{type of calibration!D4000@\codetext{\dq D4000\dq}}
  the in-band flux and the normalisation are defined as
  \begin{equation}
    \Ff(\star) \egdef \frac{1}{N_f}*\int_{\lambda_1}^{\lambda_2}{F_\nu(\star)*\df\lambda}
    \quad\text{and}\quad
    N_f \egdef \int_{\lambda_1}^{\lambda_2}{\df\lambda},
  \end{equation}
  where $F_\nu(\star)$ is the value, at frequency~$\nu$ and per unit frequency,
  of the monochromatic flux of the object,
  and $\lambda_1$ and $\lambda_2$  are the bounds of the passband.

  Currently used only to compute the Balmer break indices $D4000$
  \citep{D4000} and $D_\txt{n}4000$ \citep{Dn4000}
  from the passbands respectively contained in the files
  \{\codefile{D4000-.txt}, \codefile{D4000+.txt}\} and
  \{\codefile{Dn4000-.txt}, \codefile{Dn4000+.txt}\} of \codefile{calib_dir/};

\item[\codetext{"IRAS"}]
  \index{type of calibration!IRAS@\codetext{\dq IRAS\dq}}
  $\Ff$ is defined by \fullref*{eq:std_Flf}, but the normalisation
  is given by
  \begin{equation}
    N_f \egdef \int{T_{f, \lambda}*\frac{\lambda^{\txt{nom}}_f}{\lambda}*\df\lambda},
  \end{equation}
  where
  $\lambda^{\txt{nom}}_f$ is the nominal wavelength of the filter
  ($12*\micron$, $25*\micron$, $60*\micron$, $100*\micron$).
  The normalisation is defined so that, if $\lambda*F_\lambda(\star)$
  ($= \nu*F_\nu(\star)$) is constant, then
  \begin{equation}
    \frac{1}{N_f}*\int{F_\lambda(\star)*T_{f, \lambda}*\df\lambda} = F_\lambda(\star, \lambda = \lambda^{\txt{nom}}_f).
  \end{equation}

  Currently used for IRAS filters only.
\end{description}
The dimension of $\Ff$ is that of $F_\nu$ if the type of calibration is \codetext{"D4000"}, and that of $F_\lambda$ otherwise%
\index{type of calibration|(}%
\index{type of calibration!D4000@\codetext{\dq D4000\dq}}.

The \Emph{in-band luminosity} through filter~$f$, $\overline L_f$, is defined in the same way as
$\Ff$, but from the monochromatic luminosity (spectral power) $L_\lambda$ or $L_\nu$ instead
of the monochromatic flux $F_\lambda$ or $F_\nu$.
\paragraph{Magnitudes}
\label{sec:magnit}
The magnitude $m_f$ in filter $f$ of an object
is defined in one of the following ways, depending on the type of calibration:%
\index{type of calibration|(}%
\begin{description}

\item[\codetext{"Vega"}]
  \index{type of calibration!Vega@\codetext{\dq Vega\dq}}
  standard photometric system based on Vega.
  \begin{equation}
    m_f(\star) \egdef m_f^\txt{std}(\star)
    = -2.5*\log_{10}\ufrac{\Ff(\star)}{\Ff(\txt{Vega})} + m_f^\txt{std}(\txt{Vega}),
  \end{equation}
  where $m_f^\txt{std}(\txt{Vega})$ is the apparent magnitude adopted for Vega in the standard system.
  One has
  \begin{equation}
    m_f^\txt{std}(\txt{Vega}) = \codetext{std_Vega_def} + \mathrm{\ZPO}_f,
  \end{equation}
  where \codetext{std_Vega_def} is the default apparent magnitude of Vega (set to~$0.03$ in
  \codefile{source_dir/mod_filters_constants.f90}) and $\mathrm{\ZPO}_f$ is a possible zero-point offset for filter~$f$
  (it is read from the file containing the passband and currently set to~$0$ for all filters);

\item[\codetext{"AB"}]
  \index{type of calibration!AB@\codetext{\dq AB\dq}}
  AB system.
  \begin{align}
    m_f(\star) \egdef m_f^\AB(\star)
    &= -2.5*\log_{10}\ufrac{\int{F_\nu(\star)*T_{f, \nu}*\df\nu}}{%
      \mathrm{erg*s^{-1}*cm^{-2}*Hz^{-1}}*\int{T_{f, \nu}*\df\nu}} - 48.60
    \notag \\
    &= -2.5*\log_{10}\ufrac{\Ff(\star)}{%
      \mathrm{erg*s^{-1}*cm^{-2}*\AA^{-1}}}
    + 2.5*\log_{10}\ufrac{\mathrm{Hz}*\int{T_{f, \lambda}*\df\lambda}}{%
      \mathrm{\AA}*\int{T_{f, \nu}*\df\nu}} - 48.60
  \end{align}
  because $T_{f, \nu} = T_{f, \lambda}$ and $F_\nu*\df\nu = F_\lambda*\df\lambda$;

\item[\codetext{"TG"}]
  \index{type of calibration!TG@\codetext{\dq TG\dq}}
  Thuan \& Gunn system based on star $\mathrm{BD\,{+}17^\circ\,4708}$.
  \begin{equation}
    m_f(\star) \egdef m_f^\TG(\star)
    = -2.5*\log_{10}\ufrac{\Ff(\star)}{%
      \Ff(\mathrm{BD\,{+}17^\circ\,4708})} + 9.50\,;
  \end{equation}

\item[\codetext{"HST"}]
  \index{type of calibration!HST@\codetext{\dq HST\dq}}
  HST system.
  \begin{equation}
    m_f(\star) \egdef m_f^\HST(\star) = -2.5*\log_{10}\ufrac{\Ff(\star)}{%
      \mathrm{erg*s^{-1}*cm^{-2}*\AA^{-1}}} - 21.10.
  \end{equation}
  For a monochromatic filter, $ m_f^\HST =  m_f^\AB$;

\item[\codetext{"FOCA"}]
  \index{type of calibration!FOCA@\codetext{\dq FOCA\dq}}
  FOCA system.
  \begin{equation}
    m_f(\star) \egdef m_f^\FOCA(\star) = -2.5*\log_{10}\ufrac{\Ff(\star)}{%
      \mathrm{erg*s^{-1}*cm^{-2}*\AA^{-1}}} - 21.175\,;
  \end{equation}

\item[\codetext{"IRAS"} and \codetext{"D4000"}] magnitude not defined.
  \index{type of calibration!IRAS@\codetext{\dq IRAS\dq}}
  \index{type of calibration!D4000@\codetext{\dq D4000\dq}}
\end{description}

The following relations hold between these definitions:
\begin{equation}
  m_f^{\AB/\TG}(\star) = m_f^\txt{std}(\star) + (m_f^{\AB/\TG}[\txt{Vega}] - m_f^\txt{std}[\txt{Vega}]).
\end{equation}

\subsubsection{Outputs of \codefile{calib}}
\label{sec:calib_outputs}%
\index{calib@\codefile{calib}}%
After a few lines of general information, \codefile{calib}\index{calib@\codefile{calib}} write
in \codefile{calib_dir/calib.txt}%
\index{calib.txt@\codefile{calib.txt}} a table containing, for each filter listed in
\codefile{calib_dir/list_filters.txt}%
\index{list_filters.txt@\codefile{list_filters.txt}} (one per line), the following quantities%
\footnote{%
  \label{fn:undefined_mag}%
  When the wavelength range covered by the SED of an object does not
  fully contain that of the filter's passband, or when no magnitude is defined
  given the type of calibration of the filter (\eg~for Balmer break and IRAS filters)%
  \index{type of calibration|(}%
  \index{type of calibration!IRAS@\codetext{\dq IRAS\dq}}%
  \index{type of calibration!D4000@\codetext{\dq D4000\dq}}%
  ,
  the magnitude of the object
  is set to \codetext{undefined_mag} ($99.999$ by default; this parameter is
  defined in \codefile{source_dir/mod_filters_constants.f90}).%
}:
\begin{description}

\item[\codetext{filter_id}]
  the filter's identifier (see~\fullref{sec:filters})
  defined in the corresponding file of filter%
  \index{file of filter}
  and
  used in \codefile{colors}\index{colors@\codefile{colors}};

\item[\codetext{i_filter}]
  ordinal number of the filter in \codefile{calib_dir/list_filters.txt};

\item[\codetext{flux_band_Vega}]
  in-band flux $\Ff(\txt{Vega})$, through filter~$f$,
  of the light emitted by Vega and received on the Earth.
  The monochromatic flux of Vega on the Earth is read from file
  \codefile{calib_dir/Vega_BaSeL_SED.txt}.

  In $\mathrm{erg*s^{-1}*Hz^{-1}*cm^{-2}}$ if the type of calibration is \codetext{"D4000"};
  otherwise, in $\mathrm{erg*s^{-1}*\AA^{-1}*cm^{-2}}$;%
  \index{type of calibration|(}%
  \index{type of calibration!D4000@\codetext{\dq D4000\dq}}

\item[\codetext{filter_norm}]
  normalisation $N_f$ of the passband;

\item[\codetext{lambda_mean}]
  mean wavelength in $\AA$ of the filter ($= \lambda'_0$ in \citet{BM2012}),
  defined as
  \begin{equation}
    \codetext{lambda_mean} = \ufrac{\int{\lambda*T_{f, \lambda}*\df\lambda}}{\int{T_{f, \lambda}*\df\lambda}}\text;
  \end{equation}

\item[\codetext{lambda_eff_Vega}]
  effective wavelength in~$\AA$ of Vega.

  If the type of calibration is \codetext{"D4000"}, the effective wavelength is defined as
  \begin{equation}
    \codetext{lambda_eff(Vega)} = \ufrac{\int_{\lambda_1}^{\lambda_2}{\lambda*F_\nu(\txt{Vega})*\df\lambda}}{
      \int_{\lambda_1}^{\lambda_2}{F_\nu(\txt{Vega})*\df\lambda}}.
  \end{equation}
  Otherwise,
  \begin{equation}
    \codetext{lambda_eff(Vega)} = \ufrac{\int{\lambda*F_\lambda(\txt{Vega})*T_{f, \lambda}*\df\lambda}}{\int{F_\lambda(\txt{Vega})*T_{f, \lambda}*\df\lambda}}\text;
  \end{equation}%
  \index{type of calibration|(}%
  \index{type of calibration!D4000@\codetext{\dq D4000\dq}}

\item[\codetext{AB_Vega}]
  AB apparent magnitude of Vega;

\item[\codetext{TG_Vega}]
  Thuan \& Gunn apparent magnitude of Vega.

  The monochromatic flux of $\mathrm{BD\,{+}17^\circ\,4708}$
  on the Earth is read from file \codefile{calib_dir/BD+17d4708_SED.txt};

\item[\codetext{lum_band_Sun}]
  in-band luminosity $\overline L\vphantom{L}_f^\odot$, through filter~$f$,
  of $L_\lambda^\odot$, the monochromatic luminosity of the Sun per unit wavelength
  (read from file \codefile{calib_dir/Sun_BaSeL_SED.txt}).

  In $\mathrm{erg*s^{-1}*Hz^{-1}}$ if the type of calibration is \codetext{"D4000"};
  otherwise, in $\mathrm{erg*s^{-1}*\AA^{-1}}$;%
  \index{type of calibration|(}%
  \index{type of calibration!D4000@\codetext{\dq D4000\dq}}

\item[\codetext{mag_abs_Sun}]
  absolute magnitude of the Sun in the photometric system defined by the type of calibration%
  \index{type of calibration|(}.
  Computed from the monochromatic flux received at $10*\mathrm{pc}$ from the Sun,
  \begin{equation}
    F_\lambda^\odot = \ufrac{L_\lambda^\odot}{4*\pi*(10*\mathrm{pc})^2};
  \end{equation}

\item[\codetext{lambda_pivot}]
  pivot wavelength of the filter, defined as
  \begin{equation}
    \codetext{lambda_pivot} = \sqrt{\ufrac{\int T_{f, \lambda}*\df\lambda}{\int \lambda^{-2}*T_{f, \lambda}*\df\lambda}}.
  \end{equation}

  The mean values of $F_\lambda$ and $F_\nu$ through the passband are related by
  \begin{equation}
    \ufrac{\int{F_\nu * T_{f, \nu} * \df\nu}}{\int{T_{f, \nu} * \df\nu}}
    = \frac{\codetext{lambda_pivot}^2}{c} *
    \ufrac{\int{F_\lambda * T_{f, \lambda} * \df\lambda}}{\int{T_{f, \lambda} * \df\lambda}}
    \text;
  \end{equation}

\item[\codetext{filter_width}]
  width of the filter, defined as
  \begin{equation}
    \codetext{width} = \sqrt{\ufrac{\int{(\lambda-\codetext{lambda_mean})^2*T_{f, \lambda}*\df\lambda}}{\int T_{f, \lambda}*\df\lambda}}\text;
  \end{equation}

\item[\codetext{calib_type}]
  type of calibration%
  \index{type of calibration|(}.

\end{description}
%


\subsubsection{Outputs of \codefile{colors}}
\index{colors@\codefile{colors}}%
\paragraph{Overview}
Code \codefile{colors}\index{colors@\codefile{colors}} processes a file of spectra\index{file of spectra}
(see~\fullref{sec:run_colors}) and writes in a file of colors\index{file of colors},
after a few lines of general information, a series of tables
providing various quantities (one per column) as a function of the age of the galaxy.
Other quantities may be added to a given table or new tables may be produced
using the procedure and functions described hereafter.
To change the output of \codefile{colors}\index{colors@\codefile{colors}}, edit file
\mention*{\codefile{source_dir/incl_colors.f90}} and compile the code as
explained in \fullref{sec:compil}.

The first column in each table is the $\iage$-th~age in~Myr (variable
\codetext{data \% output_age($\iage)$}).
Other columns are added to a table through the Fortran statement
\Mention[,]{\codetext{call add_column(\placeholder{heading}, \placeholder{quantity}\optional{, \placeholder{format}})}}
where
\begin{itemize}
\item
  \placeholder{heading} is a string of characters
  choosen by the user as heading for the column;
\item
  \placeholder{quantity} is an array of reals containing the value at all ages
  of the quantity written in the column;
\item
  the optional \placeholder{format} is a string defining the Fortran-style
  format, used to output the quantity at a single age.
  A default format is used if the value of \placeholder{format}
  is not provided.
\end{itemize}
When a table is complete, the Fortran statement
\Mention{\codetext{call write_table}}
prints it in the file of colors\index{file of colors}, and the next table, if any,
may then be built.
\paragraph{Quantities available in the file of spectra}
By default, \codefile{colors}\index{colors@\codefile{colors}} outputs only a subset of the quantities directly available in the file of spectra\index{file of spectra}
(quantities \codetext{data \% \placeholder{field}} in \codefile{incl_colors.f90}; see~\fullref{tab:output_main} and, if the output of \codefile{spectra}\index{spectra@\codefile{spectra}} is detailed,
\fullref{tab:output_detailed}).
Two 
functions, \mention*{\codetext{ISM_abund}} and \mention*{\codetext{L_line}},
have also been created to
ease the access to the following quantities (through a string of characters
instead of an integer index%
\footnote{The value of the integer corresponding to a given string is computed
  by a hash function.}):
\begin{description}[.\space]

\item[Interstellar medium abundances]
  The abundance of the $\ielem$-th element in the interstellar medium
  at age $t = \codetext{data \% output_age($\iage)$}$ is given by
  \begin{equation}
    \codetext{ISM_abund($\iage$, \placeholder{elem_id})} \egdef
    \codetext{data \% ISM_abund($\iage$, $\ielem$)},
  \end{equation}
  where the identifier \placeholder{elem_id} is a string holding the value of
  \codetext{data \% elem_id($\ielem$)};

\item[Emission line luminosities]
  The normalized luminosity of the $\iline$-th emission line
  at age~$t$ is given by
  \begin{equation}
    \codetext{L_line($\iage$, \placeholder{line_id})} \egdef
    \codetext{data \% L_line($\iage$, $\iline$)},
  \end{equation}
  where the identifier \placeholder{line_id} is a string holding the value of
  \codetext{data \% line_id($\iline$)}.
  All lines are listed with their identifier and their wavelength in 
  \codefile{Cloudy_dir/list_neb_lines.txt}%
  \index{list_neb_lines.txt@\codefile{list_neb_lines.txt}}.
\end{description}
For convenience, the functions \codetext{ISM_abund} and \codetext{L_line} are overloaded:
to access the arrays containing the values of \codetext{ISM_abund($\iage$, \placeholder{elem_id})}
and \codetext{L_line($\iage$, \placeholder{line_id})} at all ages,
use \codetext{ISM_abund(\placeholder{elem_id})} and
\codetext{L_line(\placeholder{line_id})} instead
(as in the \mention{\codetext{call add_column}} statements).
\paragraph{Other quantities}
Besides the quantities already present in the file of spectra\index{file of spectra},
\codefile{colors}\index{colors@\codefile{colors}} also outputs the bolometric magnitude of the galaxy,
its in-band luminosities and absolute magnitudes through various filters, several colors, Balmer break
indices and the equivalent width of some emission lines.
These quantities are defined as follows in \codefile{colors.f90}:
\begin{description}[.\space]
\item[Bolometric magnitude]
  The normalized bolometric magnitude of the galaxy at age~$t$
  is given by 
  \begin{equation}
    \codetext{M_bol($\iage$)} = 4.75 - 2.5*\log_{10}\ufrac{\norm\Lbol(t)*\Msol}{%
      \Lsol}.
  \end{equation}

  For a galaxy with a system's mass $\Msys$, the (unnormalized) bolometric magnitude
  would be
  \begin{equation}
    \Mbol(t) = \codetext{M_bol($\iage$)} -2.5*\log_{10}\ufrac{\Msys}{\Msol}\text;
  \end{equation}

\item[In-band luminosities]
  {\emergencystretch=1em
    The normalized in-band luminosity of the galaxy in filter~$f$ at age~$t$
    (in $\mathrm{erg*s^{-1}*Hz^{-1}*\Msol^{-1}}$ if the type of calibration is \codetext{"D4000"}
    and in $\mathrm{erg*s^{-1}*\AA^{-1}*\Msol^{-1}}$ otherwise) is%
    \index{type of calibration|(}%
    \index{type of calibration!D4000@\codetext{\dq D4000\dq}}%

    \begin{equation}
      \codetext{lum_band($\iage$, \placeholder{filter_id})} =
      \norm{\overline L}_f(t),
    \end{equation}
    where \placeholder{filter_id} is the identifier of filter~$f$
    in the corresponding file of filter%
    \index{file of filter}
    and in \codefile{calib.txt}%
    \index{calib.txt@\codefile{calib.txt}},
    and $\overline L_f$ is defined in \fullref{sec:in-band}.}

  The normalized in-band luminosity of the galaxy relative to that of the Sun,
  in $\Msol^{-1}$,
  is given by
  \begin{equation}
    \codetext{lum_band_rel_Sun($\iage$, \placeholder{filter_id})}
    = \ufrac{\norm{\overline L}_f(t)}{\overline L\vphantom{L}_f^\odot}\text;
  \end{equation}

\item[Absolute magnitudes]
  The normalized absolute magnitude%
  \footnote{%
    \label{fn:undefined_mag_col}%
    Magnitudes and colors are set to \codetext{undefined_mag}
    (see \fullref{fn:undefined_mag}) if they may not be computed.}
  of the galaxy in filter~$f$,
  \codetext{magnit($\iage$, \placeholder{fil\-ter_id})},
  is given by one of the equations in \fullref{sec:magnit},
  depending on the type of calibration%
  \index{type of calibration|(},
  with
  \begin{equation}
    \Ff(t) = \ufrac{\norm{\overline L}_f(t) * \Msol}{4*\pi*(10*\mathrm{pc})^2}.
  \end{equation}

  For a galaxy with a system's mass $\Msys$, the (unnormalized) absolute magnitude
  in filter~$f$ would be
  \begin{equation}
    \mathcal{M}_f(t) = \codetext{magnit($\iage$, \placeholder{filter_id})} -2.5*\log_{10}\ufrac{\Msys}{\Msol}\text;
  \end{equation}

\item[Colors]
  Let $m_{1}$ and $m_{2}$ denote either the apparent or the absolute magnitudes of the galaxy
  in filters~$f_1$ and~$f_2$.
  The color%
  \footnoteref{fn:undefined_mag_col}
  of the galaxy at age~$t$ is given by
  \begin{equation}
    \codetext{color($\iage$, \placeholder{filter1_id}, \placeholder{filter2_id})} = m_{1}(t) - m_{2}(t),
  \end{equation}
  where the strings \placeholder{filter1_id} and
  \placeholder{filter2_id} are the identifiers of the filters~$f_1$ and~$f_2$
  in the corresponding files of filter%
  \index{file of filter}
  and in \codefile{calib.txt}%
  \index{calib.txt@\codefile{calib.txt}};

\item[Balmer break indices]
  The indices $D4000$ and $D_\txt{n}4000$ (\codetext{D4000} and \codetext{Dn4000} in
  \codefile{colors.f90}) are given by
  \begin{equation}
    \codetext{D\optional{n}4000($\iage$)} =
    \ufrac{\int_{\lambda_1^+}^{\lambda_2^+}{\norm L_\nu(t)*\df\lambda}}{%
      \int_{\lambda_1^-}^{\lambda_2^-}{\norm L_\nu(t)*\df\lambda}},
  \end{equation}
  with
  \begin{equation}
    (\lambda_1^-, \lambda_2^-, \lambda_1^+, \lambda_2^+) = \begin{cases}
      (3750, 3950, 4050, 4250)*\AA & \text{for $D4000$ \citep{D4000},} \\
      (3850, 3950, 4000, 4100)*\AA & \text{for $D_\txt{n}4000$ \citep{Dn4000}\text;}
    \end{cases}
  \end{equation}

\item[Equivalent width of emission lines]
  The equivalent width in~$\AA$ of the $\iline$-th emission line at age~$t$
  is given by
  \begin{equation}
    \codetext{eq_width($\iage$, \placeholder{line_id})} =
    \frac{\codetext{L_line($\iage$, \placeholder{line_id})}}{\norm L_\txt{cont}(t)},
  \end{equation}
  where \placeholder{line_id} is a string holding the value of
  \codetext{data \% line_id($\iline$)},
  and $\norm L_\txt{cont}(t)$ is the normalized monochromatic luminosity
  of the galaxy at age~$t$ and
  wavelength~\codetext{lambda_line($\iline$)} in the continuum;
  $\norm L_\txt{cont}(t)$ is estimated by linear interpolation
  from the values of \codetext{lum_cont($\iage$,~$\icont$)} at adjacent wavelengths.
\end{description}
By definition, colors, Balmer break indices and equivalent widths do not
depend on the normalization to the mass of the system.

The quantities \codetext{lum_band($\iage$, \placeholder{filter_id})},
\codetext{magnit($\iage$, \placeholder{fil\-ter_id})},
\codetext{color($\iage$, \placeholder{filter1_id}, \placeholder{filter2_id})}
and \codetext{eq_width($\iage$, \placeholder{line_id})}
are provided by overloaded functions:
to access the arrays containing their values at all ages,
use \codetext{lum_band(\placeholder{filter_id})},
\codetext{magnit(\placeholder{fil\-ter_id})},
\codetext{color(\placeholder{filter1_id}, \placeholder{filter2_id})}
and \codetext{eq_width(\placeholder{line_id})}
instead.
%
\clearpage
\part*{Appendices}
\phantomsection
\addcontentsline{toc}{part}{Appendices}
\begin{appendices}
\section{Typographical conventions}
\label{app:typo}
The following typographical conventions are used in this documentation:
\begin{center}
  \begin{tabular}{@{\rule[-1.5ex]{0pt}{4ex}}lp{12cm}}
    \hline
    \codetext{xyz}\index{\codetext{}} & Code text or shell command; also used for e-mail addresses.\footnotemark
    \\
    \hline
    \codefile{xyz}\index{\codefile{}} & File or directory name; also used for Web addresses.\footnoteref{fn:omitted} 
    \\
    \hline
    \texttt{\optional{xyz}}\index{\optional@$\lcorners\,\rcorners$\,}
    & Optional text.\\
    \hline
    \texttt{\cesurechar}\index{$\lnot$}
    & Symbol used to show that a string of characters has been hyphenated. \\
    \hline
    \multicolumn{2}{@{\rule[-1.5ex]{0pt}{4ex}}l}{\textbf{For dialogs with the codes:}}\\
    \keyboardinput{xyz\optional{\return}}\index{$\blacktriangleleft$}\index{\return} & Text input from the keyboard (once the 
    \keystroke{Return}\,/\,\keystroke{Enter} key is pressed).

    The  symbol \character{\return}
    is always written if the text is empty; \\
    \screenoutput{xyz}\index{$\blacktriangleright$} & Text printed on the terminal. \\
    \hline
    \skipped & Skipped lines. \\
    \hline
  \end{tabular}
\end{center}
\restoreFootnoteNumber
\stepcounter{footnote}%
\footnotetext{\label{fn:omitted}%
  \character{\guilsinglleft}, \character{\guilsinglright},
  \character{\guillemotleft} and \character{\guillemotright} are omitted
  in strings of characters mentioned (denoted by \mention{}) rather than used.
}%
The non-ascii symbols \character{$\lcorners$}, 
\character{$\rcorners$},
\character{\cesurechar},
\character{\return}, \character{$\blacktriangleleft$},
\character{$\blacktriangleright$},
\character{\guilsinglleft}, \character{\guilsinglright}, 
\character{\guillemotleft}, \character{\guillemotright} 
and \character{\skipped} should not be typed by the user!

Placeholders (\ie names of dummy files, statements, keywords or values which the user must replace 
by actual ones) are written in slanted characters.
\section{Syntax used in files of scenarios}
\index{file of scenarios}%
\subsection{Syntax used for all statements}
\label{app:syntax}
Statements are written in a Fortran~90-like manner:
\begin{description}[.\space]
\item[Line length and continuation character]
  A line cannot be longer than 132 characters.
  
  A statement can be continued on the following line by appending an ampersand 
  (\codetext{\&}) to the current line.
  It the line is cut in the middle of a string of characters, 
  a keyword or a numerical or boolean value, 
  an ampersand must also appear at the beginning of the continuation line;
\item[Multiple statements]
  Multiple statements, separated by semi-colons, can be put on the same line;
\item[End-of-line comments]
  On a line, anything following an exclamation mark is considered to be a 
  comment and is skipped, unless the exclamation mark is within a string of 
  characters.

  Blank lines are considered as comments;
\item[Block comments]
  It is also possible to comment sections of the file in a C-language manner: 
  anything between \codetext{/*} and \codetext{*/} is skipped, 
  unless \codetext{/*} is within a string of characters. 
  Block comments are not nested;
\item[Strings of characters]
  Strings of characters are 
  enclosed either in pairs of quotes (\codetext{"}) 
  or of apostrophes (\codetext{'}). 

  In a string enclosed in quotes, a quote character may be obtained 
  by typing it twice (\codetext{"{}"}).
  In a string enclosed in apostrophes, an apostrophe character may be obtained 
  by typing it twice (\codetext{'{}'});
\item[Case of letters]
  The case of letters is not significant, except in character strings;
\item[Spaces and tabulations]
  Space characters are not significant, except in character strings.
  
  Tabulations are treated as spaces.
\end{description} 
\subsection{Syntax used for assignments}
\label{app:key_val}
A file of scenarios\index{file of scenarios} contains a sequence of statements, most of them of the form 
\mention{\codetext{\usertext{key}\optional{\usertext{\codetext{(}indices\codetext{)}}}~= 
\usertext{val}}}, 
where \usertext{key} is the name of a parameter, 
\usertext{val} is either a scalar value or an array of scalar values assigned to this parameter,
and the optional \usertext{\codetext{(}indices\codetext{)}} 
provides the indices for which this assignment is done.

Depending on the parameter considered, a scalar value is either
\begin{itemize}
\item
  an integer;
\item
  a real;
\item
  a boolean, \codetext{.true.} or \codetext{.false.}
  (the case of the characters used here does not matter);
\item
  a string of characters delimited by quotes or apostrophes.
\end{itemize}
%
\subsubsection{Parameters: scalars and arrays}
\index{scalars}%
\index{arrays}%
Depending on the parameter considered, 
\usertext{key\optional{\codetext{(}indices\codetext{)}}} may either be 
a scalar or an array:
\begin{itemize}
\item
  If it is necessarily a scalar, \usertext{\codetext{(}indices\codetext{)}}
  \emph{must not} be present;
\item
  Else, \usertext{key\optional{\codetext{(}indices\codetext{)}}} \emph{may} be
  an array.

  If \mention{\usertext{\codetext{(}indices\codetext{)}}} is present,
  \usertext{indices} defines the indices for which \usertext{key} is 
  defined by \usertext{val}.
  Unless \usertext{indices} refers to a single index (see below),
  \usertext{key\codetext{(}indices\codetext{)}} is an array.

  If the \mention{\usertext{\codetext{(}indices\codetext{)}}} part is absent,
  then \usertext{key} is a scalar and
  writing \mention{\codetext{\usertext{key}~=}}
  is the same as writing \mention{\codetext{\usertext{key}(1)~=}}. 
  Note that this is not the usual convention in Fortran. 
\end{itemize}
%
\subsubsection{Indices}
Possible forms of \usertext{indices} are the following:
\begin{itemize}
\item
  A \emph{single \emph{integer} index}, \usertext{$i$};
\item 
  An \emph{index loop}:
  \begin{itemize}
  \item
    \usertext{$f$\codetext{:}$l$}: 
    all the integers $i \in \IE[f, l]$
    (the set of indices is void if $f >l$).

    Note that \usertext{key\codetext{($i$:$i$)}} is considered as an array
    of parameters 
    while \usertext{key\codetext{($i$)}} is a scalar one, although both
    refer to the same object;
  \item
    \usertext{$f$\codetext{:}$l$\codetext{:}$s$}: 
    all the integers
    $i = f + j*s$, where $j \in \IE[0, n]$ and
    $n = \left\lfloor (l-f)/s \right\rfloor$.
    (\mention{$\lfloor x\rfloor$} denotes the integer part (floor function) of $x$).

    $f$, $l$ and $s$ are integers.
    The stride $s$ must be different from $0$. 
    The signs of $l-f$ and
    $s$ must be the same; otherwise, the set of indices is void.
  \end{itemize}

  Other notations, such as \usertext{key\codetext{(:)}} 
  or \usertext{key\codetext{(:}$f$\codetext{:)}}, 
  although standard in Fortran, are not allowed in this code;
\item 
  A \emph{list of single indices} or \emph{index loops}, 
  
  \UserText[,]{\codetext{[}indices$_1$\codetext{,} indices$_2$\codetext{,} 
    $\cdots$\codetext{]}}
  where \usertext{indices$_1$}, \usertext{indices$_2$}, \etc, are either 
  single indices or index loops.
  An index can appear only once in a list.
\end{itemize}
All the indices must be larger than or equal to $1$. 
The maximal upper bound depends on the parameter considered.
%
\subsubsection{Values}
Possible forms of \usertext{val} are the following:
\begin{itemize}
  \item
    a \emph{single scalar value}, \usertext{$v$}. 

    If \usertext{\codetext{(}indices\codetext{)}} is present, then
    \[
    \forall~i\in\text{\usertext{indices}},\quad \text{\usertext{key}($i$)} = v.
    \] 
    If not, then
    \[
    \text{\usertext{key}($1$)} = v;
    \]
  \item
    an \emph{array of scalar values} of the same type, 
    \usertext{\codetext{[}$v_1$\codetext{,} $\cdots$\codetext{,} 
      $v_n$\codetext{]}},
    $n$ being the number of indices in \usertext{indices}.

    In this case, the \usertext{\codetext{(}indices\codetext{)}} part must be present and 
    must not consist in a single index.
    If \usertext{indices} contains indices $j_1,~\cdots,~j_n$ 
    in that order,
    then
    \[
    \forall~i\in\IE[1, n],\quad \text{\usertext{key}($j_i$)} = v_i.
    \]
\end{itemize}
\subsubsection{An example}
\begin{lstlisting}[name=keyval, frame=trlb]
SF_type = "ISM_mass"/*\label{line:keyvala}*/
SF_type(3:6) = "ISM_mass"/*\label{line:keyvalb}*/
SF_type(2:4:2) = "exponential"/*\label{line:keyvalc}*/
SF_expo_timescale(4:1:-2) = [2000, 500]/*\label{line:keyvald}*/
SF_ISM_timescale([1:5]) = 1000/*\label{line:keyvale}*/
SF_type(2) = "exponential"/*\label{line:keyvalf}*/
SF_type([3, 5:6]) = "none"/*\label{line:keyvalg}*/
SF_begin_time(2:4) = 1000/*\label{line:keyvalh}*/
SF_end_time([1, 4, 3]) = [500, 2000, 500]/*\label{line:keyvali}*/
\end{lstlisting}

Let $E_i$ denote the $i$-th star formation episode.
\begin{description}
\item[Line~\ref{line:keyvala}]
  \codetext{SF_type} is set to \codetext{"ISM_mass"} for $E_1$
  (SFR proportional to the ISM mass);
\item[Line~\ref{line:keyvalb}]
  The same for $E_3$ to $E_6$;
\item[Line~\ref{line:keyvalc}]
  \codetext{SF_type} is set to \codetext{"exponential"} for $E_2$ 
  (SFR exponentially decreasing or increasing with age);
  \codetext{SF_type} is also changed to \codetext{"exponential"} for $E_4$;
\item[Line~\ref{line:keyvald}]
  \codetext{SF_expo_timescale(2)} is set to $500*$Myr
  and \codetext{SF_expo_timescale(4)} to $2000*$Myr;
\item[Line~\ref{line:keyvale}]
  Parameter \codetext{SF_ISM_timescale} is set to $1000*$Myr for episodes
  $E_1$ to $E_5$, which sets \codetext{SF_type} to \codetext{"ISM_mass"}
  for $E_2$ and $E_4$ 
  (\codetext{SF_ISM_timescale} is still undefined for $E_6$);
\item[Line~\ref{line:keyvalf}]
  \codetext{SF_type} is set back to \codetext{"exponential"} for $E_2$
  and \codetext{SF_expo_timescale} recovers its previous value ($500*$Myr);
\item[Line~\ref{line:keyvalg}]
  Star formation is finally cancelled for $E_3$, $E_5$ and $E_6$;
\item[Line~\ref{line:keyvalh}]
  Star formation begins at $1*$Gyr for $E_2$, $E_3$ and $E_4$;
\item[Line~\ref{line:keyvali}]
  Star formation ends at $500*$Myr for $E_1$ and $E_3$ and at $2*$Gyr for $E_4$.
\end{description}

The final state is the following:
\begin{itemize}
\item
  {\emergencystretch=1em\relax
  $E_1$ begins at $0*$Myr (default value) and ends at $500*$Myr.
  $\codetext{SF_type(1)} =\codetext{"ISM_mass"}$
  and \codetext{SF_ISM_\-time\-scale(1)}${} = 1000*$Myr;\par}
\item
  $E_2$ and $E_4$ are consecutive to $E_1$:
  both begin at $1000*$Myr and overlap until $2000*$Myr, 
  at which age $E_4$ ends;
  $E_2$ ends at $20000*$Myr (default).
  $\codetext{SF_type(2)} = \codetext{"exponential"}$
  and $\codetext{SF_expo_timescale(2)} = 500*$Myr.
  $\codetext{SF_type(4)} = \codetext{"ISM_mass"}$
  and $\codetext{SF_ISM_timescale(4)} = 1000*$Myr; 
\item
  $E_3$ does not form stars since 
  $\codetext{SF_end_time(3)} = 500*\mathrm{Myr} < 
  \codetext{SF_begin_time(3)} = 1000*$Myr;
\item
  $E_5$ and $E_6$ do not form stars either.
\end{itemize}
\section{Playing with IMFs}
\begin{RENVOI}
  \renvoi\fullref{sec:IMF}.
\end{RENVOI}


\subsection{Adding other IMFs}
\label{app:add_IMFs}
Let us first remind that, in the code, an IMF~$\phi$ is defined by
\begin{equation}
  \phi(m) \egdef \df n/\df(\ln m),
\end{equation}
where $m$ is the initial mass of a star and $\df n$ is the number of stars,
per unit of initial mass of the SSP,
born with a mass in $\interv[m, m+\df m[$.

You may use other IMFs besides the ones already provided with the code 
(see~\fullref{sec:IMF}). 
Any additional IMF must be a continuous, piecewise power-law function.
More precisely,
\begin{itemize}
\item
  $\phi$ is continuous for all $m$ in some 
  interval $\interv[m_{\inf}, m_{\sup}]$ 
  (with $0 < m_{\inf} < m_{\sup}$)
  and null outside of this interval.
  To be in agreement with the default stellar evolutionary tracks, it
  is recommended to take $m_{\inf} \ge 0.09*\Msol$ 
  and $m_{\sup} \le 120*\Msol$;
\item
  {\emergencystretch=1em
    $\interv[m_{\inf}, m_{\sup}]$ is divided in 
    $p$ ($\ge 1$) subintervals $\interv[m_i, m_{i+1}]$,
    with $m_1 = m_{\inf}$\kern0.5pt, $m_{p+1} = m_{\sup}$ 
    and $m_i < m_{i+1}$ for all $i \in \IE[1, p]$;\par
  }
  \vspace{-\baselineskip}
\item
  \mi\begin{flalign}\md
  \forall~m \in \interv[m_i, m_{i+1}], \quad \phi(m) \propto m^{s_i}\,,
  &&
  \end{flalign}
  where the slope~$s_i$ is a constant.
\end{itemize}

To define a new IMF, create an IMF file containing the following 
informations%
\footnote{Have a look to \codefile{IMFs_dir/IMF_Kroupa.txt} 
  for example.}: 
\makeatletter
\@afterheading
\makeatother
\begin{itemize}
  \item
    zero or more comment lines (blank lines and lines where the first 
    non blank character is \mention{\codetext{!}})
    providing, for instance, the reference paper for this IMF;
  \item
    $p$ lines containing, for all integers~$i$ from $1$ to $p$,
    the value, in solar masses, of the bottom mass $m_i$
    of the $i$-th mass bin and of the slope $s_i$ in this bin.
    These values should be separated by one or more space characters;
  \item
    a last line containing the value of $m_{p+1}$ in solar masses.
\end{itemize}
Finally, add a line containing the name of this file to
\codefile{IMFs_dir/list_IMFs.txt}.

The values of $m_{\inf}$ and $m_{\sup}$ given in the IMF file
may be superseded when running \codefile{SSPs}.

Code \codefile{SSPs} will automatically ensure that, for all $i \in \IE[2, p]$,
the IMF is continuous at $m_i$ and that the IMF is normalized, \ie
\begin{equation}
  \int_{m=m_{\inf}}^{m_{\sup}} m*\phi(m)*\df(\ln m) = 1.
\end{equation}

\subsection[\protect\mention{Evolving} the IMF]{``Evolving'' the IMF}
\label{app:evolve_IMF}
The stellar initial mass function is constant in the code. 
It is however possible to mimic its evolution as a function of the 
ISM metallicity by mixing several sets of SSPs, as long as they use the same
library of stellar spectra and the same output ages.
As an example, let us consider two sets of SSPs, 
\userfile{Salpeter\codefile{_SSPs.txt}} and \userfile{RB\codefile{_SSPs.txt}},
respectively computed by \codefile{SSPs}\index{SSPs@\codefile{SSPs}} for the IMFs of \citet{Salpeter} and \citet{RB}
(from the files \codefile{IMF_Salpeter.txt} and 
\codefile{IMF_Rana_Basu.txt} in \codefile{IMF_dir/}).
The first file will look like this:
\begin{lstlisting}[frame=tbrl, numbers=none]
/*\skipped*/
Salpeter_tracks_Z0.008+.txt
Salpeter_tracks_Z0.02+.txt
/*\skipped*/
\end{lstlisting}
The second file will look the same, with \mention{\usertext{RB}} instead of 
\mention{\usertext{Salpeter}}.

{\emergencystretch=1em
  To implement a \emph{smooth} transition from the \citeauthor{Salpeter}'s IMF to the 
\citeauthor{RB}'s one between $Z = 0.008$ and $Z = 0.02$, create a
file containing all the lines in \userfile{Salpeter\codefile{_SSPs.txt}}
up to, and \emph{excluding}, \mention{\usertext{Salpeter\codetext{_tracks_Z0.02+.txt}}}, 
followed by all the lines in \userfile{RB\codefile{_SSPs.txt}}
from, and \emph{including}, \mention{\usertext{RB\codetext{_tracks_Z0.02+.txt}}}.\par}

To implement a \emph{sudden} transition from the \citeauthor{Salpeter}'s IMF to the 
\citeauthor{RB}'s one at $Z = 0.02$, create a
file containing all the lines in \userfile{Salpeter\codefile{_SSPs.txt}}
up to, and \emph{including}, \mention{\usertext{Salpeter\codetext{_tracks_Z0.02+.txt}}}, 
followed by all the lines in \userfile{RB\codefile{_SSPs.txt}}
from, and \emph{including}, \mention{\usertext{RB\codetext{_tracks_Z0.02+.txt}}}.
\section{Additional notes on the modeling}
\subsection{Relating normalized quantities to observed ones}
\label{app:normalization}
To clarify the meaning of normalized quantities and their relation to observed ones, 
let us take the following example. 
Consider a galaxy observed at wavelengths 
$\lambda'$ in the range $\interv[\lambda'_-, \lambda'_+]$ with an intensity 
($={}$spectral irradiance) $F_{\lambda'}^\txt{obs}$.
Let $z$ be the redshift of the galaxy and $D$ its luminosity distance to the observer.
The monochromatic luminosity of the galaxy in the emitter's frame is then
$L_\lambda^\txt{emit} = 4*\pi*D^2*(1+z)*F_{\lambda'}^\txt{obs}$, where $\lambda'$
is related to the emission wavelength $\lambda$ by $\lambda' = (1+z)*\lambda$.

Denote by $\smash{\norm L_\lambda^\txt{mod}}$ (quantity \codetext{lum_cont} in 
\fullref{tab:output_main}) 
the normalized monochromatic luminosity
of a model fitting best the \emph{shape} of the emitted spectrum on 
$\interv[\lambda_-, \lambda_+] \egdef \interv[\lambda'_-, \lambda'_+]/(1+z)$.
For instance, the best model may minimize 
\begin{equation}
  \chi^2 = \min_{k>0}\int_{\lambda=\lambda_-}^{\lambda_+} {\left(L_\lambda^\txt{emit} - k*\norm L_\lambda^\txt{mod}\right)^2*\df\lambda}.
\end{equation}
One has then $L_\lambda^\txt{emit} \approx k*\smash{\norm L_\lambda^\txt{mod}}$ on $\interv[\lambda_-, \lambda_+]$ for the
(model-dependent) scalar~$k$ minimizing above integral.
Since $L_\lambda^\txt{mod} = \Mref*\norm L_\lambda^\txt{mod}$,
one has $L_\lambda^\txt{mod} \approx L_\lambda^\txt{emit}$ on $\interv[\lambda_-, \lambda_+]$ if one takes $k$ as an estimate of $\Mref$.

All unnormalized quantities have to be multiplied by this factor to obtain
the \mention{real} value. For instance, an estimate of 
the mass of live stars in the galaxy is $\Mref\times\codetext{live_stars_mass}$
(see \fullref{tab:output_main}).
\subsection{Using stellar spectra available on
  too narrow a wavelength range}
\label{app:bolom_corr}
The approach described in \fullref{sec:SSPs},
to compute $\lmon_i^\LSS$ (\mention{$\lmon$} below)
is appropriate if the contribution to the bolo\-metric
luminosity of radiations emitted outside of the wavelength range covered by
the spectrum is negligible.
This is the case for the libraries of theoretical spectra we use.

To compute $\lmon$ from spectra covering a narrower wavelength range $\interv[\lambda_-,\lambda_+]$,
\eg~observed spectra, one needs the \Emph{bolometric correction} $\BC_f$
of the star in some filter~$f$.
The passband transmission of~$f$, $T_{f, \lambda}$, must be null outside of $\interv[\lambda_-,\lambda_+]$.
Then,
\begin{equation}
  \lmon = \frac{F_\lambda}{\Ff}*\frac{\Ff}{\Fbol},
\end{equation}
where $F_\lambda$ is the spectral irradiance of the star at wavelength
$\lambda$,
$\Ff \egdef \int F_\lambda*T_{f, \lambda}*\df\lambda/\mathopen{}\int T_{f, \lambda}*\df\lambda$
is its average value in filter~$f$,
and $\Fbol$ is the (bolometric) irradiance of the star.
The ratio $F_\lambda/\Ff$ is derived directly from the spectrum. 
To compute the other factor, $\Ff/\Fbol$, one uses the following relations:
\begin{itemize}
  \vspace{-\baselineskip}
\item
  \mi\begin{flalign}\md
    m_f = -2.5*\log_{10}(\Ff/\Ff^\txt{ref}) + m_f^\txt{ref},
    &&
  \end{flalign}
  \vadjust{\vspace{-0.8ex}}where $m_f$ is the apparent magnitude of the star in filter~$f$,
  $m_f^\txt{ref}$ is that of some reference source (real, \eg\ Vega, or virtual), and
  $\Ff^\txt{ref}$ is the value of $\Ff$
  for the latter;
  \vspace{-\baselineskip}
\item
  \mi\begin{flalign}\md
    m_\bol = m_f + \BC_f\,,&&
  \end{flalign}
  where $m_\bol$ is the apparent bolometric magnitude of the star;
  \vspace{-\baselineskip}
\item
  \mi\begin{flalign}\md
    m_\bol - \Mbol = 5*\log_{10}(D/[10*\txt{pc}]),&&
  \end{flalign}
  where $\Mbol$ is the absolute magnitude of the star, and $D$ is its distance;
  \vspace{-\baselineskip}
\item
  \mi\begin{flalign}\md
    \Mbol = \Mbolsol - 2.5*\log_{10}(\Lbol/\Lsol),&&
  \end{flalign}
  where $\Mbolsol = 4.74$ is the absolute bolometric magnitude of the Sun \citep{IAU_bol_mag},
  $\Lbol$ is the bolometric luminosity of the star, and $\Lsol$ is
  that of the Sun;
  \vspace{-\baselineskip}
\item
  \mi\begin{flalign}\md
    \Fbol = \Lbol/(4*\pi*D^2).&&
  \end{flalign}%
\end{itemize}
Combining these equations, one obtains
\begin{equation}
  \frac{\Ff}{\Fbol} =
  \frac{4*\pi*(10*\txt{pc})^2*\Ff^\txt{ref}}{\Lsol}*
  10^{(\BC_f-\Mbolsol)/2.5}.
\end{equation}
\subsection{Self-absorption}
\label{app:self_abs}
One can also obtain \fullref{eq:SA}, as the limit of a sequence of
iterations:
the term of order~$0$ is the original light emitted by stars and the ionized gas, attenuated by dust;
at first order, one takes into account the attenuation of energy emitted by grains
in the first processing of the original photons;
at second order, one also considers the attenuation of the energy emitted by grains
in the second processing, and so on. So,
\begin{equation}
  L_\lambda^\SA = \overline\Theta_\lambda*L_\lambda^\unatt + \sum_{k=1}^\infty L_\lambda^{\dust, k},
\end{equation}
where $L_\lambda^{\dust, k}$ is the transmitted contribution from the $k$-th processing.
With the two approximations made to model self\babelhyphen{nobreak}absorption,
one has
\begin{equation}
  L_\lambda^{\dust, k} = \overline\Theta_\lambda^{\mkern9mu\gamma}*L_\lambda^{\dust, k, \unatt},
\end{equation}
where $L_\lambda^{\dust, k, \unatt}$ is the unattenuated contribution from the $k$-th processing
and
\begin{equation}
  L_\lambda^{\dust, k, \unatt} = \beta_k*L_\lambda^{\dust, \noSA}.
\end{equation}
The energy absorbed by grains must be re-emitted, so
\begin{equation}
  \int L_\lambda^{\dust, k+1, \unatt}*\df\lambda
  = \int {\left(1-\overline\Theta_\lambda^{\mkern9mu\gamma}\right)*L_\lambda^{\dust, k, \unatt}*\df\lambda}.
\end{equation}
Therefore,
\begin{equation}
  \beta_{k+1}*\int L_\lambda^{\dust, \noSA}*\df\lambda
  = \beta_k*\int {\left(1-\overline\Theta_\lambda^{\mkern9mu\gamma}\right)*L_\lambda^{\dust, \noSA}*\df\lambda}
\end{equation}
and, since $\beta_1 = 1$,
\begin{equation}
  \beta_k = \beta_2^{k-1} \quad\text{with}\quad
  \beta_2 = \frac{\int {\left(1-\overline\Theta_\lambda^{\mkern9mu\gamma}\right)*L_\lambda^{\dust, \noSA}*\df\lambda}}{
    \int L_\lambda^{\dust, \noSA}*\df\lambda}.
\end{equation}
One finally obtains
\begin{align}
  L_\lambda^\SA &= \overline\Theta_\lambda*L_\lambda^\unatt
  + \sum_{k=1}^\infty {\overline\Theta_\lambda^{\mkern9mu\gamma}*\beta_2^{k-1}*L_\lambda^{\dust, \noSA}}
  = \overline\Theta_\lambda*L_\lambda^\unatt + \frac{\overline\Theta_\lambda^{\mkern9mu\gamma}}{1-\beta_2}*L_\lambda^{\dust, \noSA}
  \notag\\
  &= \overline\Theta_\lambda*L_\lambda^\unatt + \alpha*\overline\Theta_\lambda^{\mkern9mu\gamma}*L_\lambda^{\dust, \noSA} .
\end{align}
\section{Trees and tables of parameters}
\label{app:trees_tables}
\subsection{Preliminaries}
This section contains the following additional informations on the parameters
defining a scenario:
\begin{description}[\space]
\item[Trees]
  connecting related parameters
  (see Fig.~\ref{fig:reserv_infall_tree} to~\ref{fig:output_tree},
  p.~\pageref{fig:reserv_infall_tree} to~\pageref{fig:output_tree}).
  In these trees, the names of parameters are framed, but not their values.
  The values of parameters are shown only if the extension of the set of values
  is finite. (This typically excludes numbers and arbitrary strings such as
  filenames.)

  The name of parameters having a default value is marked with the symbol
  \character{$\defaultsymbol$}.
  When values are shown, the default value is also marked with
  \character{$\defaultsymbol$}.

  A \Emph{node} consists in a parameter name and its associated value.
  Nodes are arranged in descending order from the left to the right.
  Every node in a tree may have zero,
  one or more children (the nodes connected to their right side in trees),
  but has at most one parent (the node connected to their left side in trees).
  Arrows point from the name a parameter to the value of its parent node.

  We recall here the two rules already stated in \fullref{sec:prelim_scenarios}:
  \begin{enumerate}
  \item
    \textbf{Explicitely setting the value of a parameter implicitely assigns all
      its ancestors to consistent values%
    \footnote{So, these trees are more like drainage/river systems, with
      downstream on the left and upstream on the right.};}
  \item
    \textbf{Parameters which are not assigned in a scenario, whether explicitely
      or implicitely, retain the values they had in the previous scenario
      or remain undefined if they have no default value and have never been
      assigned.} (See \fullref{fn:rule2_exceptions}, for exceptions.)
  \end{enumerate}
\item[Tables]
  providing the physical unit, type, range and default value (if any)
  of all parameters.
  See Tables~\ref{tab:SSPs_chemical} to~\ref{tab:other_param},
  p.~\pageref{tab:SSPs_chemical} to~\pageref{tab:other_param}.

  Default values are usually either natural values, typical values, or simple
  values chosen to avoid forcing the user to make choices for other associated
  parameters\textellipsis 
  Values are usually adapted to the case where there is only one episode.
\end{description}
These informations complete those already provided in \fullref{sec:descr_param},
to \fullref{sec:other_statements}.
\subsection{Cosmology}
\label{app:cosmo_list}
\begin{RENVOI}
  \renvoi\fullref{sec:cosmo_param}.
\end{RENVOI}

\saveFootnoteNumber
\begin{longtable}{>{\TBstrut}lcccc}
  \caption{Parameters related to cosmology\captionPoint}
  \\*
  \nobreakhline
  Parameter & Unit & Type & Possible values & Default value
  \\*
  \nobreakhline\nobreakhline
  \endfirsthead
  \multicolumn{5}{@{}c@{}}{$\downarrow$}
  \\*
  \nobreakhline
  Parameter & Unit & Type & Possible values & Default value
  \\*
  \nobreakhline\nobreakhline
  \endhead
  \multicolumn{5}{@{}c@{}}{$\downarrow$}
  \\*
  \endfoot
  \nobreakhline
  \endlastfoot
  \paramname{Omega_m} & \none & real & $\in\interv[0, 1]$ & $0.308$\footnotemark
  \\
  \hline
  \paramname{H_0} & $\mathrm{km*s^{-1}*Mpc^{-1}}$ & real & $\ge 0$ & $67.8$\footnoteref{fn:Planck}
  \\
  \hline
  \paramname{form_redshift} & \none & real & $\ge 0$ & $10$
  \\
  \hline
  \paramname{CBR} & \none & boolean & \codetext{.true.}, \codetext{.false.} & \codetext{.false.}
  \\
\end{longtable}
\restoreFootnoteNumber
\stepcounter{footnote}
\footnotetext{\label{fn:Planck}%
  See \citet{Planck}.}

\vfill\break
\subsection{Single stellar populations and chemical evolution parameters}
\label{app:SSPs_chemical}%
\forbiddenBreak

\begin{RENVOI}
  \renvoi\fullref{sec:SSPs_param}, and \fullref{sec:chemical_param}.
\end{RENVOI}
\begin{table}[H]
  \caption{%
    \label{tab:SSPs_chemical}%
    Single stellar populations and chemical evolution\captionPoint}
  \begin{center}
    \begin{tabular}{>{\TBstrut}lcccc}
      \hline
      Parameter & Unit & Type & Possible values & Default value \\
      \hline\hline
      \paramname{SSPs_set} & \none & string & & undefined \\
      \hline
      \paramname{ISM_init_Z} & \none & real & $\in\interv[0, 1]$ & $0$ \\
      \hline
      \paramname{close_bin_frac} & \none & real & $\in\interv[0, 1]$ & $0.05$ \\
      \hline
    \end{tabular}
  \end{center}
\end{table}
\noindent\begin{minipage}{\linewidth}
  \noindent
  \subsection{Reservoirs and infall}
  \label{app:reserv_infall_list}%
  \forbiddenBreak

  \begin{RENVOI}
    \renvoi\fullref{sec:zones}, and \fullref{sec:reserv_infall_param}.
  \end{RENVOI}
  \forbiddenBreak
  \begin{figure}[H]
    \centerline{\includegraphics{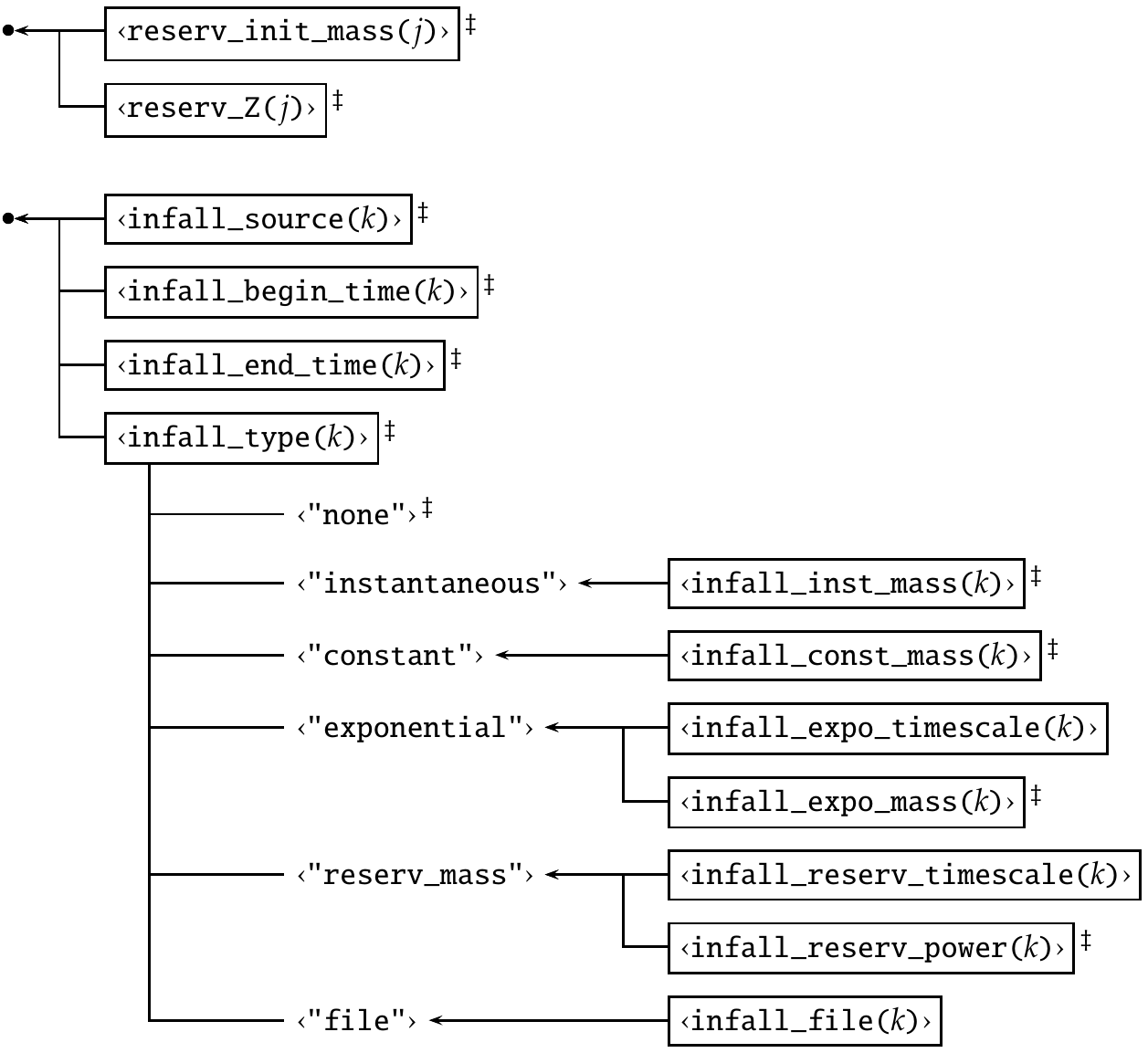}}
    \caption{\label{fig:reserv_infall_tree}%
      Trees of parameters related to the $j$-th reservoir
      and to the $k$-th episode of infall\captionPoint}
  \end{figure}
\end{minipage}
\vfill\break
\begin{longtable}{>{\TBstrut}lcccc}
  \caption{\label{tab:reserv_infall}%
    Reservoir and infall parameters\captionPoint}
  \\*
  \nobreakhline
  Parameter & Unit & Type & Possible values & Default value
  \\*
  \nobreakhline\nobreakhline
  \endfirsthead
  \multicolumn{5}{@{}c@{}}{$\downarrow$}
  \\*
  \nobreakhline
  Parameter & Unit & Type & Possible values & Default value
  \\*
  \nobreakhline\nobreakhline
  \endhead
  \multicolumn{5}{@{}c@{}}{$\downarrow$}
  \\*
  \endfoot
  \nobreakhline
  \endlastfoot
  \paramname{reserv_init_mass} & \none & array of reals & $\in\interv[0, 1]$ & $0$
  \\*
  \nobreakhline
  \paramname{reserv_Z} & \none & array of reals & $\in\interv[0, 1]$ & $0$
  \\*
  \nobreakhline
  \paramname{infall_source} & Myr & array of reals &  $\in\interv[0, 20000]$
  & $1$
  \\
  \hline
  \paramname{infall_begin_time} & Myr & array of reals &  $\in\interv[0, 20000]$ & $0$
  \\
  \hline
  \paramname{infall_end_time} & Myr & array of reals &  $\in\interv[0, 20000]$ & $20000$
  \\
  \hline
  \paramname{infall_type} & \none & array of strings &
  \begin{tabular}[t]{@{}l@{}}
    \Tstrut\codetext{"none"},\\*
    \codetext{"instantaneous"},\\*
    \codetext{"constant"},\\*
    \codetext{"exponential"},\\*
    \codetext{"reserv_mass"},\\*
    \codetext{"file"}\Bstrut
  \end{tabular}
  & \codetext{"none"}
  \\
  \hline
  \paramname{infall_inst_mass} & \none & array of reals & $\ge 0$ & $1$
  \\
  \hline
  \paramname{infall_const_mass} & \none & array of reals & $\ge 0$ & $1$
  \\
  \hline
  \paramname{infall_expo_timescale} & Myr & array of reals & $\neq 0$ & undefined
  \\
  \hline
  \paramname{infall_expo_mass} & \none & array of reals & $\ge 0$ & $1$
  \\
  \hline
  \paramname{infall_reserv_timescale} & Myr & array of reals & $> 0$ & undefined
  \\
  \hline
  \paramname{infall_reserv_power} & \none & array of reals & & $1$
  \\*
  \nobreakhline
  \paramname{infall_file} & \none & array of strings & & undefined
  \\*
\end{longtable}
\vfill\break
\subsection{Star formation}
\label{app:SF_list}

\begin{RENVOI}
  \renvoi\fullref{sec:chem_evol}, and \fullref{sec:SF_param}.
\end{RENVOI}

\vfill\break
\begin{longtable}{>{\TBstrut}lccccc}
  \caption{\label{tab:SF}%
    Star formation parameters\captionPoint}
  \\*
  \nobreakhline
  Parameter & Unit & Type & Possible values & Default value
  \\*
  \nobreakhline\nobreakhline
  \endfirsthead
  \multicolumn{5}{@{}c@{}}{$\downarrow$}
  \\*
  \nobreakhline
  Parameter & Unit & Type & Possible values & Default value
  \\*
  \nobreakhline\nobreakhline
  \endhead
  \multicolumn{5}{@{}c@{}}{$\downarrow$}
  \\*
  \endfoot
  \nobreakhline
  \endlastfoot
  \paramname{SF_begin_time} & Myr & array of reals & $\in\interv[0, 20000]$ & $0$
  \\
  \hline
  \paramname{SF_end_time} & Myr & array of reals & $\in\interv[0, 20000]$ & $20000$
  \\
  \hline
  \paramname{SF_type} & \none & array of strings &
  \begin{tabular}[t]{@{}l@{}}
    \Tstrut\codetext{"none"},\\
    \codetext{"instantaneous"},\\
    \codetext{"constant"},\\
    \codetext{"exponential"},\\
    \codetext{"peaked"},\\
    \codetext{"ISM_mass"},\\
    \codetext{"infall"},\\
    \codetext{"file"}\Bstrut
  \end{tabular}
  & \codetext{"none"}
  \\
  \hline
  \paramname{SF_inst_mass} & \none & array of reals & $\ge 0$ & $1$
  \\
  \hline
  \paramname{SF_const_mass} & \none & array of reals & $\ge 0$ & $1$
  \\
  \hline
  \paramname{SF_expo_timescale} & Myr & array of reals & $\neq 0$ & undefined
  \\
  \hline
  \paramname{SF_expo_mass} & \none & array of reals & $\ge 0$ & $1$
  \\
  \hline
  \paramname{SF_peaked_timescale} & Myr & array of reals & $> 0$ & undefined
  \\
  \hline
  \paramname{SF_peaked_mass} & \none & array of reals & $\ge 0$ & $1$
  \\
  \hline
  \paramname{SF_ISM_timescale} & Myr & array of reals & $> 0$ & undefined
  \\
  \hline
  \paramname{SF_ISM_power} & \none & array of reals & & $1$
  \\
  \hline
  \paramname{SF_ISM_threshold} & \none & array of reals & $\in\interv[0, 1]$ & $0$
  \\
  \hline
  \paramname{SF_infall_factor} & \none & array of reals & $\ge 0$ & $1$
  \\
  \hline
  \paramname{SF_file} & \none & array of strings & & undefined
  \\
  \hline
  \paramname{SF_stochastic} & \none & array of booleans
  & \codetext{.true.}, \codetext{.false.}& \codetext{.false.}
  \\
  \hline
  \paramname{SF_stoch_fluc} & \none & array of reals & & undefined
  \\
  \hline
  \paramname{SF_stoch_timescale} & Myr & array of reals & $> 0$ & undefined
  \\
  \hline
  \paramname{SF_Z_type} & \none & array of strings &
  \begin{tabular}[t]{@{}l@{}}
    \Tstrut\codetext{"consistent"},\\
    \codetext{"constant"},\\
    \codetext{"file"}\Bstrut
  \end{tabular}
  & \codetext{"consistent"}
  \\
  \hline
  \paramname{SF_Z_const_val} & \none & array of reals & $\in\interv[0, 1]$ & undefined
  \\
  \hline
  \paramname{SF_Z_file} & \none & array of strings & & undefined
  \\
  \hline
  \paramname{SF_inert_frac} & \none & array of reals & $\in\interv[0, 1]$
  & $0$
  \\
\end{longtable}
\noindent\begin{minipage}{\linewidth}
  \noindent
  \subsection{Galactic outflows}
  \label{app:outflow_list}

  \begin{RENVOI}
    \renvoi\fullref{sec:zones}, and \fullref{sec:outflow_param}.
  \end{RENVOI}

  \begin{figure}[H]
    \centerline{\includegraphics{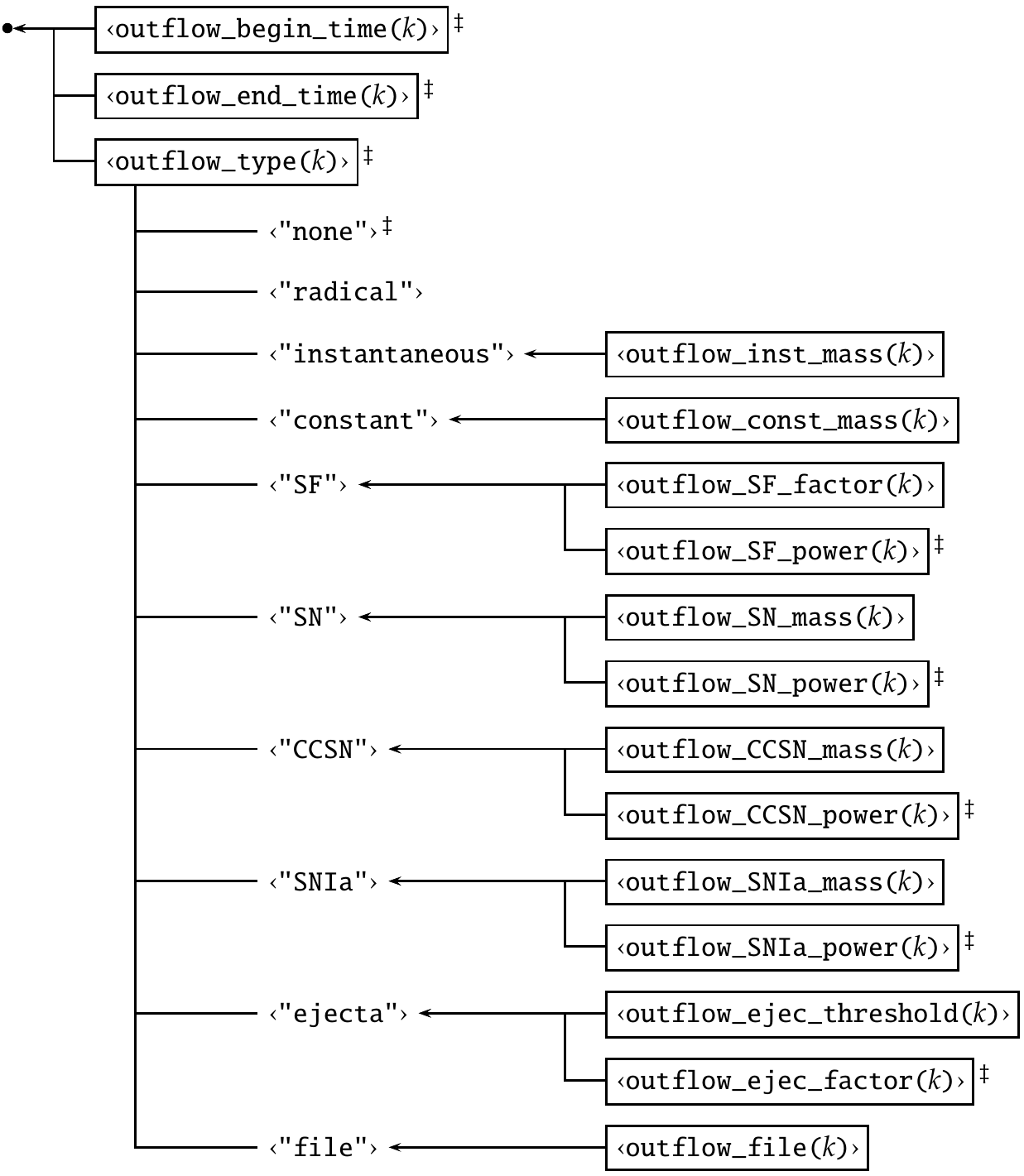}}
    \caption{\label{fig:outflow}%
      Tree of parameters related to the $k$-th episode of galactic outflow\captionPoint}
  \end{figure}
\end{minipage}

\newpage
\begin{longtable}{>{\TBstrut}lcccc}
  \caption{\label{tab:outflow}%
    Outflow parameters\captionPoint}
  \\*
  \nobreakhline
  Parameter & Unit & Type & Possible values & Default value
  \\*
  \nobreakhline\nobreakhline
  \endfirsthead
  \multicolumn{5}{@{}c@{}}{$\downarrow$}
  \\*
  \nobreakhline
  Parameter & Unit & Type & Possible values & Default value
  \\*
  \nobreakhline\nobreakhline
  \endhead
  \multicolumn{5}{@{}c@{}}{$\downarrow$}
  \\*
  \endfoot
  \nobreakhline
  \endlastfoot
  \paramname{outflow_begin_time} & Myr & array of reals & $\in\interv[0, 20000]$
  & $0$
  \\
  \hline
  \paramname{outflow_end_time} & Myr & array of reals & $\in\interv[0, 20000]$
  & $20000$
  \\
  \hline
  \paramname{outflow_type} & \none & array of strings &
  \begin{tabular}[t]{@{}l@{}}
    \Tstrut\codetext{"none"},\\
    \codetext{"radical"},\\
    \codetext{"instantaneous"},\\
    \codetext{"constant"},\\
    \codetext{"SF"},\\
    \codetext{"SN"},\\
    \codetext{"CCSN"},\\
    \codetext{"SNIa"},\\
    \codetext{"ejecta"},\\
    \codetext{"file"}\Bstrut
  \end{tabular}
  & \codetext{"none"}
  \\
  \hline
  \paramname{outflow_inst_mass} & \none & array of reals & $\ge 0$ & undefined
  \\
  \hline
  \paramname{outflow_const_mass} & \none & array of reals & $\ge 0$ & undefined
  \\
  \hline
  \paramname{outflow_SF_factor} & \none & array of reals & $\ge 0$ & undefined
  \\
  \hline
  \paramname{outflow_SF_power} & \none & array of reals & $\ge 0$ & $0$
  \\
  \hline
  \paramname{outflow_SN_mass} & $\Msol$ & array of reals & $\ge 0$
  & undefined
  \\
  \hline
  \paramname{outflow_SN_power} & \none & array of reals & $\ge 0$ & $0$
  \\
  \hline
  \paramname{outflow_CCSN_mass} & $\Msol$ & array of reals & $\ge 0$
  & undefined
  \\
  \hline
  \paramname{outflow_CCSN_power} & \none & array of reals & $\ge 0$ & $0$
  \\
  \hline
  \paramname{outflow_SNIa_mass} & $\Msol$ & array of reals & $\ge 0$
  & undefined
  \\
  \hline
  \paramname{outflow_SNIa_power} & \none & array of reals & $\ge 0$ & $0$
  \\
  \hline
  \paramname{outflow_ejec_threshold} & \none & array of reals & & undefined
  \\
  \hline
  \paramname{outflow_ejec_factor} & \none & array of reals & $\ge 0$ & $1$
  \\
  \hline
  \paramname{outflow_file} & \none & array of strings & & undefined
  \\
\end{longtable}
\vfill\break
\subsection{Dust evolution}
\label{app:dust_evol_list}

\begin{RENVOI}
  \renvoi\fullref*{sec:dust_evol_basic}, \fullref*{sec:dust_evol_Dwek}, and
  \fullref{sec:dust_evol_param}.
\end{RENVOI}

\begin{figure}[H]
  \centerline{\includegraphics{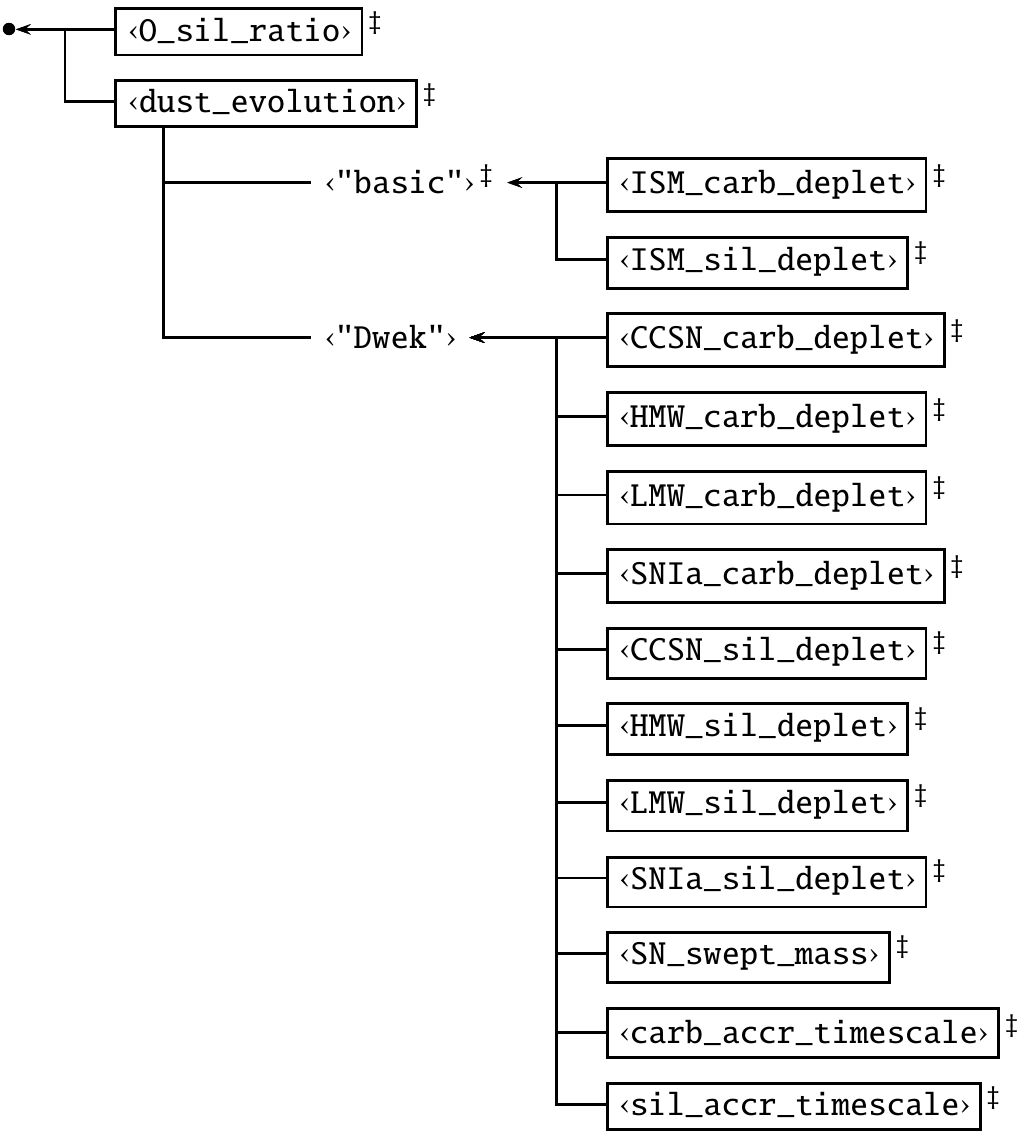}}
  \caption{Tree of parameters related to dust evolution\captionPoint}
\end{figure}

\newpage
\saveFootnoteNumber
\begin{longtable}{>{\TBstrut}lcccc}
  \caption{Parameters related to dust evolution\captionPoint}
  \\*
  \nobreakhline
  Parameter & Unit & Type & Possible values & Default value
  \\*
  \nobreakhline\nobreakhline
  \endfirsthead
  \multicolumn{5}{@{}c@{}}{$\downarrow$}
  \\*
  \nobreakhline
  Parameter & Unit & Type & Possible values & Default value
  \\*
  \nobreakhline\nobreakhline
  \endhead
  \multicolumn{5}{@{}c@{}}{$\downarrow$}
  \\*
  \endfoot
  \nobreakhline
  \endlastfoot
  \paramname{O_sil_ratio} & \none & real & $\ge 0$ & $1$\footnotemark
  \\
  \nobreakhline
  \paramname{dust_evolution} & \none & boolean &
  \begin{tabular}[t]{@{}l@{}}
    \Tstrut\codetext{"basic"}, \\
    \codetext{"Dwek"}\Bstrut
  \end{tabular}
  & \codetext{"basic"}
  \\
  \hline
  \paramname{ISM_carb_deplet} & \none & real & $\in\interv[0, 1]$ & $1/3$
  \\
  \hline
  \paramname{ISM_sil_deplet} & \none & real & $\in\interv[0, 1]$ & $1$
  \\
  \hline
  \paramname{CCSN_carb_deplet} & \none & real & $\in\interv[0, 1]$ & $0.5$\footnotemark
  \\
  \hline
  \paramname{CCSN_sil_deplet} & \none & real & $\in\interv[0, 1]$ & $0.8$\footnoteref{fn:Dwek}
  \\
  \hline
  \paramname{HMW_carb_deplet} & \none & real & $\in\interv[0, 1]$ & $1$\footnoteref{fn:Dwek}
  \\
  \hline
  \paramname{HMW_sil_deplet} & \none & real & $\in\interv[0, 1]$ & $1$\footnoteref{fn:Dwek}
  \\
  \hline
  \paramname{LMW_carb_deplet} & \none & real & $\in\interv[0, 1]$ & $1$\footnoteref{fn:Dwek}
  \\
  \hline
  \paramname{LMW_sil_deplet} & \none & real & $\in\interv[0, 1]$ & $1$\footnoteref{fn:Dwek}
  \\
  \hline
  \paramname{SNIa_carb_deplet} & \none & real & $\in\interv[0, 1]$ & $0.5$\footnoteref{fn:Dwek}
  \\
  \hline
  \paramname{SNIa_sil_deplet} & \none & real & $\in\interv[0, 1]$ & $0.8$\footnoteref{fn:Dwek}
  \\
  \hline
  \paramname{SN_swept_mass} & $\Msol$ & real & $\ge 0$ & $400$\footnotemark
  \\
  \hline
  \paramname{carb_accr_timescale} & Myr & real & $\ge 0$ & $250$\footnotemark
  \\*
  \nobreakhline
  \paramname{sil_accr_timescale} & Myr & real & $\ge 0$ & $250$\footnoteref{fn:accr_timescale}
  \\
\end{longtable}
\restoreFootnoteNumber
\stepcounter{footnote}
\footnotetext{Value from \citet{Dwek98}, sec.~5.2.
  \citet{McKinnon+} recommend a lower value.}
\stepcounter{footnote}
\footnotetext{\label{fn:Dwek}%
  Values from \citet{Dwek98}, sec.~5.2.}
\stepcounter{footnote}
\footnotetext{%
  $\approx \langle m_{\txt{d}}\rangle_{\txt{MW}}/Z_{\txt{d}, \txt{MW}}$:
  see \citet{Dwek+Cherchneff}, after their eq.~(9).}
\stepcounter{footnote}
\footnotetext{\label{fn:accr_timescale}%
  See \citet{McKinnon+}, sec.~3.2.}
\kern-2\parindent
\noindent\begin{minipage}{\textwidth}
  \noindent
  \subsection{Dust attenuation and emission}
  \label{app:dust_transfer_list}

  \begin{RENVOI}
    \renvoi\fullref{sec:diffuse_attenuation}, \fullref{sec:GSD}, and \fullref{sec:dust_transfer_param}.
  \end{RENVOI}

  \noindent  \begin{figure}[H]
    \stdParbox
    \noindent\begin{minipage}{\textwidth}
      \centerline{\includegraphics{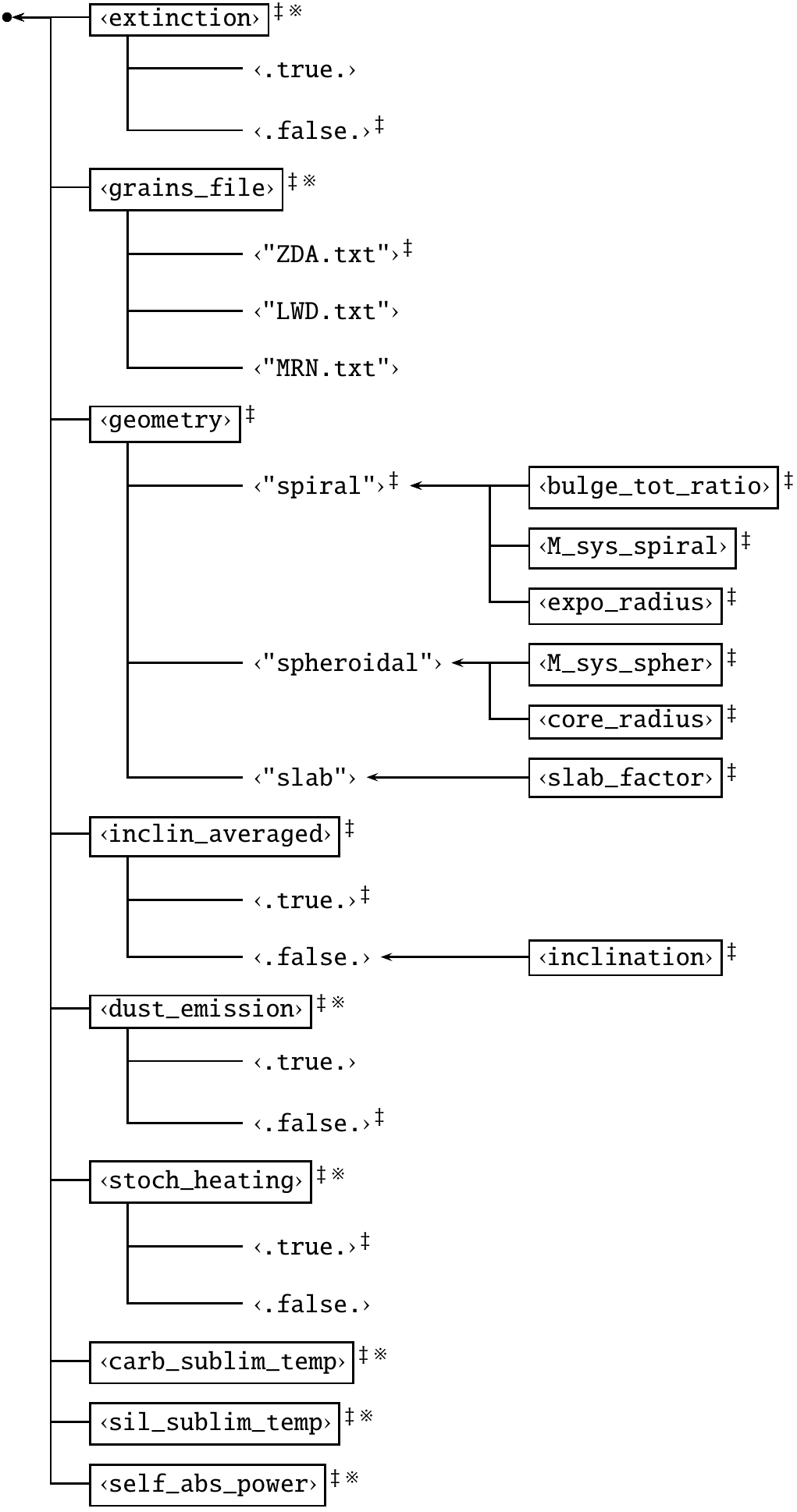}}
      \makeatletter
      \unskip
      \begingroup
      \def\thefootnote{\textreferencemark}%
      \long\def\@makefntext#1{%
        \restoreindent
        \noindent
        \leftskip=\footnotelabelwidth
        \leavevmode\llap{\thefootnote\quad}#1%
      }%
      \noindent\@mpfootnotetext{Explicitely assigning to some value, say \usertext{val},
        a parameter marked with the symbol \character{\thefootnote},
        say \usertext{param}, automatically sets to \usertext{val} the parameters
        \codetext{\usertext{param}_SFC} and \codetext{\usertext{param}_DISM},
        respectively related either to star-forming clouds or to the diffuse medium.
        For instance, the explicit assignment 
        \mention{$\codetext{dust_emission} = \codetext{.true.}$}
        entails both $\codetext{dust_emission_SFC} = \codetext{.true.}$ and
        $\codetext{dust_emission_DISM} = \codetext{.true.}$.

        The parameters \codetext{\usertext{*}_SFC} and \codetext{\usertext{*}_DISM}
        may be assigned directly to define the scenario more finely.
        They are not shown in the tree for the sake of clarity.
        All of them are listed in \fullref{sec:dust_transfer_param}, and in
        Table~\ref{tab:dust_transfer} below.%
      }%
      \endgroup
      \addtocounter{mpfootnote}{-1}%
      \makeatother
    \end{minipage}
    \caption{\label{fig:dust_transfer}%
      Tree of parameters related to the attenuation by dust grains and to their emission\captionPoint}
  \end{figure}
\end{minipage}

\newpage
\saveFootnoteNumber
\begin{longtable}{>{\TBstrut}lcccc}
  \caption{\label{tab:dust_transfer}%
    Parameters related to the attenuation by dust grains and to their emission\captionPoint}
  \\*
  \nobreakhline
  Parameter & Unit & Type & Possible values & Default value
  \\*
  \nobreakhline\nobreakhline
  \endfirsthead
  \multicolumn{5}{@{}c@{}}{$\downarrow$}
  \\*
  \nobreakhline
  Parameter & Unit & Type & Possible values & Default value
  \\*
  \nobreakhline\nobreakhline
  \endhead
  \multicolumn{5}{@{}c@{}}{$\downarrow$}
  \\*
  \endfoot
  \nobreakhline
  \endlastfoot
  \paramname{extinction} & \none & boolean & \codetext{.true.}, \codetext{.false.} & \codetext{.false.}
  \\
  \paramname{extinction_SFC} & $=$ & $=$ & $=$ & $=$
  \\
  \paramname{extinction_DISM} & $=$ & $=$ & $=$ & $=$
  \\
  \hline
  \paramname{grains_file} & \none & string &
  \begin{tabular}[t]{@{}l@{}}
    \Tstrut\codetext{"ZDA.txt"}, \\
    \codetext{"LWD.txt"}, \\
    \codetext{"MRN.txt"}\Bstrut
  \end{tabular}
  & \codetext{"ZDA.txt"}
  \\
  \paramname{grains_file_SFC} & $=$ & $=$ & $=$ & $=$
  \\
  \paramname{grains_file_DISM} & $=$ & $=$ & $=$ & $=$
  \\
  \hline
  \paramname{geometry} & \none & string &
  \begin{tabular}[t]{@{}l@{}}
    \Tstrut\codetext{"spiral"}, \\
    \codetext{"spheroidal"}, \\
    \codetext{"slab"}\Bstrut
  \end{tabular}
  & \codetext{"spiral"}\\
  \hline
  \paramname{inclin_averaged} & \none & boolean & \codetext{.true.}, \codetext{.false.} & \codetext{.true.}
  \\
  \hline
  \paramname{inclination} & degrees & real & $\in\interv[0, 90]$ &  $0$
  \\
  \hline
  \paramname{bulge_tot_ratio} & \none & real & $\in\interv[0, 1]$ & $1/7$
  \\
  \hline
  \paramname{M_sys_spiral} & $\Msol$ & real & $\ge 0$ & $7\times10^{10}$
  \\
  \hline
  \paramname{expo_radius} & parsecs & real & $> 0$ & $3.5\times10^3$
  \\
  \hline
  \paramname{M_sys_spher} & $\Msol$ & real & $\ge 0$ & $2.79\times10^{11}$
  \\
  \hline
  \paramname{core_radius} & parsecs & real & $> 0$ & $192$
  \\
  \hline
  \paramname{slab_factor} & H~atoms/cm$^2$ & real & $\ge 0$ & $6.8\times10^{21}$
  \\
  \hline
  \paramname{dust_emission} & \none & boolean & \codetext{.true.}, \codetext{.false.}& \codetext{.false.}
  \\
  \paramname{dust_emission_SFC} & $=$ & $=$ & $=$ & $=$
  \\
  \paramname{dust_emission_DISM} & $=$ & $=$ & $=$ & $=$
  \\
  \hline
  \paramname{stoch_heating} & \none & boolean & \codetext{.true.}, \codetext{.false.}& \codetext{.true.}
  \\
  \paramname{stoch_heating_SFC} & $=$ & $=$ & $=$ & $=$
  \\
  \paramname{stoch_heating_DISM} & $=$ & $=$ & $=$ & $=$
  \\
  \hline
  \paramname{carb_sublim_temp} & kelvins & real & $> 0$ & $1750$\footnotemark
  \\
  \paramname{carb_sublim_temp_SFC} & $=$ & $=$ & $=$ & $=$
  \\
  \paramname{carb_sublim_temp_DISM} & $=$ & $=$ & $=$ & $=$
  \\
  \hline
  \paramname{sil_sublim_temp} & kelvins & real & $> 0$ & $1400$\footnoteref{fn:sublim_temp}
  \\
  \paramname{sil_sublim_temp_SFC} & $=$ & $=$ & $=$ & $=$
  \\
  \paramname{sil_sublim_temp_DISM} & $=$ & $=$ & $=$ & $=$
  \\
  \hline
  \paramname{self_abs_power} & \none & real & $\in\interv[0, 1]$ & $0$
  \\
  \paramname{self_abs_power_SFC} & $=$ & $=$ & $=$ & $=$
  \\
  \paramname{self_abs_power_DISM} & $=$ & $=$ & $=$ & $=$
  \\
\end{longtable}
\restoreFootnoteNumber
\stepcounter{footnote}
\footnotetext{\label{fn:sublim_temp}%
  \citet{Laor+Draine}, sec.~2.3, 3\textsuperscript{rd}~paragraph.}

\noindent\begin{minipage}{\textwidth}
  \noindent
  \subsection{Star-forming clouds and nebular emission}
  \label{app:cloud_list}

  \begin{RENVOI}
    \renvoi\fullref{sec:cloud_neb}, and \fullref{sec:cloud_neb_param}.
  \end{RENVOI}
  \begin{figure}[H]
    \noindent\begin{minipage}{\textwidth}
      \centerline{\includegraphics{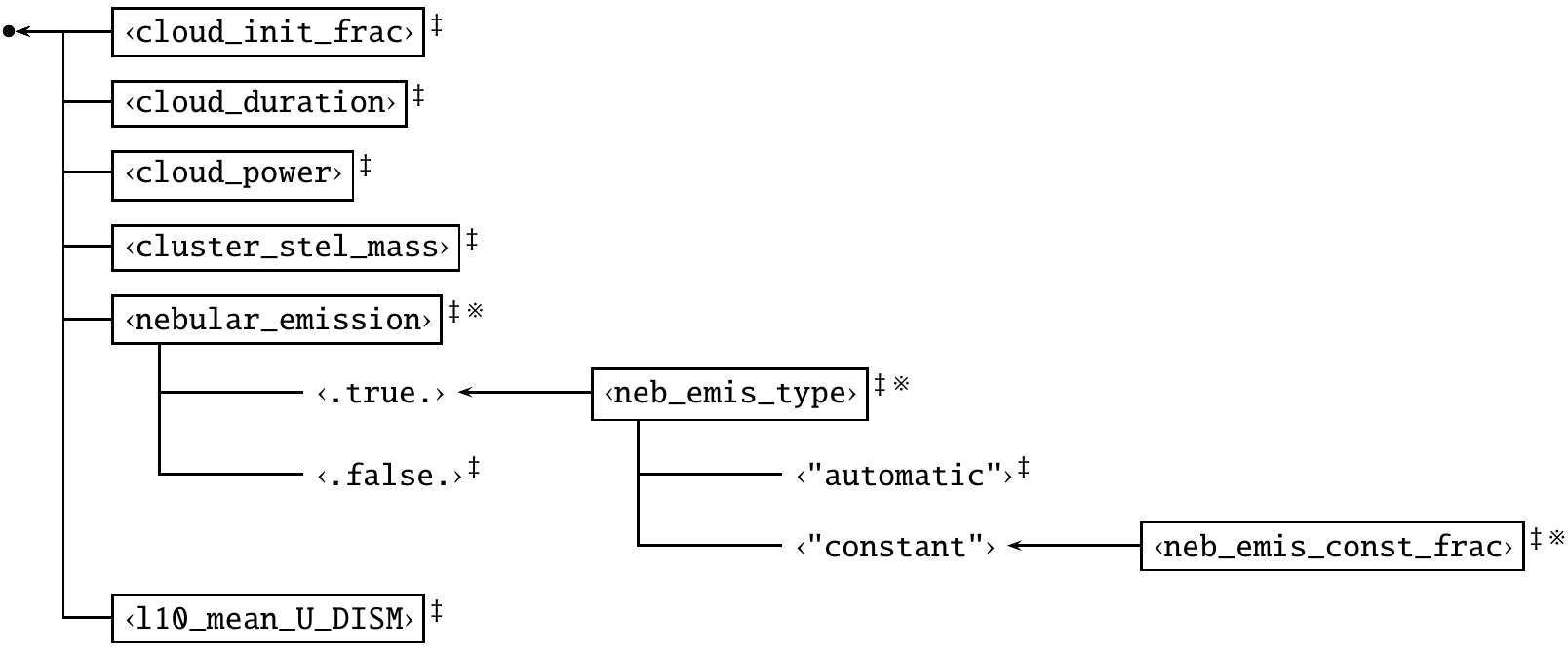}}
      \makeatletter
      \unskip
      \begingroup
      \def\thefootnote{\textreferencemark}%
      \long\def\@makefntext#1{%
        \restoreindent
        \noindent
        \leftskip=\footnotelabelwidth
        \leavevmode\llap{\thefootnote\quad}#1%
      }%
      \@mpfootnotetext{See footnote to \fullref{fig:dust_transfer}.
        The parameters \codetext{\usertext{*}_SFC} and \codetext{\usertext{*}_DISM}
        are listed in \fullref{sec:cloud_neb_param}, and in
        Table~\ref{tab:cloud_neb} below.%
      }%
      \endgroup
      \addtocounter{mpfootnote}{-1}%

      \makeatother
    \end{minipage}
    \caption{\label{fig:cloud_neb_tree}%
      Tree of parameters related to star-forming clouds and nebular emission\captionPoint}
  \end{figure}

\end{minipage}

\saveFootnoteNumber
\begin{longtable}{>{\TBstrut}lcccc}
  \caption{\label{tab:cloud_neb}%
    Parameters for star-forming clouds and nebular emission\captionPoint}
  \\*
  \nobreakhline
  Parameter & Unit & Type & Possible values & Default value
  \\*
  \nobreakhline\nobreakhline
  \endfirsthead
  \multicolumn{5}{@{}c@{}}{$\downarrow$}
  \\*
  \nobreakhline
  Parameter & Unit & Type & Possible values & Default value
  \\*
  \nobreakhline\nobreakhline
  \endhead
  \multicolumn{5}{@{}c@{}}{$\downarrow$}
  \\*
  \endfoot
  \nobreakhline
  \endlastfoot
  \paramname{cloud_init_frac} & \none & real & $\in\interv[0, 1]$ & $1$
  \\
  \hline
  \paramname{cloud_duration} & Myr & real & $\ge 0$ & $10$
  \\
  \hline
  \paramname{cloud_power} & \none & real & $\ge 0$ & $1$
  \\
  \hline
  \paramname{cluster_stel_mass} & $\Msol$ & real & $\ge 0$ & $10^4$
  \\
  \hline
  \paramname{nebular_emission} & \none & boolean & \codetext{.true.}, \codetext{.false.} & \codetext{.false.}
  \\
  \paramname{nebular_emission_SFC} & $=$ & $=$ & $=$ & $=$  
  \\
  \paramname{nebular_emission_DISM} & $=$ & $=$ & $=$ & $=$  
  \\
  \hline
  \paramname{neb_emis_type} & \none & string &
  \begin{tabular}[t]{@{}l@{}}
    \Tstrut\codetext{"automatic"},\\*
    \codetext{"constant"}
    \Bstrut
  \end{tabular}
  & \codetext{"automatic"}
  \\
  \paramname{neb_emis_type_SFC} & $=$ & $=$ & $=$ & $=$  
  \\
  \paramname{neb_emis_type_DISM} & $=$ & $=$ & $=$ & $=$  
  \\
  \hline
  \paramname{neb_emis_const_frac} & \none & real & $\in\interv[0, 1]$ & $0.7$
  \\
  \paramname{neb_emis_const_frac_SFC} & $=$ & $=$ & $=$ & $=$  
  \\
  \paramname{neb_emis_const_frac_DISM} & $=$ & $=$ & $=$ & $=$  
  \\
  \hline
  \paramname{l10_mean_U_DISM} & \none & real & & $-3.5$\footnotemark
  \\
\end{longtable}
\restoreFootnoteNumber
\stepcounter{footnote}
\footnotetext{%
  See \citet{U_DISM} and \citet{Dopita+2006}.}
\subsection{Parameters for output files}
\label{app:output_list}

\begin{longtable}{>{\TBstrut}lcccc}
  \caption{Parameters for output files\captionPoint}
  \\*
  \nobreakhline
  Parameter & Unit & Type & Possible values & Default value
  \\*
  \nobreakhline\nobreakhline
  \endfirsthead
  \multicolumn{5}{@{}c@{}}{$\downarrow$}
  \\*
  \nobreakhline
  Parameter & Unit & Type & Possible values & Default value
  \\*
  \nobreakhline\nobreakhline
  \endhead
  \multicolumn{5}{@{}c@{}}{$\downarrow$}
  \\*
  \endfoot
  \nobreakhline
  \endlastfoot
  \paramname{spectra_output} & \none & string &
  \begin{tabular}[t]{@{}l@{}}
    \Tstrut\codetext{"none"},\\*
    \codetext{"basic"},\\*
    \codetext{"detailed"}\Bstrut
  \end{tabular}
  & \codetext{"basic"}
  \\
  \nobreakhline
  \paramname{RF_output} & \none & boolean & \codetext{.true.}, \codetext{.false.} & \codetext{.false.}
  \\
  \hline
  \paramname{sublim_output} & \none & boolean & \codetext{.true.}, \codetext{.false.} & \codetext{.false.}
  \\
  \hline
  \paramname{spectra_file} & \none & string &
  & See \fullref{sec:spectra_file}.
  \\
  \hline
  \paramname{prefix} & \none & string &
  & \codetext{""}
  \\
  \hline
  \paramname{stamp_time} & \none & boolean & \codetext{.true.}, \codetext{.false.}
  & \codetext{.false.}
  \\
  \hline
  \paramname{overwrite} & \none & boolean & \codetext{.true.}, \codetext{.false.}
  & \codetext{.false.}
  \\
  \hline
  \paramname{ages_file} & \none & string &
  & \codetext{"spectra_ages.txt"}
  \\
  \hline
  \paramname{grain_temp_output} & \none & boolean &
  \codetext{.true.}, \codetext{.false.} & \codetext{.false.}
  \\
  \hline
  \paramname{grain_temp_file} & \none & string &
  & See \fullref{sec:grain_files}.
  \\
  \hline
  \paramname{grain_SED_output} & \none & boolean &
  \codetext{.true.}, \codetext{.false.} & \codetext{.false.}
  \\
  \hline
  \paramname{grain_SED_file} & \none & string &
  & See \fullref{sec:grain_files}.
  \\
  \hline
  \paramname{grain_output_SFC} & \none & boolean &
  \codetext{.true.}, \codetext{.false.} & \codetext{.true.}
  \\
  \hline
  \paramname{grain_output_DISM} & \none & boolean &
  \codetext{.true.}, \codetext{.false.} & \codetext{.true.}
  \\
  \hline
  \paramname{grain_output_min_age} & Myr & string &
  $\ge 0$ & $0$
  \\
  \hline
  \paramname{grain_output_max_age} & Myr & string &
  $\ge 0$ & $+\infty$ (!)
  \\
  \hline
  \paramname{grain_output_min_size} & $\micron$ & string &
  $\ge 0$ & $0$
  \\
  \nobreakhline
  \paramname{grain_output_max_size} & $\micron$ & string &
  $\ge 0$ & $+\infty$ (!)
  \\
\end{longtable}

\subsection{Other parameters}

\begin{longtable}{>{\TBstrut}lcccc}
  \caption{%
    \label{tab:other_param}%
    Other parameters\captionPoint}
  \\*
  \nobreakhline
  Parameter & Unit & Type & Possible values & Default value
  \\*
  \nobreakhline\nobreakhline
  \endfirsthead
  \multicolumn{5}{@{}c@{}}{$\downarrow$}
  \\*
  \nobreakhline
  Parameter & Unit & Type & Possible values & Default value
  \\*
  \nobreakhline\nobreakhline
  \endhead
  \multicolumn{5}{@{}c@{}}{$\downarrow$}
  \\*
  \endfoot
  \nobreakhline
  \endlastfoot
  \paramname{verbosity} & \none & integer & See \fullref{sec:verbosity}. & $0$
  \\
  \hline
  \paramname{check_only} & \none & boolean
  & \codetext{.true.}, \codetext{.false.}& \codetext{.false.}
  \\
  \hline
  \paramname{seed} & \none & array of integers & \multicolumn{2}{c}{See \fullref{sec:random}.}
  \\
\end{longtable}
\end{appendices}
\clearpage
\phantomsection
\bibliographystyle{aa}
\bibliography{references,references2}
\clearpage
\phantomsection
\printindex
\clearpage
\bgroup
\protected\def\captionPoint{}%
\phantomsection
\listoffigures
\phantomsection
\listoftables
\egroup
\phantomsection
\addcontentsline{ptc}{section}{Detailed table of contents}
\bgroup
\makeatletter
\def\notInToc#1{}%
\makeatother
\tableofcontents
\egroup
\clearpage
\stopcontents[sommaire]
\addcontentsline{toc}{section}{A galactic study of long ago}
\begin{figure}[H]
  \begin{center}
    \includegraphics[bb=60 250 1200 1600, width=0.8\textwidth, clip]{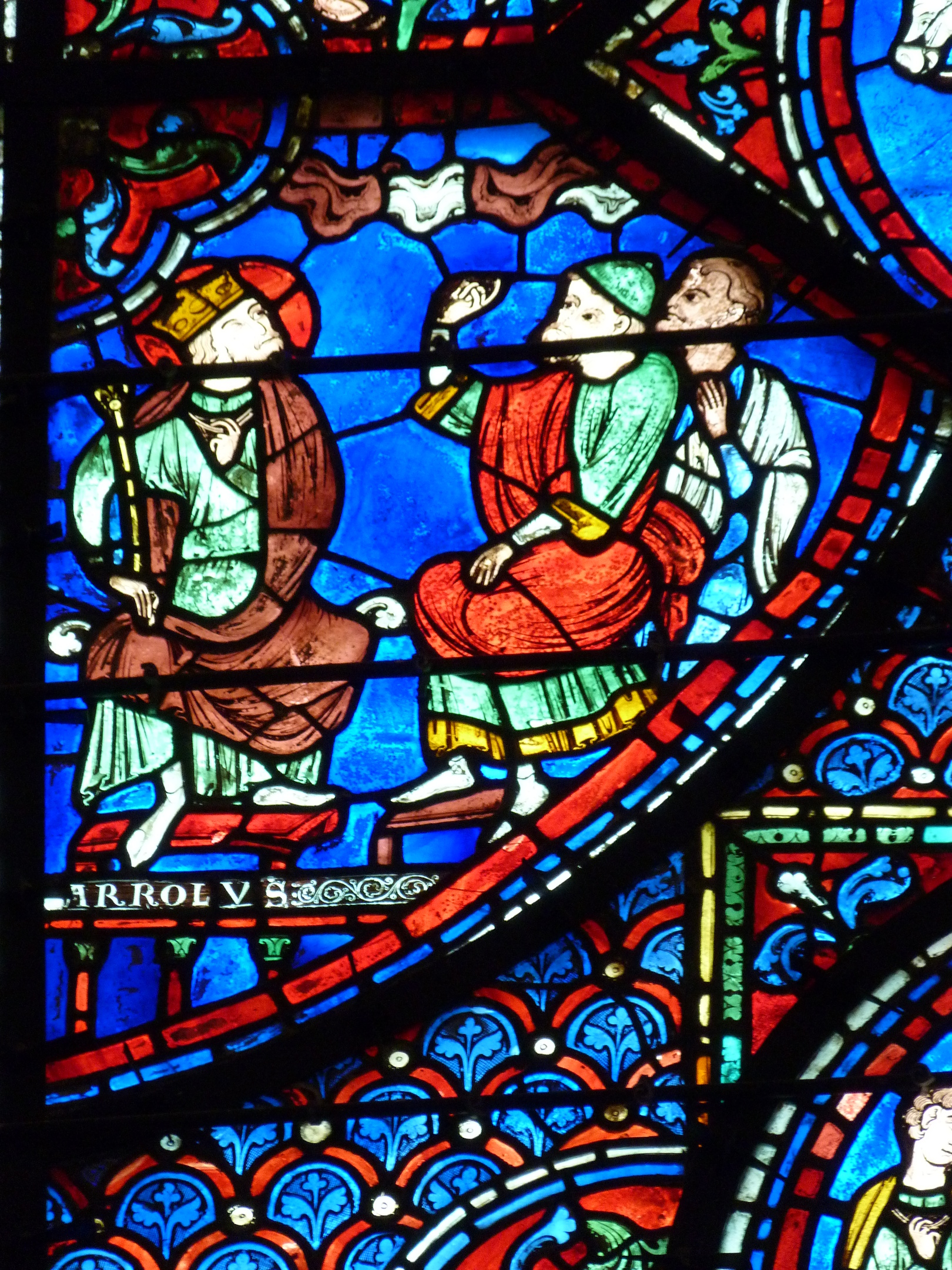}
    \caption[Charlemagne's advisers show him the Milky Way\captionPoint]{%
      Charlemagne's advisers show him the Milky Way (cathedral of Chartres, 
      13\textsuperscript{th} century)\captionPoint}
  \end{center}
\end{figure}
\vfill
\prgsymb
\vfill
\begin{center}
  \emph{Ce travail est dédié à mes filles, les plus belles œuvres auxquelles
  j'ai contribué.}
\end{center}
\begin{flushright}
  M.~F.
\end{flushright}
\vfill
\resumecontents[sommaire]
\end{document}